\documentclass[preprint, review, sort&compress, 10pt]{elsarticle}

\usepackage{amssymb}
\usepackage{amsthm}
\usepackage{graphicx}
\usepackage{subcaption}
\usepackage{amsmath}
\usepackage{amsfonts}
\usepackage{booktabs}
\usepackage{algorithm}
\usepackage{algpseudocode}
\usepackage{caption}
\usepackage{microtype}
\usepackage{csquotes}
\usepackage{xcolor}
\usepackage{tikz}
\usetikzlibrary{arrows.meta,positioning,shapes.geometric,fit,calc}

\definecolor{femblue}{RGB}{226,238,252}
\definecolor{passcgreen}{RGB}{228,244,232}
\definecolor{lossorange}{RGB}{252,239,218}
\definecolor{outputgray}{RGB}{241,243,245}
\definecolor{hiddenblue}{RGB}{210,229,248}
\definecolor{linegray}{RGB}{90,96,102}
\usepackage{hyperref}
\usepackage{cleveref}
\usepackage{comment}
\usepackage{color}
\usepackage{microtype}
\usepackage[margin=1.90cm]{geometry}

\theoremstyle{remark}
\newtheorem{remark}{Remark}

\begin{document}
	
	\begin{frontmatter}
		
		
		
		\title{A physics-informed SUPG-stabilized finite element framework with shock-capturing for simulating inviscid high-speed flows around a cylinder}

		\author[Slymn1]{S\"uleyman Cengizci\corref{cor1}}
		\address[Slymn1]{Department of Computer Technologies, Antalya Bilim University, Antalya 07190, Turkey}
		\cortext[cor1]{Corresponding author}
		\ead{suleyman.cengizci@antalya.edu.tr}
		
		
		
		\author[omur]{Ömür Uğur}
		\address[omur]{Institute of Applied Mathematics, Middle East Technical University, Ankara 06800, Turkey}
		\ead{ougur@metu.edu.tr}
		

		
		\begin{abstract}
			This study presents a hybrid computational framework for simulating
			non-reacting inviscid high-speed flows of nitrogen gas (N$_2$) around a
			circular cylinder. Owing to the strongly convection-dominated nature of the
			compressible Euler equations, the compressible-flow
			streamline-upwind/Petrov--Galerkin (SUPG) formulation is combined with the
			YZ$\beta$ shock-capturing technique to stabilize the finite element
			discretization in the presence of strong discontinuities. Building upon the
			stabilized solution, a physics-informed neural network (PINN) is employed as a
			post-processing correction stage
			(\underline{P}INN-\underline{A}ugmented \underline{S}UPG with
			\underline{S}hock-\underline{C}apturing---PASSC). The network is anchored to
			the finite element solution through a shock-weighted data-consistency loss,
			while the governing equations are enforced in a conservative space--time
			control-volume form supplemented by macroscopic conservation windows, an
			entropy-admissibility penalty, and the boundary conditions of the underlying
			problem. Two-dimensional simulations are performed for free-stream Mach
			numbers ranging from $2.0$ to $12.0$, and the results are assessed against
			analytical normal-shock and stagnation relations, the semi-empirical Billig
			correlation, and reference solutions from the literature. The correction is designed to improve the numerical representation of shocks by reducing localized discretization-induced oscillations and mesh-scale serrations while preserving the agreement of the stabilized solution with the analytical and semi-empirical reference quantities. Across the Mach-number range considered, the correction remains closely
			anchored to the stabilized solution while substantially reducing
			discretization-induced oscillatory features and preserving the principal
			shock and global flow structures. At the three hypersonic Mach numbers, for
			which the stabilized stand-off distances already lie within approximately
			$2\%$ of the Billig correlation, the correction introduces only small
			absolute changes in the detected shock position. The stagnation-pressure
			ratios obtained with both approaches remain within approximately $2.5\%$ of
			the corresponding Rayleigh pitot values over the complete Mach-number range,
			with the correction reducing the stagnation-pressure discrepancy from
			approximately $1.6\%$ to $0.7\%$ at $M_\infty=5.0$.
			For the fixed network architecture and training budget employed here, the
			observed PASSC training time is approximately independent of the Mach number,
			corresponding to about $43\%$ of the measured finite element wall-clock cost
			at $M_\infty=2.0$ and about $8\%$ of the estimated finite element cost at
			$M_\infty=12.0$. The full
			implementation, including the stabilized FEM solver, the correction network,
			and reproduction scripts, is publicly available.
		\end{abstract}
		
		
		
		
		
	\end{frontmatter}
	
	\section{Introduction}
	High-speed flows, including supersonic and hypersonic regimes, are central to
	many aerospace and defense applications. Although there is no universally
	accepted threshold separating supersonic and hypersonic conditions, flows at
	approximately five times the speed of sound and above are commonly classified
	as hypersonic. At such speeds, strong shock waves, severe aerodynamic heating,
	and high-temperature gas effects become increasingly important and must be
	accounted for in the design and analysis of high-speed vehicles. Experimental
	investigation of these regimes is challenging and costly, particularly because
	of the limitations associated with hypersonic wind-tunnel testing. Consequently,
	computational fluid dynamics (CFD) has become an essential tool for studying
	the complex flow physics encountered under supersonic and hypersonic
	conditions. For a comprehensive discussion of hypersonic aerodynamics and
	high-temperature gas dynamics, see~\cite{Anderson2019}.
	
	For more than the last four decades, the methods used in CFD and reported in the literature have mostly been stabilized methods. Standard discretization methods yield numerical instabilities (spurious oscillations) in the simulation of flows with high Reynolds and Mach numbers and in the presence of shocks, necessitating stabilized formulations. One of the most established stabilized finite element methods is the
	streamline-upwind/Petrov--Galerkin (SUPG) formulation. The method was first introduced for incompressible flow computations in a 1979 ASME paper~\cite{Hughes79a}. Later, in 1982, the work was extended with further studies and examples by the same authors and published as a journal article~\cite{Brooks82a}. The compressible version of the SUPG formulation, which is called the \enquote{compressible-flow SUPG,} was first introduced in the context of conservation variables in a NASA technical report~\cite{Tezduyar82a} in 1982 and published as an AIAA paper~\cite{Tezduyar83a} the following year. In 1984, a more comprehensive version, with additional examples, was published in a journal article~\cite{Hughes84a}. Afterward, various compressible-flow methods similar to the SUPG were developed, such as the Taylor--Galerkin method~\cite{Donea84a} and the streamline-diffusion method~\cite{Johnson84a}. The compressible-flow SUPG method introduced in 1982 is now called \enquote{$\text{(SUPG)}_{82}$.} 
	
	The $\text{(SUPG)}_{82}$ formulation was initially used without a
	shock-capturing mechanism, and test simulations revealed the need for special
	treatment in regions containing shocks. In 1987~\cite{Hughes87a}, the
	$\text{(SUPG)}_{82}$ formulation was reformulated in terms of the entropy
	variables and augmented with a shock-capturing term, yielding more satisfactory
	results. In an ASME paper~\cite{LeBeau91a} published in 1991, the
	$\text{(SUPG)}_{82}$ formulation was supplemented with a shock-capturing term
	very similar to that introduced in~\cite{Hughes87a}; the added term included a
	shock-capturing parameter now denoted by $\delta_{91}$. The set of stabilization
	parameters, denoted by $\tau$, used with the $\text{(SUPG)}_{82}$ formulation
	introduced in~\cite{Tezduyar82a,Tezduyar83a,Hughes84a}, is now denoted by
	$\tau_{82}$. The stabilized formulation for the advection--diffusion--reaction
	equation introduced in~\cite{Tezduyar86a} included a shock-capturing term and
	a stabilization parameter that accounted for the interaction between the
	shock-capturing and SUPG terms. Thus, the shock-capturing mechanism does not
	increase the SUPG stabilization when the advection and shock directions
	coincide. In~\cite{LeBeau91a}, the definition of $\tau_{82}$ was slightly
	modified. Although the $\text{(SUPG)}_{82}$ formulation underwent some minor
	modifications in subsequent years, it was typically used with the same
	shock-capturing parameter, $\delta_{91}$, until 2004.
	
	In 2004, new strategies for determining the stabilization and shock-capturing parameters in the $\text{(SUPG)}_{82}$ were introduced by Tezduyar~\cite{Tezduyar02j, Tezduyar04m}. These new stabilization parameters are today referred to as \enquote{$\tau_{04}$.} The new shock-capturing parameters introduced can be divided into two categories: the style of discontinuity-capturing directional dissipation (DCDD)~\cite{Tezduyar02a, Tezduyar04m, Rispoli05a} and the residual-based YZ$\beta$~\cite{Tezduyar02j, Tezduyar04m} shock-capturing. We focus on the YZ$\beta$ technique throughout this work, and some of its highlights include that it is easier to calculate the YZ$\beta$ shock-capturing parameter than $\delta_{91}$, the sharpness parameter $\beta$, embedded in the definition, offers flexibility for computing flows with mild or sharp shocks, and the technique yields more accurate results than $\delta_{91}$ does. Further details can be found in~\cite{Tezduyar04o, Tezduyar05d, Tezduyar05f}.

	Studies on hypersonic aerodynamics date back to the late 1940s, and a
	substantial body of experimental, theoretical, and computational work has
	since accumulated on the subject. Interest in CFD, particularly in high-speed
	flow research, increased significantly with the introduction of the stabilized
	formulations briefly reviewed in the preceding paragraphs, which originated in
	the late 1970s for incompressible flows and were extended to the
	compressible-flow setting in the early 1980s. The compressible-flow SUPG
	formulation based on entropy variables was introduced in~\cite{Hughes86} and
	was subsequently applied to the aerothermal design of hypersonic
	vehicles~\cite{mallet88} and to turbulent and thermochemical nonequilibrium
	flows~\cite{chalot92}. The compressible-flow SUPG methods based on the
	conservation and entropy variables were compared in subsonic, transonic, and
	supersonic flows in~\cite{LeBeau92a}, where the two formulations were found to
	yield nearly identical results. The compressible-flow SUPG formulation
	of~\cite{LeBeau91a} was employed in parallel computations
	in~\cite{Tezduyar93a,Tezduyar94b}, and was later combined with the
	Enhanced-Discretization Interface-Capturing Technique (EDICT)~\cite{Tezduyar97a}
	for more accurate shock representations~\cite{Mittal97b}. Subsequently, in a
	series of papers, Kirk and co-workers~\cite{kirk1,kirk2,kirk5} examined
	SUPG-based stabilized formulations for hypersonic flow simulations, including
	thermochemical nonequilibrium flows with the Spalart--Allmaras turbulence
	closure~\cite{kirk7}.

	In more recent years, the compressible-flow SUPG method was used in several challenging computations. In~\cite{Xu17a}, its pressure-primitive variables version, together with a shock-capturing term and the Arbitrary Lagrangian--Eulerian (ALE) method, was developed and applied to the simulation of a single-stage gas turbine. Weakly-enforced essential boundary conditions and sliding-interface formulations for the compressible-flow equations were added in~\cite{Xu17a} to enable gas turbine simulations with boundary-layer flow and stator--rotor interaction. Within the FUN3D flow solver, a SUPG-stabilized finite element discretization was implemented as a linkable library for solving a variety of flow computations, including hypersonic flows~\cite{Anderson2018}. In~\cite{Kozak20a}, the compressible-flow SUPG formulation was applied to study the off-design performance of the gas turbine stage, and in~\cite{Xu19a}, it was used to simulate the full-rotorcraft aerodynamics. In~\cite{Bazilevs20b}, based on the developments in~\cite{Xu17a}, improvements were introduced to the shock-capturing term, weakly-enforced essential boundary conditions, and sliding-interface operators to increase the accuracy and robustness of the ALE-based compressible-flow formulation, especially when used in combination with Isogeometric Analysis (IGA)~\cite{Hughes05a,Bazilevs06a}. In~\cite{Korobenko21a}, the formulation was again employed together with a shock-capturing mechanism for high-speed flow simulations. The methodology was subsequently extended to hypersonic thermochemical nonequilibrium flows using a residual-based shock-capturing operator and a five-species, two-temperature air model~\cite{Codoni2022}. More recently, SUPG formulations have also been employed for heat-flux prediction in hypersonic flows~\cite{Codoni2023} and for assessing the performance of the Spalart--Allmaras turbulence model within a pressure-primitive stabilized finite element framework for hypersonic turbulent flows~\cite{Verma2026}. Apart from the SUPG-based formulations, an edge-based finite element
	discretization was proposed for simulating a wide range of flows,
	including two- and three-dimensional thermochemical nonequilibrium
	computations~\cite{Seguin2019,Gao2019}. In the discontinuous Galerkin
	(DG) setting, a GPU-accelerated high-order implicit DG method with
	shock capturing was developed for large eddy simulation of hypersonic
	flows~\cite{Terrana2020}, and shock-capturing strategies for DG
	approximations of hypersonic flows in thermochemical nonequilibrium
	have also been investigated~\cite{VanHeyningen2023shock}; a recent
	review of DG methods for hypersonic flows can be found
	in~\cite{Hoskin2024}.

	In parallel with the remarkable progress achieved in stabilized finite element methods, recent years have witnessed rapid advances in scientific machine learning for the numerical solution of partial differential equations (PDEs). Among these developments, Physics-Informed Neural Networks (PINNs), originally introduced by Raissi et al.~\cite{Raissi2019}, have emerged as one of the most influential approaches. Unlike conventional supervised neural networks, PINNs incorporate the governing differential equations directly into the loss function through automatic differentiation, thereby enabling the approximation of PDE solutions while simultaneously satisfying the underlying physical laws together with the prescribed initial and boundary conditions. Owing to their mesh-free nature and flexibility, PINNs have been successfully applied to a broad spectrum of forward and inverse problems arising in computational science and engineering~\cite{Karniadakis2021, Mitusch2021, Grossmann2024, Jin2021}. In the high-speed flow context in particular, physics-informed and
	physics-constrained networks have recently been employed for predicting
	hypersonic flow fields and aerodynamic heating, where positional
	encodings are introduced to improve the representation of shocks and
	other steep gradients~\cite{Dai2025}, for the real-time reconstruction
	of transient temperature fields in hypersonic thermal protection
	systems~\cite{Wei2026}, for surrogate modeling of rarefied gas dynamics,
	including hypersonic flow around a cylinder~\cite{Roohi2026}, and for
	augmenting continuum transport closures and wall models in
	transition-continuum hypersonic regimes~\cite{Nair2026}.
	
	Despite these encouraging developments, standalone PINNs still encounter
	considerable challenges when applied to strongly convection-dominated
	problems involving thin boundary layers, shocks, and other sharp
	solution features. Since the network is required to learn the entire
	solution manifold directly from the governing equations, the
	optimization process often becomes computationally demanding and may
	require prohibitively large numbers of training epochs before accurately
	resolving discontinuities. Moreover, the well-known spectral bias of
	deep neural networks tends to favor smooth solution components, making
	the accurate representation of steep gradients considerably more
	difficult than in classical numerical methods; although remedies such as
	high-frequency positional encodings~\cite{Dai2025} alleviate this
	tendency, resolving strong shocks remains substantially harder for
	standalone networks than for shock-capturing discretizations.
	Consequently, although PINNs constitute a powerful computational
	paradigm, they are not yet regarded as a practical replacement for
	mature stabilized finite element formulations for challenging high-speed
	compressible flow simulations.
	
	On the other hand, stabilized finite element methods such as the SUPG-YZ$\beta$ formulation provide numerically robust and physically meaningful approximations even for highly convection-dominated flows. Their principal limitation is not numerical instability, but rather the additional artificial dissipation intentionally introduced to suppress spurious oscillations. While this stabilization successfully eliminates nonphysical oscillations, it may also smear shocks and other localized flow structures, particularly on relatively coarse meshes. Recovering these sharp solution features using mesh refinement alone generally requires substantially finer spatial and temporal discretizations, resulting in significantly increased computational cost. Motivated by these complementary strengths and limitations, an attractive alternative is not to replace stabilized finite element methods with PINNs, but rather to combine both methodologies within a unified computational framework. In such a hybrid strategy, the stabilized FEM solution provides
	a reliable, substantially oscillation-free approximation of the governing
	equations, whereas the PINN is anchored to this stabilized solution through
	a data-consistency loss and refines it under the constraints of the governing
	equations and boundary conditions. Consequently, instead of learning the
	solution from the governing equations and associated constraints alone, the
	neural network remains closely tied to the stabilized approximation and
	focuses on reducing the excessive numerical diffusion introduced by the
	stabilization procedure. This substantially simplifies the learning task by
	avoiding the need for the network to reconstruct the global flow structure
	from scratch.
	
	Building upon our recently introduced PASSC (\underline{P}INN-\underline{A}ugmented \underline{S}UPG with \underline{S}hock-\underline{C}apturing) framework for steady convection-dominated transport problems~\cite{Cengizci2026_pinn_stationary} and its subsequent extension to transient convection-dominated problems~\cite{cengizci_pinn_transient}, the present study further extends the PASSC methodology to the numerical simulation of supersonic and hypersonic inviscid flows governed by the compressible Euler equations. To that end, the flow is assumed to consist solely of nitrogen gas, while chemical reactions and viscous effects are neglected. Owing to the strongly convection-dominated nature of the compressible Euler
	equations, a SUPG formulation enhanced with the YZ$\beta$ shock-capturing
	technique is employed to obtain numerically stable and substantially
	oscillation-free approximations capable of capturing strong shocks. Although this stabilized formulation effectively suppresses spurious oscillations, the artificial dissipation introduced for stabilization may smear shocks and other localized flow structures, particularly on relatively coarse meshes. To alleviate this drawback, the stabilized finite element solution is further enhanced through a physics-informed neural network acting as a post-processing corrector. Rather than being trained from scratch on the governing equations alone, the PINN is anchored to the SUPG-YZ$\beta$ solution through a data-consistency loss while enforcing the governing equations and boundary conditions. Consequently, the proposed hybrid framework aims to recover sharper shock
	profiles and reduce the excessive numerical diffusion introduced by the
	stabilization procedure while remaining anchored to the underlying stabilized
	finite element approximation.

	The main objectives and contributions of this study include the following:
	1) To develop a hybrid PINN-assisted SUPG-YZ$\beta$ FEM framework for
	physics-informed post-processing of inviscid high-speed compressible flow
	simulations;
	2) To investigate the ability of the physics-informed post-processing
	correction to improve the representation of shocks---suppressing localized
	spurious oscillations and excessive numerical diffusion---while preserving,
	and where possible improving, the accuracy of the stabilized approximation
	with respect to analytical and semi-empirical reference quantities;
	3) To demonstrate that the open-source FEniCS and PyTorch ecosystems can be
	seamlessly coupled to implement advanced stabilized finite element
	formulations and hybrid FEM--PINN methodologies; and
	4) To provide a fully open implementation of the proposed framework,
	including the stabilized FEM solver, the correction network, and the
	reproduction scripts, so that the reported results can be independently
	verified and the framework reused.

	
	\section{Governing equations}
	\label{sec2}
	Consider a spatial domain $\Omega\subset \mathbb{R}^{n_{\text{sd}}}$ with boundary $\Gamma$ over a time interval $I_{t}=[0,t_\text{{f}}]$, where $n_\text{sd}$ denotes the spatial dimension, and $t_\text{{f}}$ is the final time. The Euler equations of compressible flows can be given as follows:
	\begin{align}
		\frac{\partial \rho}{\partial t} + \nabla \cdot \left(\rho\mathbf{u} \right) &=  0 &\text{in} \enspace \Omega \times I_{t}, \label{mass} \\
		\frac{\partial \left( \rho \mathbf{u}\right)}{\partial t} + \nabla \cdot \left(\rho \mathbf{u} \otimes \mathbf{u} \right)+ \nabla p &=  \mathbf{0} &\text{in} \enspace \Omega \times I_{t}, \label{mom} \\
		\frac{\partial \left( \rho e\right)}{\partial t} + \nabla \cdot \left(\rho \mathbf{u} h\right) &=  0 & \quad  \text{in} \enspace \Omega \times I_{t},  \label{energy} 
	\end{align}
	where $\mathbf{U}=[\rho,\rho\mathbf{u},\rho e]^{T}$ is the vector of conservation variables, and $\rho$ is the fluid density. In the case of $n_\text{sd}=2$, the velocity vector can be given as $\mathbf{u}=[u_{1},u_{2}]^{T}$. The terms $e$ and $h$ denote the specific total energy and specific total enthalpy, respectively, and are defined as
	\begin{equation} \label{totener}
		e=e_\text{int}+\frac{1}{2}\Vert\mathbf{u}\Vert^{2},
	\end{equation}
	\begin{equation} \label{totenth}
		h=e+\frac{p}{\rho},
	\end{equation}
	where $e_\text{int}=c_{v}T$. The first term on the right-hand side of Eq.~\eqref{totener}, i.e., $e_\text{int}$, denotes the internal energy, and the second term is the kinetic energy. With $T$ representing the absolute temperature, the pressure, $p$, is given as follows: 
	\begin{equation}
		p=\left(\gamma-1 \right)\rho e_\text{int}=\rho R_{\text{N}_{2}} T, 
	\end{equation}
	where $\gamma= c_{p}/c_{v}$ is the ratio of specific heats at constant pressure and volume, assumed to be given as $\gamma=1.4$, and $R_{\text{N}_{2}}=296.8$ J/(kg$\cdot$K) is the ideal gas constant for nitrogen. For completeness, the specific heats are defined as $c_{v}=R_{\text{N}_{2}}/(\gamma-1)$ and $c_{p}=\gamma R_{\text{N}_{2}}/(\gamma-1)$.  Then, the temperature is
	\begin{equation}
		T= \frac{\gamma-1}{{R_{\text{N}_{2}}}}e_\text{int}=\frac{\gamma-1}{R_{\text{N}_{2}}}\left(e- \frac{1}{2}\Vert \mathbf{u}\Vert^{2}\right),
	\end{equation}
	the speed of sound ($c$) satisfies the following equation
	\begin{equation} \label{speed}
		c^{2}=\gamma R_{\text{N}_{2}} T,
	\end{equation}
	and the Mach number is defined as
	\begin{equation}
		M= \frac{\Vert \mathbf{u}\Vert}{c},
	\end{equation}
	where the norm $\Vert \cdot \Vert$ corresponds to the standard Euclidean norm throughout the manuscript.
	
	The system given by Eqs.~\eqref{mass}--\eqref{energy} can be expressed in a more compact way, in a vector form in 2D, as follows:
	\begin{equation} \label{compact}
		\frac{\partial \mathbf{U}}{\partial t} +  \frac{\partial \mathbf{F}_{1}}{\partial x_{1}} + \frac{\partial \mathbf{F}_{2}}{\partial x_{2}} = \mathbf{S}.
	\end{equation}
	The vector of conservation variables and the flux vectors are defined as
	\begin{equation}
		\mathbf{U}=\begin{bmatrix}
			\rho \\ \rho u_{1} \\ \rho u_{2} \\ \rho e
		\end{bmatrix}, \quad
		\mathbf{F}_{i}=\begin{bmatrix}
			\rho u_{i} \\ \rho u_{i}u_{1} + \delta_{i1} p \\ \rho u_{i}u_{2} + \delta_{i2} p  \\ \rho u_{i}h 
		\end{bmatrix},
	\end{equation}
	where $i=1,2$, and $\delta_{ij}$ are the components of the identity matrix $\mathbf{I}$. The source/sink vector $\mathbf{S}$ is set to a zero-vector throughout this study, since no source terms or chemical reactions are considered. Then, Eq.~\eqref{compact} can be given in a quasi-linear form as follows:
	\begin{equation}
		\label{eulersystem}
		\frac{\partial \mathbf{U}}{\partial t} + \mathbf{A}_{1} \frac{\partial \mathbf{U}}{\partial x_{1}} + \mathbf{A}_{2}\frac{\partial \mathbf{U}}{\partial x_{2}} = \mathbf{0},
	\end{equation}
	where the matrices $\mathbf{A}_{1}$ and $\mathbf{A}_{2}$ are defined as
	\begin{equation} \label{Jacob}
		\mathbf{A}_{1}= \frac{\partial \mathbf{F}_{1}}{\partial \mathbf{U}}, \quad
		\mathbf{A}_{2}= \frac{\partial \mathbf{F}_{2}}{\partial \mathbf{U}}.
	\end{equation}
	These matrices are explicitly given as~\cite{Tezduyar82a, Hughes84a}
	\begin{equation}
		\label{JacobianAx}
		\mathbf{A}_{1} =  
		\begin{bmatrix}
			0 & 1 & 0 & 0 \\
			(\gamma-1)\Vert \mathbf{u} \Vert^{2}/2-u_{1}^{2} & (3-\gamma)u_{1} & (1-\gamma)u_{2} & \gamma-1 \\
			-u_{1}u_{2} & u_{2} & u_{1} & 0 \\
			(\gamma-1)\Vert \mathbf{u} \Vert^{2}u_{1}/2-u_{1}h & h-(\gamma-1)u_{1}^{2} & (1-\gamma)u_{1}u_{2} & \gamma u_{1}
		\end{bmatrix}, 
	\end{equation}
	\begin{equation}
		\label{JacobianAy}
		\mathbf{A}_{2} =  
		\begin{bmatrix}
			0 & 0 & 1 & 0 \\
			-u_{1}u_{2} & u_{2} &  u_{1} & 0 \\
			(\gamma-1)\Vert \mathbf{u} \Vert^2/2-u_{2}^{2} & (1-\gamma)u_{1} & (3-\gamma)u_{2} & \gamma-1 \\
			(\gamma-1)\Vert \mathbf{u} \Vert^2u_{2}/2-u_{2}h & (1-\gamma)u_{1}u_{2} & h-(\gamma-1) u_{2}^{2} & \gamma u_{2}
		\end{bmatrix}.
	\end{equation}
	
	The boundary and initial conditions associated with Eq.~\eqref{eulersystem} are in the form of $\mathbf{U}\left(\mathbf{x},t\right)=\mathbf{G}\left(\mathbf{x},t\right)$ for an inflow boundary, $\mathbf{u} \cdot \mathbf{n} = 0$ for a slip surface, and $\mathbf{U}\left(\mathbf{x},0\right) = \mathbf{U}_{0}$, where $\mathbf{G}$ and $\mathbf{U}_{0}$ are given functions. Here, $\mathbf{n}=[n_1, n_2]$ is the outward-oriented unit normal vector.
	
	\section{Compressible-flow SUPG formulation and YZ$\beta$ shock-capturing}
	\label{sec3}
	
	The discrete unknowns are the primitive variables, collected in the vector
	\begin{equation}
		\mathbf{Q}=\left[\rho,\,\mathbf{u},\,T\right]^{T},
	\end{equation}
	and approximated with continuous, piecewise-linear mixed finite elements on triangular meshes. The conservative variables and the Euler fluxes are evaluated from the
	primitive unknowns, i.e., $\mathbf{U}=\mathbf{U}\!\left(\mathbf{Q}\right)$
	and $\mathbf{F}_{i}=\mathbf{F}_{i}\!\left(\mathbf{Q}\right)$, through the
	constitutive relations given in Section~\ref{sec2}. The finite-dimensional trial function set and test function space are
	defined as follows:
	\begin{align}
		\mathcal{S}^{h}_{\mathbf{Q}} &= \{\mathbf{Q}^{h} : \mathbf{Q}^{h}\in
		[\mathcal{H}^{1h}(\Omega)]^{n_\text{sd}+2}, \;
		\mathbf{Q}^{h} \doteq \mathbf{Q}_{\infty} \enspace \text{on} \enspace
		\Gamma_{\text{IN}}\},\\
		\mathcal{V}^{h}_{\mathbf{Q}} &= \{\mathbf{W}^{h} : \mathbf{W}^{h}\in
		[\mathcal{H}^{1h}(\Omega)]^{n_\text{sd}+2}, \;
		\mathbf{W}^{h} \doteq \mathbf{0} \enspace \text{on} \enspace
		\Gamma_{\text{IN}}\},
	\end{align}
	where $\Gamma_{\text{IN}}\subset\Gamma$ denotes the supersonic/hypersonic inflow boundary, on which all primitive variables are prescribed strongly and set to the free-stream values collected in $\mathbf{Q}_{\infty}$ (see Section~\ref{sec5}). Here, the symbol \enquote{$\doteq$} indicates that the essential boundary condition is imposed in the interpolation (nodal) sense, following the convention of~\cite{Tezduyar91c}. The finite element space $\mathcal{H}^{1h}$ is defined by
	\begin{equation} \label{femspace}
		\mathcal{H}^{1h}(\Omega)=\{\Phi^{h} : \Phi^{h} \in \mathcal{C}^{0}(\overline{\Omega}), \; \Phi^{h}|_{\Omega^{e}} \in \mathcal{P}_{1}(\Omega^{e}), \; \forall \Omega^{e} \in \mathcal{T}^{h} \}.
	\end{equation}
	Here, $\mathcal{P}_{1}(\Omega^{e})$ represents the set of linear polynomials
	over element $\Omega^{e}$, $\mathcal{T}^{h}$ is the set of (triangular)
	elements resulting from the finite element discretization (triangulation) of
	the computational domain $\Omega$, and $\mathcal{C}^{0}(\overline{\Omega})$
	is the space of continuous functions on $\overline{\Omega}$. The triangulation is assumed to be conforming, with
	\[
	\overline{\Omega}
	=
	\bigcup_{\Omega^e\in\mathcal{T}^h}\overline{\Omega^e},
	\qquad
	\operatorname{int}(\Omega^e)\cap
	\operatorname{int}(\Omega^{e'})=\varnothing
	\quad\text{for } e\neq e'.
	\]
	
	The Galerkin part of the formulation is written in the conservative (divergence) form and integrated by parts, so that the boundary integrals carry the physical Euler normal fluxes. The semi-discrete SUPG formulation of Eq.~\eqref{eulersystem}, complemented
	with YZ$\beta$ shock-capturing, then reads as follows: find
	$\mathbf{Q}^{h} \in \mathcal{S}^{h}_{\mathbf{Q}}$ such that for all test functions
	$\mathbf{W}^{h} \in \mathcal{V}^{h}_{\mathbf{Q}}$,
	\begin{multline} \label{supg}
		\int_{\Omega} \mathbf{W}^{h} \cdot \frac{\partial \mathbf{U}^{h}}{\partial t}\,d\Omega
		- \int_{\Omega} \left( \mathbf{F}^{h}_{1}\cdot\frac{\partial \mathbf{W}^{h}}{\partial x_{1}} + \mathbf{F}^{h}_{2}\cdot\frac{\partial \mathbf{W}^{h}}{\partial x_{2}}\right) d\Omega
		+ \int_{\Gamma\setminus\Gamma_{\text{CYL}}} \left(\mathbf{F}^{h}_{1}\,n_{1}+\mathbf{F}^{h}_{2}\,n_{2}\right)\cdot \mathbf{W}^{h}\,d\Gamma
		+ \int_{\Gamma_{\text{CYL}}} \mathbf{F}^{\text{wall}}_{\mathbf{n}}\cdot \mathbf{W}^{h}\,d\Gamma \\
		+ \sum_{e=1}^{n_\text{el}}\int_{\Omega^{e}} \tau_{\text{SUPG}} \left( \frac{\partial \mathbf{W}^{h}}{\partial x_{1}} \mathbf{A}^{h}_{1} +
		\frac{\partial \mathbf{W}^{h}}{\partial x_{2}} \mathbf{A}_{2}^{h} \right ) \cdot \mathbf{R}\!\left(\mathbf{U}^{h}\right) d\Omega \\
		+ \sum_{e=1}^{n_\text{el}}\int_{\Omega^{e}} \nu_{\text{SHOC}} \left(\frac{\partial \mathbf{W}^{h}}{\partial x_{1}} \cdot \frac{\partial \mathbf{U}^{h}}{\partial x_{1}} + \frac{\partial \mathbf{W}^{h}}{\partial x_{2}} \cdot \frac{\partial \mathbf{U}^{h}}{\partial x_{2}}  \right) d\Omega=0,
	\end{multline}
	where $\mathbf{U}^{h} = \mathbf{U}\!\left(\mathbf{Q}^{h}\right)$,
	$\mathbf{F}^{h}_{i} = \mathbf{F}_{i}\!\left(\mathbf{Q}^{h}\right)$,
	$n_\text{el}$ is the number of elements in $\mathcal{T}^{h}$,
	$\Gamma_{\text{CYL}}$ denotes the cylinder surface, and
	\begin{equation} \label{strongres}
		\mathbf{R}\!\left(\mathbf{U}^{h}\right) = \frac{\partial \mathbf{U}^{h}}{\partial t} + \mathbf{A}^{h}_{1} \frac{\partial \mathbf{U}^{h}}{\partial x_{1}} +  \mathbf{A}^{h}_{2} \frac{\partial \mathbf{U}^{h}}{\partial x_{2}}
	\end{equation}
	is the strong residual of the governing equations. The vector $\mathbf{F}^{\text{wall}}_{\mathbf{n}}$ appearing in the boundary integral over the cylinder surface is the physical inviscid slip-wall flux introduced in Section~\ref{sec5} (see Eq.~\eqref{slip}); it imposes the zero-normal-velocity (slip) condition weakly and consistently, without any penalty parameter. On the remaining boundary portions, the consistent Euler normal flux is retained; on the inflow boundary, where the essential conditions are imposed strongly, the test functions vanish and the corresponding boundary term is inert.
	
	\begin{remark}
		The choice of $\mathcal{P}_{1}$ elements is motivated by both computational
		cost and consistency with the stabilization framework. Higher-order elements
		can improve the formal accuracy of the approximation in smooth regions, but
		they increase the computational cost considerably and, in the presence of
		strong shocks, tend to produce more pronounced oscillations, thereby placing
		greater demands on the stabilization and shock-capturing operators. Moreover,
		the element-level quantities $\tau_{\mathrm{SUPG}}$, $\nu_{\mathrm{SHOC}}$,
		and $h_{\mathrm{JGN}}$ employed here are constructed directly from the
		piecewise-constant gradients of the $\mathcal{P}_{1}$ basis functions.
	\end{remark}
	
	\begin{remark}
		In the stabilized formulation~\eqref{supg}, the first line represents the standard Galerkin finite element formulation in conservative form, including the boundary flux terms. The first and second lines together correspond to the compressible-flow SUPG formulation, and, with the additional term for discontinuity-capturing in the last line, the formulation represents the SUPG-based stabilized formulation enhanced with YZ$\beta$ shock-capturing. Noting that we deal with compressible-flow SUPG, we call it the \enquote{compressible-flow SUPG-YZ$\beta$ formulation.}
	\end{remark}
	
	The stabilization and shock-capturing parameters, $\tau_{\text{SUPG}}$ and $\nu_{\text{SHOC}}$, are constructed in the style of~\cite{Tezduyar04o, Tezduyar05d, Tezduyar05f}. The stabilization parameter is based on the advective-acoustic form of the $\tau_{\text{SUGN1}}$ time scale:
	\begin{equation} \label{tausugn1}
		\tau_{\text{SUPG}}= \left(\sum_{a=1}^{n_{\text{en}}}\Big( c^h\,\vert\mathbf{j}\cdot \nabla N_{a}\vert + \vert\mathbf{u}^{h}\cdot \nabla N_{a}\vert \Big) \right)^{-1},
	\end{equation}
	where $c^h$ is the local acoustic speed, $n_\text{en}$ is the number of element nodes, $N_{a}$ is the interpolation function associated with node $a$, and $\mathbf{j}$ is the unit vector aligned with the density gradient defined in Eq.~\eqref{jdef} below. The same scalar value of $\tau_{\text{SUPG}}$ is used for all four conservation equations, i.e., $\boldsymbol{\tau}_{\text{SUPG}}=\tau_{\text{SUPG}}\,\mathbf{I}$.

	The shock-capturing parameter, $\nu_{\text{SHOC}}$, is based on the YZ$\beta$ technique~\cite{Tezduyar04o, Tezduyar05d, Tezduyar05f} and is given, in the scaled form employed in this work, as
	\begin{equation} \label{shoc}
		\nu_{\text{SHOC}}
		=
		C_{\text{YZ}\beta}\,
		\big\Vert\mathbf{Y}^{-1}\mathbf{Z}\big\Vert
		\left(
		\sum_{i=1}^{n_{\text{sd}}}
		\Big\Vert
		\mathbf{Y}^{-1}
		\frac{\partial \mathbf{U}^{h}}{\partial x_{i}}
		\Big\Vert^{2}
		\right)^{\frac{\beta}{2}-1}
		\big\Vert\mathbf{Y}^{-1}\mathbf{U}^{h}\big\Vert^{1-\beta}
		\left(\frac{h_{\text{SHOC}}}{2}\right)^{\beta},
	\end{equation}
	where $C_{\text{YZ}\beta}$ is a nondimensional scaling constant.
	
	\begin{remark}
		Note that the original YZ$\beta$ formulation does not contain the 
		scaling constant $C_{\text{YZ}\beta}$. Here, a user-defined scaling 
		constant is introduced to adjust the overall magnitude of the 
		shock-capturing viscosity, allowing the amount of discontinuity-capturing 
		dissipation to be tuned for the strong shocks encountered in the present 
		high-speed flow computations. The value $C_{\text{YZ}\beta} = 15$ was determined through numerical 
		experiments. Smaller values were 
		observed to admit localized overshoots, whereas substantially larger 
		values led to visibly smeared shock profiles.
	\end{remark}
	
	For the sharpness parameter $\beta=2$ used in our computations, the scaled-gradient factor in Eq.~\eqref{shoc} has a zero exponent and drops out exactly, so that the implemented expression reduces to
	\begin{equation} \label{shocbeta2}
		\nu_{\text{SHOC}}=C_{\text{YZ}\beta}\,\frac{\big\Vert\mathbf{Y}^{-1}\mathbf{Z} \big\Vert}{\big\Vert\mathbf{Y}^{-1}\mathbf{U}^{h}\big\Vert}\left(\frac{h_{\text{SHOC}}}{2} \right)^{2}.
	\end{equation}
	The diagonal scaling matrix is defined as
	\begin{equation} \notag
		\mathbf{Y}=   \begin{bmatrix}
			(U_{1})_{\text{ref}} & 0 & 0 & 0  \\
			0 & (U_{2})_{\text{ref}} & 0 & 0 \\
			0 & 0 & (U_{3})_{\text{ref}} & 0 \\
			0 & 0 & 0 & (U_{4})_{\text{ref}}
		\end{bmatrix},
	\end{equation}
	where the diagonal entries $(U_1)_{\text{ref}}, \ldots, (U_4)_{\text{ref}}$
	represent reference values, such as the free-stream conditions, for the
	corresponding entries in the vector of conservation variables $\mathbf{U}$,
	i.e., $\rho$, $\rho u_1$, $\rho u_2$, and $\rho e$, respectively.

	\begin{remark}
		Note that since the $x_2$-component of the free-stream velocity is assumed
		to be zero in our computations, we set $(U_3)_{\text{ref}} =
		(U_2)_{\text{ref}}$ in the scaling matrix $\mathbf{Y}$.
	\end{remark}
	
	The vector $\mathbf{Z}$ of YZ$\beta$ can be defined in its advective-form as
	\begin{equation} \label{Z_stationary}
		\mathbf{Z}=\mathbf{A}_{1}^{\textit{h}} \frac{\partial \mathbf{U}^{h}}{\partial x_{1}}
		+ \mathbf{A}_{2}^{\textit{h}}\frac{\partial \mathbf{U}^{h}}{\partial x_{2}}.
	\end{equation}
	It can also be defined in the residual-form as follows:
	\begin{equation} \label{Z_time_dependent}
		\mathbf{Z}= \frac{\partial \mathbf{U}^{h}}{\partial t} + \mathbf{A}_{1}^{\textit{h}} \frac{\partial \mathbf{U}^{h}}{\partial x_{1}}
		+ \mathbf{A}_{2}^{\textit{h}}\frac{\partial \mathbf{U}^{h}}{\partial x_{2}}.
	\end{equation}
	In our computations, we adopt the advective-form given by Eq.~\eqref{Z_stationary}, evaluated from the previously converged time level and held fixed during the nonlinear iterations of the current time step (see Section~\ref{sec4}).
	
	The element length scale, $h_{\text{SHOC}}$, is calculated from
	\begin{equation} \label{elementlen}
		h_{\text{SHOC}}=h_{\text{JGN}}=2\left(\sum_{a=1}^{n_\text{en}}\vert\mathbf{j}\cdot \nabla N_{a}\vert \right)^{-1},
	\end{equation}
	where the vector $\mathbf{j}$ is the unit vector in the direction of the gradient of the density function:
	\begin{equation} \label{jdef}
		\mathbf{j}= \frac{\nabla \rho^{h}}{\Vert\nabla \rho^{h}\Vert}.
	\end{equation}
	
	\begin{remark}
		As described by Tezduyar et al.~\cite{Tezduyar04o,Tezduyar05d,Tezduyar05f}, the sharpness parameter $\beta$ embedded in Eq.~\eqref{shoc} can be set as $\beta=1$ for mild shocks and $\beta=2$ for sharper shocks. Here, we set it as $\beta=2$.
	\end{remark}
	
	\begin{remark} \label{rem:freezing}
		In the implementation, the stabilization and shock-capturing data are computed element-by-element from the previously converged solution and represented as piecewise-constant (per-element) fields: $\tau_{\text{SUPG}}$ and $h_{\text{SHOC}}$ are assembled directly from the $\mathcal{P}_{1}$ basis-function gradients of each triangle, while $\nu_{\text{SHOC}}$ is obtained by projecting Eq.~\eqref{shocbeta2} onto the space of piecewise constants and clipping any negative or non-finite values to zero. At the initially uniform free-stream state, where $\nabla\rho^{h}\approx\mathbf{0}$ and Eq.~\eqref{jdef} is undefined, the direction $\mathbf{j}$ falls back to the local flow direction (and, in the fully quiescent limit, to the $x_{1}$-direction) until a density gradient develops. All of these quantities are then held fixed throughout the nonlinear iterations of the current time step.
	\end{remark}
	
	\begin{remark}
		Originally developed for compressible-flow simulations, the SUPG-YZ$\beta$ formulation has subsequently been extended to a broad range of convection-dominated problems, including arterial drug delivery applications~\cite{bazilevs_yzb_2007}, Burgers'-type equations at high Reynolds numbers~\cite{Cengizci2023amc}, shallow-water equations~\cite{Cengizci2023zamm}, natural and mixed convection~\cite{Cengizci24a,Cengizci24b,Cengizci2024c}, financial option pricing models~\cite{Cengizci2024_heston}, haptotaxis-driven tumor growth models~\cite{Cengizci2026}, chemically reactive transport problems~\cite{cengizci22}, magnetohydrodynamic duct flows at high Hartmann numbers~\cite{Cengizci2024_mhd}, three-dimensional convection-dominated elliptic problems~\cite{Cengizci2024_3d}, time-fractional convection--diffusion equations~\cite{Cengizci2024_fractional}, and coupled reaction--convection--diffusion systems~\cite{Cengizci2023_rcd}. Collectively, these studies demonstrate that the SUPG-YZ$\beta$ methodology constitutes a robust and versatile stabilization framework for a wide spectrum of convection-dominated problems.
	\end{remark}

	\section{Temporal discretization}
	\label{sec4}
	The time-marching is performed with a first-order, semi-implicit backward Euler scheme. As we step from time level $n$ to $n+1$, the temporal derivative is approximated as
	\begin{equation}
		\frac{\partial \mathbf{U}}{\partial t} \approx \frac{\mathbf{U}_{n+1}-\mathbf{U}_{n}}{\varDelta t_{n}},
	\end{equation}
	and the Galerkin flux terms and the solution gradients in the stabilized formulation~\eqref{supg} are expressed at time level $n+1$. The Euler flux Jacobians $\mathbf{A}_{1}$ and $\mathbf{A}_{2}$ appearing in the SUPG test operator and in the strong residual~\eqref{strongres}, together with the stabilization and shock-capturing data $\tau_{\text{SUPG}}$, $\nu_{\text{SHOC}}$, and $h_{\text{SHOC}}$ (see Remark~\ref{rem:freezing}), are frozen at the previously converged time level $n$. Accordingly, the fully discrete strong residual employed in the SUPG term reads
	\begin{equation} \label{discres}
		\mathbf{R}_{n+1}\!\left(\mathbf{U}^{h}\right) = \frac{\mathbf{U}^{h}_{n+1}-\mathbf{U}^{h}_{n}}{\varDelta t_{n}} + \mathbf{A}^{h}_{1,n} \frac{\partial \mathbf{U}^{h}_{n+1}}{\partial x_{1}} +  \mathbf{A}^{h}_{2,n} \frac{\partial \mathbf{U}^{h}_{n+1}}{\partial x_{2}}.
	\end{equation}
	Since the conservative variables and fluxes remain nonlinear functions of the primitive unknowns $\mathbf{Q}^{h}_{n+1}$, a nonlinear algebraic system is solved at every time step by (damped) Newton--Raphson iterations, in which the consistent Jacobian of the discrete residual is generated by automatic (symbolic) differentiation of the variational form. The linear system arising at each Newton iteration is solved with the
	multifrontal massively parallel sparse direct solver MUMPS~\cite{Amestoy2001}.
	
	The nominal time-step size is determined from the Courant--Friedrichs--Lewy (CFL) condition based on the free-stream state and the smallest cell diameter of the mesh, $h_{\min}$:
	\begin{equation}
		\label{dtnominal}
		\varDelta t_{0}
		=
		\frac{C_{\varDelta t}\,h_{\min}}
		{\lVert \mathbf{u}_{\infty}\rVert + c_{\infty}},
	\end{equation}
	where $C_{\varDelta t}$ is a user-specified Courant number and
	$c_{\infty}=\sqrt{\gamma R_{\mathrm{N}_2}T_{\infty}}$ denotes the free-stream speed of sound.
	Around this nominal value, an adaptive time-stepping strategy is employed.
	If the Newton iterations fail to converge at the current time step, or if the
	resulting solution violates the admissibility (positivity) bounds
	$\rho^{h}\geq\rho_{\min}$ and $T^{h}\geq T_{\min}$, the step is rejected, the time-step size is halved, and the nonlinear solve is repeated. At each retry, the sequence of Newton relaxation (damping) factors
	$\left(1.0,\,0.5,\,0.25,\,0.1\right)$ is attempted.
	The time-step reduction is applied at most $10$ times, with the time-step size bounded from below by $10^{-4}\varDelta t_{0}$.
	Conversely, whenever the Newton iterations converge readily (within $5$ iterations at full relaxation), the time-step size is increased by a factor of $1.10$, up to the nominal value $\varDelta t_{0}$.
	The Courant numbers employed in the computations are reported in Section~\ref{sec6}.
	
	\section{Computational domain and boundary conditions}
	\label{sec5}
	
\begin{figure}[htb]
		\centering
		\begin{tikzpicture}[scale=0.82]
			\draw[thick,-] (-8,0) -> (-8,6);
			\draw[thick,-] (-8,6) -> (4,6);
			\draw[thick,-] (4,6) -> (4,0);
			\draw[thick,-] (4,0) -> (-8,0);
			\draw[thick,->] (-8,0) -> (-7,0);
			\draw[thick,->] (-8,1) -> (-7,1);
			\draw[thick,->] (-8,2) -> (-7,2);
			\draw[thick,->] (-8,3) -> (-7,3);
			\draw[thick,->] (-8,4) -> (-7,4);
			\draw[thick,->] (-8,5) -> (-7,5);
			\draw[thick,->] (-8,6) -> (-7,6);
			\draw[thick,->] (-8,6) -> (-8,7.0); 
			\draw[thick,->] (4,0) -> (5.0,0); 
			\node[left, above] at (-8.5,7) {$x_{2}$};
			\node[right] at (5,-0.5) {$x_{1}$};
			\node[right] at (-7,4.5) {$\mathbf{u}_{\infty}$};
			\draw[red, thick,-](-5,3) circle (0.25);
			\node[left, right=-1cm] at (-8,6) {$1.0$};
			\node[left, right=-1cm] at (-8,0) {$0.0$};
			\node[left, below=0cm] at (-8,0) {$0.0$};
			\node[left, right=-1cm] at (-8,3) {$0.5$};
			\node[right, right=0.2cm] at (-4,3) {$0.1$};
			\draw[dashed] (-8,3) -> (-5,3);
			\draw[thick,-, blue] (-4.7,3.25) -> (-4,3.25);
			\draw[thick,-, blue] (-4.7,2.75) -> (-4,2.75);
			\draw[thick,<->, green] (-4,2.79) -> (-4,3.21);
			\draw[dashed] (-8,3) -> (-5,3);
			\draw[dashed] (-5,3) -> (-5,0);
			\node[below=2mm] at (-5,0) {$0.5$};
			\node[below=2mm] at (4,0) {$2.0$};
		\end{tikzpicture}
		\caption{Computational domain with free-stream velocity vector $\mathbf{u}_{\infty}$. All dimensions are in meters (m).} \label{fig:domain}
	\end{figure}
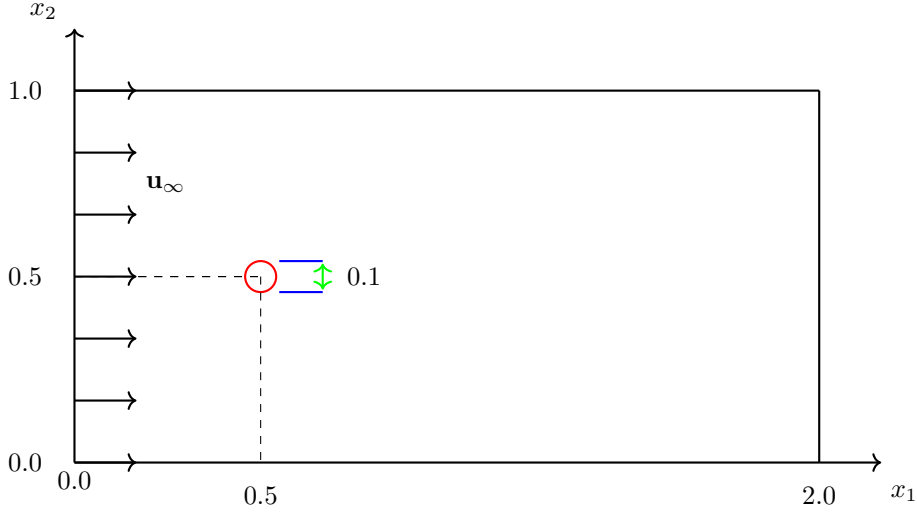

	The computational domain is shown in Figure~\ref{fig:domain}. At the
	supersonic/hypersonic inlet, all primitive flow variables
	$\left(\rho,\mathbf{u},T\right)$ are prescribed strongly and set to their
	free-stream values. Unlike the inflow boundary, no essential boundary condition
	is specified at the outlet since all flow characteristics leave the flow
	domain; there, the consistent Euler normal flux arising from the integration
	by parts in Eq.~\eqref{supg} is retained (a \enquote{do-nothing} treatment). For a detailed discussion, see~\cite{blazek, Cengizci_thesis}. The top and bottom boundaries are placed sufficiently far from the cylinder and are treated in the same do-nothing manner, i.e., no essential condition is imposed on them and the consistent Euler normal flux is retained in the corresponding boundary integrals. The initial conditions are also set to the free-stream conditions.
	
	On the cylinder surface, the inviscid slip (zero-normal-velocity) condition, $\mathbf{u}\cdot\mathbf{n}=0$, is imposed weakly and in a penalty-free manner through the physical inviscid wall flux. Substituting the slip condition into the Euler normal flux $\mathbf{F}_{1}n_{1}+\mathbf{F}_{2}n_{2}$ yields
	\begin{equation} \label{slip}
		\mathbf{F}^{\text{wall}}_{\mathbf{n}}=\begin{bmatrix}
			0 \\ p\, n_{1} \\ p\, n_{2} \\ 0
		\end{bmatrix},
	\end{equation}
	which is used in the boundary integral over $\Gamma_{\mathrm{CYL}}$ in the
	stabilized formulation given by Eq.~\eqref{supg}. This treatment enforces
	exactly zero normal mass and energy fluxes through the wall while retaining
	the pressure contribution to the momentum equations. Unlike the penalty-based
	treatment adopted in~\cite{Cengizci_thesis}, the present formulation is
	entirely penalty-free and therefore does not require the selection of any
	user-defined penalty parameter.
	
	\section{The PASSC framework}
	\label{sec:PASSC}
	
	The stabilized finite element formulation presented in the previous sections provides robust and substantially oscillation-free approximations for compressible flows. Nevertheless, like many residual-based stabilization techniques, it introduces a certain amount of artificial numerical diffusion, which may smear sharp flow features, such as shock waves and contact discontinuities. To alleviate this drawback, we employ the PASSC framework, in which a physics-informed neural network is trained as a post-processing stage. Rather than being trained from scratch on the governing equations alone, the PINN is anchored to the stabilized finite element solution through a shock-weighted data-consistency loss and refines it under the constraints of the governing equations---enforced in a shock-consistent, space--time control-volume form, supplemented by macroscopic conservation constraints and an entropy-admissibility penalty---and the boundary conditions. Consequently, the corrected solution remains anchored to the underlying
	SUPG-YZ$\beta$ approximation, while the physics-informed post-processing
	targets the excessive numerical diffusion introduced by the stabilization.
	The main components of the PASSC framework are described in the following
	subsections.

	\subsection{Network representation, nondimensionalization, and the implied correction}
	\label{sec:correction}
	Let $\mathbf{U}^{h}$ denote the stabilized finite element approximation obtained using the SUPG-YZ$\beta$ formulation presented in the previous sections. In the PASSC framework, the neural network represents the flow field in nondimensional primitive variables. Introducing the coordinate-scaling length $L=1$~m and the free-stream reference state, the nondimensional variables are defined as
	\begin{equation} \label{eq:nondim}
		\rho^{*}=\frac{\rho}{\rho_{\infty}}, \quad
		u_1^{*}=\frac{u_{1}}{\Vert\mathbf{u}_{\infty}\Vert}, \quad
		u_2^{*}=\frac{u_{2}}{\Vert\mathbf{u}_{\infty}\Vert}, \quad
		T^{*}=\frac{T}{T_{\infty}}, \quad
		x_{i}^{*}=\frac{x_{i}}{L}, \quad
		t^{*}=\frac{t\,\Vert\mathbf{u}_{\infty}\Vert}{L},
	\end{equation}
	so that the nondimensional pressure and specific total energy read
	\begin{equation} \label{eq:nondim_state}
		p^{*}=\frac{\rho^{*}T^{*}}{\gamma M_\infty^{2}}, \qquad
		e^{*}=\frac{T^{*}}{\gamma M_\infty^{2}\left(\gamma-1\right)}+\frac{1}{2}\left(\left(u_1^{*}\right)^{2}+\left(u_2^{*}\right)^{2}\right),
	\end{equation}
	where $M_\infty$ is the free-stream Mach number. The network defines the mapping
	\begin{equation} \label{eq:network_map}
		\mathbf{Q}_{\theta}:\;\mathbf{z}=\left(x_{1}^{*},x_{2}^{*},t^{*}\right)\;\longmapsto\;
		\left(\rho^{*},\,u_1^{*},\,u_2^{*},\,T^{*}\right)^{T},
	\end{equation}
	parameterized by the trainable parameters $\theta$. Predicting the primitive variables allows the data-consistency and boundary-condition losses to act directly on the network outputs, whereas the governing-equation residual is assembled in conservative variables and enforced in a space--time control-volume sense (see Section~\ref{sec:loss}) to remain consistent with the Rankine--Hugoniot jump conditions across shocks.
	
	The enhanced PASSC solution is obtained by first converting the nondimensional
	network output back to dimensional primitive variables using the reference
	scales in Eq.~\eqref{eq:nondim}, and then reassembling the conservative
	variables through the constitutive relations of Section~\ref{sec2}:
	\begin{equation}
		\mathbf{U}^{\mathrm{PASSC}}
		=
		\mathbf{U}\!\left(\rho_{\infty}\rho^{*}_{\theta},
		\Vert\mathbf{u}_{\infty}\Vert u^{*}_{1,\theta},
		\Vert\mathbf{u}_{\infty}\Vert u^{*}_{2,\theta},
		T_{\infty}T^{*}_{\theta}\right).
		\label{eq:passc_reassembly}
	\end{equation}
	The correction delivered by the PASSC stage is then implied by the decomposition
	\begin{equation}
		\boldsymbol{\delta}_{\theta}
		=
		\mathbf{U}^{\mathrm{PASSC}}
		-
		\mathbf{U}^{h},
		\label{eq:passc_decomposition}
	\end{equation}
	with the components
	\begin{equation}
		\boldsymbol{\delta}_{\theta}
		=
		\begin{bmatrix}
			\delta_{\rho}\\
			\delta_{\rho u_1}\\
			\delta_{\rho u_2}\\
			\delta_{\rho e}
		\end{bmatrix}
		\label{eq:correction_vector}
	\end{equation}
	corresponding to the density, the $x_1$- and $x_2$-momentum densities, and the total energy density, respectively. 
	
	Unlike conventional PINN formulations, which attempt to reconstruct the complete solution field solely from the governing equations, the proposed framework initializes the network at the free-stream state and anchors it to the stabilized finite element solution through a data-consistency loss. During the initial data-dominant stage of the optimization, the network prediction is rapidly driven toward the stabilized finite element solution $\mathbf{U}^{h}$. As the physics contribution is subsequently strengthened, the implied correction $\boldsymbol{\delta}_{\theta}$ in Eq.~\eqref{eq:passc_decomposition} is primarily concentrated in regions where artificial numerical diffusion has smeared sharp flow features, rather than constituting an independent reconstruction of the flow field. Since the SUPG-YZ$\beta$ approximation already captures the global flow structure while providing a robust and substantially oscillation-free solution, the learning task is significantly simplified. Consequently, the neural network can focus on recovering sharp flow features
	that are excessively smeared by the stabilization while remaining anchored
	to the underlying stabilized finite element approximation.
	
	\subsection{Neural network architecture}
	\label{sec:architecture}
	
	The mapping in Eq.~\eqref{eq:network_map} is approximated using a fully
	connected feed-forward neural network. The network takes the nondimensional
	spatiotemporal coordinates
	$\mathbf{z}=\left(t^{*},x_{1}^{*},x_{2}^{*}\right)$
	as its input and predicts the nondimensional primitive field
	$\left(\rho^{*},u_1^{*},u_2^{*},T^{*}\right)^{T}$.
	To improve the representation of localized high-frequency flow features,
	the input coordinates are first mapped into a higher-dimensional space using
	a random Fourier feature embedding, in which the fixed projection matrix is
	drawn from a Gaussian distribution with scale $\sigma=4$. The resulting
	feature vector is then propagated through a sequence of fully connected
	residual blocks comprising linear layers, SiLU activation functions, and
	Layer Normalization. Residual (skip) connections are incorporated throughout
	the network to facilitate gradient propagation and improve optimization
	stability. A narrowing layer followed by a final linear layer produces the
	four-dimensional output of Eq.~\eqref{eq:network_map}. The bias of the output
	layer is initialized at the free-stream state,
	$\left(\rho^{*},u_1^{*},u_2^{*},T^{*}\right)=\left(1,1,0,1\right)$,
	and the output weights are scaled down at initialization so that the network
	prediction starts close to the free-stream field and training proceeds by
	learning localized deviations from it.

	\subsection{Physics-informed loss function}
	\label{sec:loss}
	
	The trainable parameters $\theta$ of the neural network are determined by minimizing a physics-informed loss function that simultaneously enforces the governing equations and the prescribed boundary conditions, and anchors the network prediction to the stabilized FEM approximation. The overall objective function consists of three complementary components, namely a data-consistency loss, a physics loss, and a boundary-condition loss, and is defined as
	\begin{equation}
		\mathcal{L}
		=
		w_\text{data}\,\mathcal{L}_\text{data}
		+
		w_\text{pde}\,\mathcal{L}_\text{pde}
		+
		w_\text{bc}\,\mathcal{L}_\text{bc},
		\label{eq:total_loss}
	\end{equation}
	where $w_\text{data}$, $w_\text{pde}$, and $w_\text{bc}$ are positive weighting coefficients that balance the relative contributions of the corresponding loss terms during the optimization process. Instead of fixed weights, a continuous curriculum is employed: the weights are prescribed at a small number of keyframes, expressed as fractions of the epoch budget, and are linearly interpolated between consecutive keyframes, so that the training signal contains no discontinuous weight jumps. The keyframes used in this work are
	\begin{equation}
		\left(\varphi;\; w_\text{data},\, w_\text{pde},\, w_\text{bc}\right)
		\in
		\left\{
		\left(0;\,1,\,0.01,\,0.1\right),\;
		\left(0.2;\,1,\,0.1,\,0.3\right),\;
		\left(0.4;\,1,\,0.25,\,0.6\right),\;
		\left(0.7;\,1,\,0.5,\,1\right),\;
		\left(1;\,1,\,1,\,1\right)
		\right\},
		\label{eq:weight_schedule}
	\end{equation}
	where $\varphi\in[0,1]$ denotes the elapsed fraction of the epoch budget. The data weight is kept at $w_\text{data}=1$ throughout the training (sustained data anchoring), and the boundary weight is ramped up to $w_\text{bc}=1$ over the first $70\%$ of the training and held thereafter. The physics weight, in contrast, is ramped up continuously over the entire training, from $w_\text{pde}=0.01$ at the start to full parity with the data weight, $w_\text{pde}=1$, at the end. This curriculum reflects the division of labor in the framework: the early, data-dominant phase establishes the finite element anchor, after which the authority of the physics term grows steadily, so that the final phase of the training is governed in equal parts by fidelity to the anchor and by the discrete conservation statements described below. Granting the physics term full authority is admissible only because the residual it minimizes is trustworthy near discontinuities---it is evaluated in a shock-consistent space--time control-volume form, reinforced by macroscopic conservation windows and an entropy-admissibility penalty---whereas a pointwise strong-form residual, being ill-defined across shocks, would reward smearing of the discontinuity if weighted this aggressively. The same schedule is used for all free-stream Mach numbers considered.

	The data-consistency loss anchors the network prediction to the stabilized finite element approximation. It is evaluated in the nondimensional primitive variables over the mesh vertices of the last $K_{s}$ stored solution snapshots (with $K_{s}=10$ in this work):
	\begin{equation}
		\mathcal{L}_\text{data}
		=
		\frac{1}{4\,N_\text{data}}
		\sum_{i=1}^{N_\text{data}}
		\omega_i
		\left\|
		\mathbf{Q}_{\theta}(\mathbf{z}_i)
		-
		\mathbf{Q}^{h}(\mathbf{z}_i)
		\right\|_2^2,
		\label{eq:data_loss}
	\end{equation}
	where $\mathbf{Q}^{h}$ denotes the nondimensionalized primitive finite element solution, $N_\text{data}$ is the total number of space--time training points assembled from the stored snapshots, and $\omega_i \geq 1$ are per-node weights that concentrate the data authority on the shock region. Here, and in the physics and inflow terms below, the squared norms are additionally averaged over the four solution components, so that the reported loss values correspond to componentwise mean-squared errors, consistent with the released implementation.
	
	The weights $\omega_i$ are constructed once per snapshot from the finite element density-gradient magnitude $\Vert\nabla\rho^{h}\Vert$ (a numerical-schlieren field, evaluated at the mesh vertices from the piecewise-linear interpolant). For each snapshot, a relative shock indicator is first formed as
	\begin{equation}
		s_i
		=
		\min\!\left(
		\frac{\Vert\nabla\rho^{h}\Vert_i}{g_\text{ref}},\,1
		\right),
		\label{eq:shock_indicator}
	\end{equation}
	where $g_\text{ref}$ is the $99$th percentile of $\Vert\nabla\rho^{h}\Vert$ over the vertices of that snapshot. The percentile reference (rather than the maximum) is deliberate: normalizing by the single steepest node would boost only that node while leaving the shock feet---precisely the region most prone to smearing---barely weighted, and would also be fragile to isolated gradient spikes; with the percentile reference and the clip in Eq.~\eqref{eq:shock_indicator}, the entire shock transition band saturates near the full weight. The indicator field is then regularized morphologically on the mesh-edge graph: one erosion pass (minimum propagation along the mesh edges) suppresses isolated single-node spikes, after which $n_\text{d}+1$ dilation passes (maximum propagation) restore the eroded ring and extend the boosted band by approximately $n_\text{d}$ local mesh spacings beyond the discrete gradient support, with $n_\text{d}=8$ in this work. The dilation is essential: without it, the optimizer tends to deposit spurious low-amplitude deviations in the thin strip immediately outside the finite element shock band, where the data penalty would otherwise be cheapest. The final weights read
	\begin{equation}
		\omega_i
		=
		1 + \lambda\, \tilde{s}_i,
		\qquad
		\lambda = 4,
		\label{eq:shock_weights}
	\end{equation}
	where $\tilde{s}_i$ denotes the morphologically processed indicator. No global rescaling is applied, so smooth regions retain exactly unit weight (full data authority everywhere) while the shock band receives up to a fivefold boost. The motivation for the weighting is that shock-transition vertices constitute only a small fraction of the mesh: under a plain mean-squared error, the optimizer can smear the shock at a negligible data cost, whereas the up-weighting reprices this trade directly. Since the weights depend only on the fixed finite element snapshots, they are computed once before training and remain constant across epochs.

	The governing equations are enforced through a space--time control-volume formulation of the Euler residual. With $\mathbf{U}^{*}$, $\mathbf{F}^{*}_{1}$, and $\mathbf{F}^{*}_{2}$ denoting the nondimensional conservative variables and Euler fluxes evaluated from $\mathbf{Q}_{\theta}$ through Eq.~\eqref{eq:nondim_state}, the pointwise (strong-form) residual reads
	\begin{equation}
		\mathbf{R}^\text{PASSC}
		=
		\frac{\partial \mathbf{U}^{*}}{\partial t^{*}}
		+
		\frac{\partial \mathbf{F}^{*}_1}{\partial x^{*}_1}
		+
		\frac{\partial \mathbf{F}^{*}_2}{\partial x^{*}_2},
		\label{eq:euler_residual}
	\end{equation}
	where
	\[
	\mathbf{R}^\text{PASSC}
	=
	\left(
	R^\text{PASSC}_{\rho},
	R^\text{PASSC}_{\rho u_1},
	R^\text{PASSC}_{\rho u_2},
	R^\text{PASSC}_{\rho e}
	\right)^{T}
	\]
	denotes the residual vector corresponding to the conservation of mass, the two momentum equations, and the total energy equation, respectively. Rather than evaluating Eq.~\eqref{eq:euler_residual} pointwise through automatic differentiation, the residual is enforced in its integral form over compact space--time boxes. Consider a square spatial control volume $\omega_i$ of half-width $h_i$ centered at $(x^{*}_{1,i},x^{*}_{2,i})$, extended in time over $[t^{*}_i-k_i,\,t^{*}_i+k_i]$, and let
	\begin{equation}
		\left\langle \mathbf{U}^{*}\right\rangle_{\omega_i}\!(t^{*})
		=
		\frac{1}{4h_i^{2}}\int_{\omega_i}\mathbf{U}^{*}\!\left(\mathbf{x}^{*},t^{*}\right)\mathrm{d}\Omega
		\label{eq:cell_average}
	\end{equation}
	denote the cell average of the conservative state. The space--time control-volume residual is then defined as
	\begin{equation}
		\overline{\mathbf{R}}^\text{PASSC}\!\left(\mathbf{z}_i\right)
		=
		\frac{\left\langle \mathbf{U}^{*}\right\rangle_{\omega_i}\!\left(t^{*}_i+k_i\right)
			-
			\left\langle \mathbf{U}^{*}\right\rangle_{\omega_i}\!\left(t^{*}_i-k_i\right)}{2\,k_i}
		+
		\frac{1}{4\,h_i^{2}}
		\oint_{\partial\omega_i}
		\left(
		\mathbf{F}^{*}_1\, n_1 + \mathbf{F}^{*}_2\, n_2
		\right)
		\mathrm{d}s\,\bigg|_{t^{*}_i},
		\label{eq:weak_residual}
	\end{equation}
	where $\mathbf{n}=(n_1,n_2)$ is the outward unit normal on
	$\partial\omega_i$. The two temporal faces of the space--time box
	contribute the difference of the cell averages, while the four spatial
	faces contribute the flux circulation evaluated at the temporal
	midpoint. Accordingly, Eq.~\eqref{eq:weak_residual} constitutes a
	space--time control-volume residual in which the lateral flux
	contribution is approximated by the midpoint rule in time. The boundary
	flux integral is evaluated using an $n_g$-point Gauss--Legendre
	quadrature on each edge, with $n_g=11$, while the cell averages on the
	temporal faces are computed using a $4\times4$ tensor-product Gauss
	rule. The temporal half-width is tied to the spatial one through
	$k_i=\max\!\left(10^{-3}T_w,\min(h_i,0.45\,T_w)\right)$, where $T_w$
	denotes the length of the nondimensional training window. The lower bound
	prevents numerical cancellation in the temporal-face difference, whereas
	the upper bound keeps the time slab within the training window; away from
	these bounds, $k_i=h_i$, so the boxes are approximately isotropic in the
	nondimensional coordinates.
	
	In the computations reported here, every term in
	Eq.~\eqref{eq:weak_residual} is evaluated through derivative-free
	forward passes of the network; no spatial or temporal derivatives of
	the network outputs are computed by automatic differentiation in the
	physics loss. Two properties of this construction are particularly
	useful. First, replacing the pointwise temporal derivative by the
	difference of cell averages over the two temporal faces yields a
	residual that remains well defined in the presence of moving
	discontinuities and incorporates the shock-speed contribution entering
	the Rankine--Hugoniot balance. Together with the midpoint approximation
	of the lateral flux, this provides a shock-consistent integral residual
	without requiring pointwise differentiation across a discontinuity.
	In contrast, the strong-form residual~\eqref{eq:euler_residual} is not
	classically defined at a discontinuity, and its direct pointwise
	minimization may favor excessively smooth representations of shocks.
	Second, for near-steady fields, quadrature errors in the cell averages
	largely cancel in the difference between the two temporal faces because
	the same quadrature rule is employed at both time levels. Consequently,
	a moderate quadrature order is sufficient for the cell averages even
	when a steep internal layer intersects the control volume. The
	$n_g=11$ edge quadrature provides sufficiently fine sampling of the
	finite-element-scale shock profiles over the range of control-volume
	sizes employed in the present computations. For smooth solutions, as
	$h_i,k_i\rightarrow0$, Eq.~\eqref{eq:weak_residual} approaches the
	strong-form residual with
	$\mathcal{O}(h_i^2)+\mathcal{O}(k_i^2)$ consistency under the spatial
	quadrature and temporal midpoint approximations, so that the two
	residual formulations retain the same nondimensional scaling.
	
	For robustness in the presence of strong shocks, each residual
	component is passed through the smooth odd saturation
	\[
	R \mapsto
	R_{\max}^{\mathrm{PASSC}}
	\tanh\!\left(
	\frac{R}{R_{\max}^{\mathrm{PASSC}}}
	\right),
	\qquad
	R_{\max}^{\mathrm{PASSC}}=10.
	\]
	Within the unsaturated range, the residual value and its gradient are
	essentially preserved, whereas for larger magnitudes the gradient
	decays smoothly rather than being abruptly truncated. Thus, unlike a
	hard clamp, the saturation retains a nonzero learning signal in regions
	where the residual is large, including the vicinity of strong shocks.

	The control-volume centers are resampled at every optimization step, and the half-widths $h_i$ are drawn log-uniformly from the multiscale range $\left[5\times10^{-4},\,2\times10^{-2}\right]$, partitioned into $S=4$ log-equal bands (strata) with equal sample counts, so that every scale is guaranteed to be present in every batch. Writing $\mathcal{M}_s$ for the componentwise mean-squared saturated residual over stratum $s$, the control-volume part of the physics loss reads
	\begin{equation}
		\mathcal{L}_\text{cv}
		=
		\frac{1}{S}
		\sum_{s=1}^{S}
		\beta_s\,\mathcal{M}_s,
		\label{eq:physics_loss}
	\end{equation}
	where the balance weights $\beta_s$ are proportional to the inverse of an
	exponential moving average (EMA) of $\mathcal{M}_s$ (decay $0.99$), normalized to unit
	mean and clipped to $[1/4,\,4]$. The rationale is that the residuals of the smallest control volumes are systematically largest near shocks---inside the finite-thickness numerical shock layer the flux balance cannot vanish and saturates at the bound---so under a plain mean they would dominate the gradient and drown the reducible large-scale conservation errors that actually determine the shock position. The inverse-EMA weighting grants each scale comparable gradient authority, while the unit-mean normalization and the clipping keep the magnitude of $\mathcal{L}_\text{cv}$ comparable to the plain mean (uniform weights recover it exactly). Near the domain boundaries, each half-width is capped by the local clearance so that no control volume crosses a boundary: control volumes are excluded from a band of nondimensional width $0.02$ adjacent to the outer boundaries, where the boundary-condition and data losses govern the solution, while near the cylinder only a thin pad of width $5\times10^{-3}$ is excluded and the half-widths are permitted to shrink down to a floor of $2.5\times10^{-3}$; the temporal centers are drawn so that the space--time box remains inside the training window. The reduced pad and the near-wall floor are essential at the higher Mach numbers, where the bow shock hugs the body closely and the physics loss must remain active within the thin shock layer.

	The largest control-volume half-width, $2\times10^{-2}$, is small relative
	to the flow domain, whereas the quantity that pins the shock position is
	integral conservation over volumes spanning the shock end to end: a control
	volume containing the entire shock relates the pre- and post-shock states
	directly through the Rankine--Hugoniot balance. Two inexpensive constructions
	supply this large-scale information.
	First, at every optimization step, eight macro flux windows---axis-aligned
	rectangles with independently sampled half-widths drawn log-uniformly from an
	initial range $\left[0.05,\,0.35\right]$ and rejection-sampled to lie strictly
	clear of the cylinder (so that the edge circulation constitutes the complete
	boundary flux, with no wall-flux surface integral required) and inside the
	outer exclusion band---are evaluated with the same space--time residual, using
	a $32$-point edge quadrature and $8\times8$ cell averages, since their edges
	may be crossed by the full shock; their componentwise mean-squared saturated
	residual, $\mathcal{L}_{\text{macro}}$, is added to the physics loss with
	unit weight.
	Second, every fourth physics evaluation additionally lays four rectangular
	patches, with half-widths drawn from an initial range
	$\left[0.03,\,0.12\right]$ and sampled with the same cylinder-clear rejection,
	subdivides each into a $4\times4$ array of equal tiles, and evaluates the
	control-volume residual on every tile. Adjacent tiles share their faces with
	identical quadrature points, so in the equal-area average of the unsaturated
	tile residuals the interior fluxes cancel exactly, and the tile average
	coincides with the patch-level residual evaluated with a composite
	($4\times n_g$-point per side) boundary quadrature. Both the per-tile
	mean-squared saturated residual, $\mathcal{L}_{\text{tile}}$, and that of the
	saturated patch aggregates, $\mathcal{L}_{\text{agg}}$, enter the loss with
	unit weight: the tiles enforce the local balance, while the telescoped
	aggregate enforces conservation over the whole patch without the saturation
	nonlinearity interfering with the interior cancellation.

	Weak solutions of the Euler equations are not unique, and nothing in a pure
	flux-balance loss distinguishes physically admissible shocks from
	entropy-violating (expansion-shock) configurations; the same observation
	motivated the thermodynamically consistent PINN formulation of
	Patel et al.~\cite{Patel2022}, in which integral conservation statements
	are supplemented with entropy conditions to select the physically relevant
	weak solution. Introducing the specific entropy
	$s^{*}=\ln T^{*}-\left(\gamma-1\right)\ln\rho^{*}$
	(normalized by $c_v$; the additive constant is immaterial to the balance),
	the implementation evaluates the logarithms using
	$\max(\rho^{*},10^{-6})$ and $\max(T^{*},10^{-6})$ to maintain numerical
	well-posedness during training. The entropy-balance residual is then assembled
	with exactly the same space--time control-volume machinery and from the same
	network evaluations as the conservative residual, without requiring additional
	network forward passes:
	\begin{equation}
		\overline{R}_s\!\left(\mathbf{z}_i\right)
		=
		\frac{\left\langle \rho^{*}s^{*}\right\rangle_{\omega_i}\!\left(t^{*}_i+k_i\right)
			-
			\left\langle \rho^{*}s^{*}\right\rangle_{\omega_i}\!\left(t^{*}_i-k_i\right)}{2\,k_i}
		+
		\frac{1}{4\,h_i^{2}}
		\oint_{\partial\omega_i}
		\rho^{*}s^{*}\left(\mathbf{u}^{*}\cdot\mathbf{n}\right)
		\mathrm{d}s\,\bigg|_{t^{*}_i}.
		\label{eq:entropy_residual}
	\end{equation}
	Admissible weak solutions satisfy $\overline{R}_s\geq0$ in the distributional sense---across a physical compressive shock the entropy production is strictly positive---so only the negative part is penalized, through the squared hinge $\mathcal{L}_\text{ent}=\mathrm{mean}\left[\max\!\left(0,\,-\overline{R}_s-\varepsilon_s\right)^{2}\right]$ with the tolerance $\varepsilon_s=10^{-3}$, a small deadband that prevents the $\mathcal{O}(h^{2})$ quadrature noise of smooth regions, which fluctuates around zero, from being spuriously penalized. The hinge leaves admissible shocks entirely untouched while excluding the entropy-destroying branch of weak solutions.

	Collecting the components above, the physics loss evaluated at each optimization step reads
	\begin{equation}
		\mathcal{L}_\text{pde}
		=
		\mathcal{L}_\text{cv}
		+
		w_\text{ent}\,\mathcal{L}_\text{ent}
		+
		\mathcal{L}_\text{macro}
		+
		\left(
		\mathcal{L}_\text{tile}
		+
		\mathcal{L}_\text{agg}
		\right),
		\qquad
		w_\text{ent}=0.1,
		\label{eq:pde_total}
	\end{equation}
	where the tiling terms $\mathcal{L}_\text{tile}$ and $\mathcal{L}_\text{agg}$ are present on every fourth evaluation and vanish otherwise, and $N_c=2{,}048$ interior control volumes are used per step for $\mathcal{L}_\text{cv}$ and $\mathcal{L}_\text{ent}$.

	The boundary conditions enforced on the network mirror the boundary specification of the finite element formulation exactly. They are imposed softly, on randomly sampled boundary points, through
	\begin{equation}
		\mathcal{L}_\text{bc}
		=
		\underbrace{\frac{1}{4\,N_\text{in}}\sum_{i=1}^{N_\text{in}}
			\left\|\mathbf{Q}_{\theta}(\mathbf{z}_i)-\mathbf{Q}^{*}_{\infty}\right\|_2^2}_{\text{supersonic/hypersonic inflow}}
		\;+\;
		\underbrace{\frac{1}{N_\text{c,cyl}}\sum_{i=1}^{N_\text{c,cyl}}
			\left(\mathbf{u}^{*}(\mathbf{z}_i)\cdot\mathbf{n}_i\right)^2}_{\text{cylinder slip}},
		\label{eq:boundary_loss}
	\end{equation}
	where $\mathbf{Q}^{*}_{\infty}=\left(1,1,0,1\right)^{T}$ is the nondimensional free-stream state. At the supersonic/hypersonic inflow, the full free-stream state is imposed, matching the inflow Dirichlet conditions of the finite element solver; on the cylinder surface, the slip condition $\mathbf{u}\cdot\mathbf{n}=0$ is enforced, consistent with the penalty-free wall treatment of Section~\ref{sec5}. No condition is required at the supersonic outflow. The top and bottom outer boundaries deliberately carry no boundary-condition penalty in the PASSC loss: in the finite element formulation, these boundaries receive only the plain normal Euler flux after integration by parts---a free (do-nothing) treatment---and the network must refine the same boundary-value problem as the anchor solution it is trained on; imposing an additional constraint there would drive the correction toward a different problem than the one solved by the stabilized discretization. In each optimization step, $256$ spatial points are sampled on each of the
	inflow and cylinder boundaries and evaluated at five randomly sampled time
	levels within the training window, yielding
	$N_{\mathrm{in}}=N_{\mathrm{c,cyl}}=1280$ space--time boundary samples
	for the corresponding terms in Eq.~\eqref{eq:boundary_loss}. Both the spatial
	boundary points and the sampled time levels are refreshed at every optimization
	step.
	
	The objective function given by Eq.~\eqref{eq:total_loss} therefore anchors the network to the stabilized FEM solution---with the data authority concentrated where the shock is most prone to smearing---enforces the governing conservation laws in a shock-consistent, entropy-admissible space--time control-volume sense across a hierarchy of length scales, together with the boundary conditions of the underlying problem, and enables the network to suppress the localized and dispersive spurious oscillations introduced by the numerical discretization.

	\subsection{Training procedure}
	\label{sec:training}
	
	The PASSC stage is executed after the stabilized FEM solution has been obtained. First, the SUPG-YZ$\beta$ finite element formulation presented in Section~\ref{sec3} is employed to compute a numerically stable approximation of the compressible Euler equations, and the primitive vertex fields are stored as solution snapshots at every accepted time step. The nondimensional training data are then assembled from the last $K_{s}$ snapshots, which constitute the data anchor of the network, and the shock-based data weights of Eq.~\eqref{eq:shock_weights} are computed once from these snapshots.
	
	The trainable parameters are updated with the AdamW optimizer, using the
	momentum parameters
	$\left(\beta_{1},\beta_{2}\right)=\left(0.9,0.999\right)$
	and a weight decay of $10^{-7}$. The base learning rate is set to
	$8\times10^{-5}$ and scaled with the square root of the mini-batch size
	relative to a reference batch of $256$. Thus, for the mini-batch size of
	$16{,}384$ used in the present computations, the initial learning rate is
	$6.4\times10^{-4}$. It is subsequently reduced by a plateau-based scheduler
	with a reduction factor of $0.9$, a patience of $150$ epochs, and a minimum
	learning rate of $10^{-6}$.
	At every optimization step, a mini-batch of $16{,}384$ data points, a fresh
	set of $N_c=2{,}048$ interior control volumes, together with the macro windows
	and, every fourth physics evaluation, the tiling patches of
	Section~\ref{sec:loss}, and freshly sampled boundary points are used. The
	parameter gradients are clipped to a unit norm. Training is performed for
	$3{,}000$ epochs using the continuously interpolated loss weights of
	Eq.~\eqref{eq:weight_schedule}.
	Numerical safeguards are applied to non-finite loss and gradient values:
	if a mini-batch objective becomes non-finite, the data-loss component alone
	is used for that step; the resulting objective is capped at $100$ before
	backpropagation; and parameter updates associated with non-finite gradients
	are skipped.
	
	Model selection and learning-rate scheduling are both driven by the unweighted,
	schedule-independent evaluation metric
	\begin{equation}
		\mathcal{E}
		=
		\mathcal{L}_{\text{data}}
		+
		\mathcal{L}_{\text{pde}}
		+
		\mathcal{L}_{\text{bc}},
		\label{eq:eval_metric}
	\end{equation}
	averaged over each epoch, and the parameter set attaining the lowest value of
	$\mathcal{E}$ is retained as the final model. Selecting on the weighted
	objective~\eqref{eq:total_loss} would be inappropriate under a ramped schedule:
	since the physics and boundary weights grow during training, the same raw
	residual level yields a smaller weighted total early on, which would
	systematically favor early, weakly physics-weighted parameter states. The
	unweighted metric~\eqref{eq:eval_metric} compares all epochs on the same
	footing and also prevents the plateau detection of the scheduler from being
	confounded by the weight ramps. Through this optimization process, the network is driven to refine localized flow features affected by the artificial numerical diffusion of the stabilized discretization while remaining anchored to the finite element solution and progressively improving consistency with the governing equations and boundary conditions.

	\subsection{Overall algorithm}
	\label{sec:algorithm}
	
	The overall computational workflow of the proposed PASSC framework is
	summarized in Algorithm~\ref{alg:passc_euler}. The methodology consists of
	two sequential stages. First, the transient compressible Euler equations are
	solved using the stabilized SUPG-YZ$\beta$ finite element formulation to
	obtain a robust and substantially oscillation-free numerical solution, whose
	primitive vertex fields are stored as snapshots. Subsequently, a
	physics-informed neural network, anchored to these snapshots through the
	shock-weighted data-consistency loss, is trained to refine localized flow
	features affected by the artificial numerical diffusion of the stabilized
	discretization while enforcing the space--time control-volume form of the
	governing equations---including the macroscopic conservation and
	entropy-admissibility constraints---together with the boundary conditions.
	
	\begin{algorithm}[!ht]
		\caption{PASSC framework for the compressible Euler equations}
		\label{alg:passc_euler}
		\small
		\begin{algorithmic}[1]
			
			\Require
			Finite element mesh $\mathcal{T}^{h}$, initial condition $\mathbf{U}_{0}$,
			boundary data, time interval $[0,t_{\text{f}}]$, Courant number
			$C_{\varDelta t}$ and adaptive time-stepping controls, number of anchor
			snapshots $K_{s}$, shock-weighting parameters $(\lambda,n_{\text{d}})$,
			control-volume, macro-window, and tiling sampling parameters,
			entropy-penalty parameters $(w_{\text{ent}},\varepsilon_s)$, neural network
			architecture, epoch budget, and loss-weight
			keyframes~\eqref{eq:weight_schedule}.
			
			\Ensure
			Enhanced PASSC solution $\mathbf{U}^{\mathrm{PASSC}}$.
			
			\State Solve the transient compressible Euler equations using the
			SUPG-YZ$\beta$ stabilized finite element method
			(Sections~\ref{sec3}--\ref{sec5}), storing the primitive vertex fields at
			every accepted time step.
			
			\State Assemble the nondimensional training data from the last $K_{s}$
			primitive snapshots $\mathbf{Q}^{h}$ of the stabilized finite element
			solution, and compute the shock-based data weights $\omega_i$ of
			Eq.~\eqref{eq:shock_weights} from the snapshot density-gradient fields.
			
			\State Initialize the PASSC neural network with trainable parameters
			$\theta$, with the output bias set to the free-stream state.
			
			\Repeat
			
			\State Sample a data mini-batch, boundary points, scale-stratified
			space--time control volumes with clearance-capped half-widths $(h_i,k_i)$,
			macro flux windows, and---every fourth physics evaluation---face-sharing
			tiling patches.
			
			\State Evaluate the network prediction $\mathbf{Q}_{\theta}$ and compute
			the shock-weighted data loss $\mathcal{L}_{\text{data}}$ and the boundary
			loss $\mathcal{L}_{\text{bc}}$.
			
			\State Assemble the space--time control-volume residuals of
			Eqs.~\eqref{eq:weak_residual} and~\eqref{eq:entropy_residual} through Gauss
			quadrature of the edge fluxes and temporal-face cell averages, apply the
			smooth saturation, and form the physics loss
			$\mathcal{L}_{\text{pde}}$ of Eq.~\eqref{eq:pde_total} with the inverse-EMA
			scale balance.
			
			\State Form the total loss
			$\mathcal{L}
			=w_{\text{data}}\mathcal{L}_{\text{data}}
			+w_{\text{pde}}\mathcal{L}_{\text{pde}}
			+w_{\text{bc}}\mathcal{L}_{\text{bc}}$
			with the linearly interpolated keyframe weights.
			
			\State Update the trainable parameters $\theta$ using the AdamW optimizer
			(with gradient clipping and learning-rate scheduling on the unweighted
			metric~\eqref{eq:eval_metric}).
			
			\Until{the epoch budget is exhausted}
			
			\State Restore the parameters attaining the lowest epoch-averaged
			unweighted metric~\eqref{eq:eval_metric} and reassemble the enhanced
			dimensional conservative solution $\mathbf{U}^{\mathrm{PASSC}}$ according
			to Eq.~\eqref{eq:passc_reassembly}.
			
		\end{algorithmic}
	\end{algorithm}
	
	Algorithm~\ref{alg:passc_euler} highlights the sequential nature of the proposed PASSC methodology. Unlike conventional PINN approaches that seek to approximate the solution field by directly enforcing the governing equations and associated constraints, the proposed framework first computes a stabilized approximation using the SUPG-YZ$\beta$ finite element method and subsequently employs the neural network as a physics-informed post-processing stage anchored to that approximation. Consequently, the network prediction remains closely tied to the stabilized solution, and the correction it delivers, Eq.~\eqref{eq:passc_decomposition}, is primarily associated with the artificial numerical diffusion introduced by the stabilization procedure rather than with the complete solution field. This substantially simplifies the learning task by avoiding the need for the network to reconstruct the global flow structure from scratch.

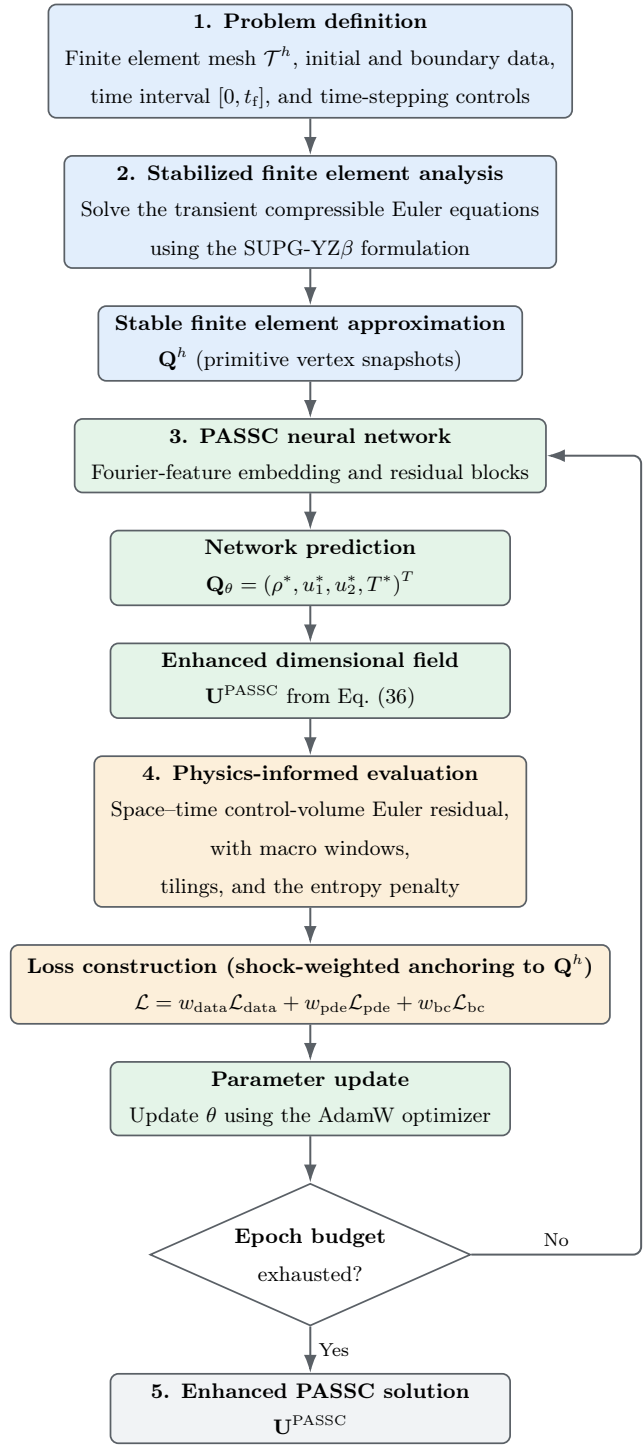
\begin{figure}[!ht]
		\centering
		
		\begin{tikzpicture}[
			scale=0.88,
			transform shape,
			node distance=5.5mm,
			font=\small,
			baseblock/.style={
				rectangle,
				rounded corners=3pt,
				draw=linegray,
				line width=0.65pt,
				align=center,
				minimum width=6.0cm,
				minimum height=9.5mm,
				inner xsep=7pt,
				inner ysep=4pt
			},
			femblock/.style={
				baseblock,
				fill=femblue
			},
			passcblock/.style={
				baseblock,
				fill=passcgreen
			},
			lossblock/.style={
				baseblock,
				fill=lossorange
			},
			outputblock/.style={
				baseblock,
				fill=outputgray
			},
			decision/.style={
				diamond,
				draw=linegray,
				line width=0.65pt,
				fill=white,
				align=center,
				aspect=2.25,
				inner xsep=6pt,
				inner ysep=2pt
			},
			flowarrow/.style={
				-{Latex[length=2.4mm,width=1.7mm]},
				line width=0.75pt,
				draw=linegray
			},
			looparrow/.style={
				-{Latex[length=2.4mm,width=1.7mm]},
				line width=0.75pt,
				draw=linegray,
				rounded corners=3pt
			}
			]
			
			
			\node[femblock] (problem)
			{\textbf{1. Problem definition}\\[-1pt]
				Finite element mesh $\mathcal{T}^{h}$, initial and boundary data,\\
				time interval $[0,t_{\text{f}}]$, and time-stepping controls};
			
			\node[femblock, below=of problem] (fem)
			{\textbf{2. Stabilized finite element analysis}\\[-1pt]
				Solve the transient compressible Euler equations\\
				using the SUPG-YZ$\beta$ formulation};
			
			\node[femblock, below=of fem] (uh)
			{\textbf{Stable finite element approximation}\\[-1pt]
				$\mathbf{Q}^{h}$ (primitive vertex snapshots)};
			
			\node[passcblock, below=of uh] (network)
			{\textbf{3. PASSC neural network}\\[-1pt]
				Fourier-feature embedding and residual blocks};
			
			\node[passcblock, below=of network] (correction)
			{\textbf{Network prediction}\\[-1pt]
				$\mathbf{Q}_{\theta}=\left(\rho^{*},u_1^{*},u_2^{*},T^{*}\right)^{T}$};
			
			\node[passcblock, below=of correction] (corrected)
			{\textbf{Enhanced dimensional field}\\[-1pt]
				$\mathbf{U}^{\mathrm{PASSC}}$
				from Eq.~\eqref{eq:passc_reassembly}};
			
			\node[lossblock, below=of corrected] (residual)
			{\textbf{4. Physics-informed evaluation}\\[-1pt]
				Space--time control-volume Euler residual,\\
				with macro windows,\\
				tilings, and the entropy penalty};
			
			\node[lossblock, below=of residual] (loss)
			{\textbf{Loss construction (shock-weighted anchoring to $\mathbf{Q}^{h}$)}\\[-1pt]
				$\mathcal{L}
				=
				w_\text{data}\mathcal{L}_\text{data}
				+
				w_\text{pde}\mathcal{L}_\text{pde}
				+
				w_\text{bc}\mathcal{L}_\text{bc}$};
			
			\node[passcblock, below=of loss] (optimizer)
			{\textbf{Parameter update}\\[-1pt]
				Update $\theta$ using the AdamW optimizer};
			
			\node[decision, below=7mm of optimizer] (criterion)
			{\textbf{Epoch budget}\\exhausted?};
			
			\node[outputblock, below=7mm of criterion] (output)
			{\textbf{5. Enhanced PASSC solution}\\[-1pt]
				$\mathbf{U}^{\mathrm{PASSC}}$};
			
			
			\draw[flowarrow] (problem) -- (fem);
			\draw[flowarrow] (fem) -- (uh);
			\draw[flowarrow] (uh) -- (network);
			\draw[flowarrow] (network) -- (correction);
			\draw[flowarrow] (correction) -- (corrected);
			\draw[flowarrow] (corrected) -- (residual);
			\draw[flowarrow] (residual) -- (loss);
			\draw[flowarrow] (loss) -- (optimizer);
			\draw[flowarrow] (optimizer) -- (criterion);
			
			
			\draw[flowarrow]
			(criterion) --
			node[right, font=\footnotesize]{Yes}
			(output);
			
			\draw[looparrow]
			(criterion.east)
			-- ++(2.55,0)
			node[above, midway, font=\footnotesize]{No}
			|- (network.east);
			
		\end{tikzpicture}
		
		\caption{Computational workflow of the proposed PASSC framework for the
			transient compressible Euler equations.}
		\label{fig:passc_workflow}
		
	\end{figure}

	The principal stages of the proposed PASSC framework are further illustrated in
	Figure~\ref{fig:passc_workflow}. The workflow emphasizes that the stabilized
	finite element approximation constitutes the primary numerical solution, whereas
	the neural network is employed solely as a physics-informed post-processing
	stage. During training, the network prediction is anchored to the stored
	SUPG-YZ$\beta$ snapshots through the shock-weighted data loss, while the physics
	and boundary losses enforce the governing equations and boundary conditions.
	
	The mapping introduced in Section~\ref{sec:correction} is approximated
	using a fully connected feed-forward neural network whose architecture,
	depicted in Figure~\ref{fig:nn-architecture}, follows the network design
	employed in our previous PASSC studies for steady~\cite{Cengizci2026_pinn_stationary}
	and transient~\cite{cengizci_pinn_transient} convection-dominated problems.
	The network takes the nondimensional spatiotemporal coordinates
	$\mathbf{z}=(t^{*},x_1^{*},x_2^{*})$ as its input and predicts the
	nondimensional primitive field
	$\mathbf{Q}_{\theta}
	= (\rho^{*},\,u_1^{*},\,u_2^{*},\,T^{*})^{T}$,
	from which the enhanced conservative field
	$\mathbf{U}^{\mathrm{PASSC}}=\mathbf{U}\!\left(\mathbf{Q}_{\theta}\right)$
	of Eq.~\eqref{eq:passc_reassembly} is assembled.
	%
	\usetikzlibrary{positioning,arrows.meta,calc,shapes.geometric,decorations.pathreplacing,patterns}
	\def\nh{128}
	\def\nr{8}
	\def\nF{24}
	\def\inputdim{3}
	\def\fourierdim{51}
	\def\narrowdim{64}
	\def\outputdim{4}
	
	\begin{figure}[!htb]
		\centering
		\resizebox{0.95\textwidth}{!}{
			\begin{tikzpicture}[
				>=Stealth,
				neuron/.style={circle, draw, minimum size=11pt, inner sep=0pt, line width=0.5pt},
				input neuron/.style={neuron, fill=blue!20, draw=blue!50},
				fourier neuron/.style={neuron, fill=violet!18, draw=violet!50},
				hidden neuron/.style={neuron, fill=cyan!15, draw=cyan!50},
				res neuron/.style={neuron, fill=green!15, draw=green!45!black},
				narrow neuron/.style={neuron, fill=red!12, draw=red!45},
				output neuron/.style={neuron, fill=red!22, draw=red!65, minimum size=17pt},
				sum node/.style={circle, draw=gray!70, fill=gray!8, minimum size=19pt,
					inner sep=1pt, line width=0.7pt, font=\scriptsize},
				conn/.style={line width=0.12pt, draw=black!12},
				skip conn/.style={line width=0.8pt, draw=orange!65!black, dashed,
					-{Stealth[length=3.5pt]}},
				assemble conn/.style={line width=0.7pt, draw=gray!60,
					-{Stealth[length=4pt]}},
				anchor conn/.style={line width=0.8pt, draw=gray!55, dashed,
					-{Stealth[length=4pt]}},
				layer label/.style={font=\small\bfseries, align=center},
				sublabel/.style={font=\scriptsize, text=gray!65!black, align=center},
				]
				
				\def\neuronsp{0.50}
				\def\layersep{2.0}
				
				\def\xI{0}
				\node[input neuron] (I-1) at (\xI, 1.2)  {\tiny$t^{*}$};
				\node[input neuron] (I-2) at (\xI, 0)    {\tiny$x_1^{*}$};
				\node[input neuron] (I-3) at (\xI, -1.2) {\tiny$x_2^{*}$};
				
				\def\xF{2.0}
				\foreach \i in {1,...,4} {
					\pgfmathsetmacro{\yy}{2.1 - (\i-1)*0.55}
					\node[fourier neuron] (F-t\i) at (\xF, \yy) {};
				}
				\node[font=\normalsize, text=gray!60] (F-dots) at (\xF, 0) {$\vdots$};
				\foreach \i in {1,...,4} {
					\pgfmathsetmacro{\yy}{-0.55 - (\i-1)*0.55}
					\node[fourier neuron] (F-b\i) at (\xF, \yy) {};
				}
				
				\def\xH{4.0}
				\foreach \i in {1,...,5} {
					\pgfmathsetmacro{\yy}{2.65 - (\i-1)*0.55}
					\node[hidden neuron] (H-t\i) at (\xH, \yy) {};
				}
				\node[font=\normalsize, text=gray!60] (H-dots) at (\xH, 0) {$\vdots$};
				\foreach \i in {1,...,5} {
					\pgfmathsetmacro{\yy}{-0.55 - (\i-1)*0.55}
					\node[hidden neuron] (H-b\i) at (\xH, \yy) {};
				}
				
				\foreach \blk in {1,2,3} {
					\pgfmathsetmacro{\xR}{4.0 + \blk * 2.0}
					\foreach \i in {1,...,5} {
						\pgfmathsetmacro{\yy}{2.65 - (\i-1)*0.55}
						\node[res neuron] (R\blk-t\i) at (\xR, \yy) {};
					}
					\node[font=\normalsize, text=gray!60] (R\blk-dots) at (\xR, 0) {$\vdots$};
					\foreach \i in {1,...,5} {
						\pgfmathsetmacro{\yy}{-0.55 - (\i-1)*0.55}
						\node[res neuron] (R\blk-b\i) at (\xR, \yy) {};
					}
				}
				
				\def\xEll{9.0}
				\node[font=\Large\bfseries, text=green!40!black]
				at (\xEll, 0) {$\cdots$};
				
				\def\xN{12.6}
				\foreach \i in {1,...,4} {
					\pgfmathsetmacro{\yy}{1.85 - (\i-1)*0.55}
					\node[narrow neuron] (N-t\i) at (\xN, \yy) {};
				}
				\node[font=\normalsize, text=gray!60] (N-dots) at (\xN, 0) {$\vdots$};
				\foreach \i in {1,...,4} {
					\pgfmathsetmacro{\yy}{-0.55 - (\i-1)*0.55}
					\node[narrow neuron] (N-b\i) at (\xN, \yy) {};
				}
				
				\def\xO{14.6}
				\node[output neuron] (O-1) at (\xO, 1.35)  {\tiny$\rho^{*}$};
				\node[output neuron] (O-2) at (\xO, 0.45)  {\tiny$u_1^{*}$};
				\node[output neuron] (O-3) at (\xO, -0.45) {\tiny$u_2^{*}$};
				\node[output neuron] (O-4) at (\xO, -1.35) {\tiny$T^{*}$};
				
				
				\foreach \i in {1,2,3} {
					\foreach \j in {1,...,4} {
						\draw[conn] (I-\i) -- (F-t\j);
						\draw[conn] (I-\i) -- (F-b\j);
					}
				}
				
				\foreach \src in {t1,t2,t3,t4,b1,b2,b3,b4} {
					\foreach \dst in {t1,t2,t3,t4,t5,b1,b2,b3,b4,b5} {
						\draw[conn] (F-\src) -- (H-\dst);
					}
				}
				
				\foreach \src in {t1,t2,t3,t4,t5,b1,b2,b3,b4,b5} {
					\foreach \dst in {t1,t2,t3,t4,t5,b1,b2,b3,b4,b5} {
						\draw[conn] (H-\src) -- (R1-\dst);
					}
				}
				
				\foreach \src in {t1,t2,t3,t4,t5,b1,b2,b3,b4,b5} {
					\foreach \dst in {t1,t2,t3,t4,t5,b1,b2,b3,b4,b5} {
						\draw[conn] (R1-\src) -- (R2-\dst);
					}
				}
				
				\foreach \src in {t1,t2,t3,t4,t5,b1,b2,b3,b4,b5} {
					\foreach \dst in {t1,t2,t3,t4,t5,b1,b2,b3,b4,b5} {
						\draw[conn] (R2-\src) -- (R3-\dst);
					}
				}
				
				\foreach \src in {t1,t2,t3,t4,t5,b1,b2,b3,b4,b5} {
					\foreach \dst in {t1,t2,t3,t4,b1,b2,b3,b4} {
						\draw[conn] (R3-\src) -- (N-\dst);
					}
				}
				
				\foreach \src in {t1,t2,t3,t4,b1,b2,b3,b4} {
					\foreach \dst in {1,2,3,4} {
						\draw[conn, line width=0.25pt, draw=black!20]
						(N-\src) -- (O-\dst);
					}
				}
				
				\draw[skip conn]
				([yshift=3pt]H-t1.north) to[out=70, in=110]
				([yshift=3pt]R1-t1.north);
				
				\draw[skip conn]
				([yshift=3pt]R1-t1.north) to[out=70, in=110]
				([yshift=3pt]R2-t1.north);
				
				\draw[skip conn]
				([yshift=3pt]R2-t1.north) to[out=70, in=110]
				([yshift=3pt]R3-t1.north);
				
				\draw[decorate,
				decoration={brace, amplitude=7pt, raise=4pt},
				green!50!black, line width=0.7pt]
				(5.5, 3.6) -- (10.5, 3.6)
				node[midway, above=13pt, font=\small\bfseries,
				text=green!40!black]
				{$\times\;\nr$ Residual Blocks};
				
				%
				\def\xS{16.4}
				
				\node[sum node] (SUM) at (\xS, 0)
				{\scriptsize Assembly};
				
				\foreach \dst in {1,2,3,4} {
					\draw[assemble conn, draw=red!45] (O-\dst) -- (SUM);
				}
				
				\node[font=\scriptsize, text=gray!55!black, align=center]
				(UH) at (\xS, -1.9)
				{$\mathbf{Q}^{h}$ snapshots\\[-2pt]
					{\tiny (SUPG-YZ$\beta$)}};
				
				\draw[anchor conn]
				(UH) --
				node[right, font=\tiny, text=gray!55!black]
				{$\mathcal{L}_{\text{data}}$}
				(SUM);
				
				\node[font=\small, text=black!75,
				right=8pt of SUM, align=left]
				(UP) {$\mathbf{U}^{\mathrm{PASSC}}$};
				
				\draw[assemble conn] (SUM) -- (UP);
				
				\def\labelY{-3.6}
				
				\node[layer label, text=blue!60!black]
				at (\xI, \labelY) {Input};
				\node[sublabel] at (\xI, \labelY - 0.4)
				{$\mathbb{R}^{\inputdim}$};
				\node[sublabel] at (\xI, \labelY - 0.75)
				{$(t^{*}, x_1^{*}, x_2^{*})$};
				
				\node[layer label, text=violet!60!black]
				at (\xF, \labelY) {Fourier};
				\node[sublabel] at (\xF, \labelY - 0.4)
				{$\mathbb{R}^{\fourierdim}$};
				\node[sublabel] at (\xF, \labelY - 0.75)
				{$[\mathbf{z};\sin;\cos]$};
				
				\node[layer label, text=cyan!45!black]
				at (\xH, \labelY) {Input Layer};
				\node[sublabel] at (\xH, \labelY - 0.4)
				{$\nh$ neurons};
				\node[sublabel] at (\xH, \labelY - 0.75)
				{SiLU};
				
				\node[layer label, text=green!40!black]
				at (8, \labelY) {Hidden Layers};
				\node[sublabel] at (8, \labelY - 0.4)
				{$\nh$ neurons/block};
				\node[sublabel] at (8, \labelY - 0.75)
				{Skip $+$ LayerNorm $+$ SiLU};
				
				\node[layer label, text=red!55!black]
				at (\xN, \labelY) {Narrowing};
				\node[sublabel] at (\xN, \labelY - 0.4)
				{$\narrowdim$ neurons};
				\node[sublabel] at (\xN, \labelY - 0.75)
				{SiLU};
				
				\node[layer label, text=red!65!black]
				at (\xO, \labelY) {Output};
				\node[sublabel] at (\xO, \labelY - 0.4)
				{$\outputdim$ neurons};
				\node[sublabel] at (\xO, \labelY - 0.75)
				{$\mathbf{Q}_{\theta}(\mathbf{z})$};
				
				\node[layer label, text=gray!55!black]
				at (\xS + 0.7, \labelY) {Assembly};
				\node[sublabel] at (\xS + 0.7, \labelY - 0.4)
				{Eq.~\eqref{eq:passc_reassembly}};
				\node[sublabel] at (\xS + 0.7, \labelY - 0.75)
				{$\mathbf{U}^{\mathrm{PASSC}}$};
				
		\end{tikzpicture}}
		
		\caption{Schematic of the PASSC network architecture employed
			for the compressible Euler equations
			(shown for $n_h=\nh$, $n_r=\nr$, $n_\mathrm{F}=\nF$, $n_\mathrm{sd}=2$).
			The nondimensional spatiotemporal input
			$\mathbf{z}=(t^{*},x_1^{*},x_2^{*})\in\mathbb{R}^{\inputdim}$
			is mapped through a random Fourier feature embedding to
			$\boldsymbol{\varphi}(\mathbf{z})\in\mathbb{R}^{\fourierdim}$,
			projected to a hidden dimension of $\nh$ via the input layer with SiLU
			activation, and then processed by $\nr$ residual blocks---each comprising
			two fully connected layers, a skip (residual) connection, and layer
			normalization. A narrowing layer ($\nh\!\to\!\narrowdim$) followed by a
			four-neuron linear output layer, whose bias is initialized at the
			free-stream state, produces the nondimensional primitive field
			$\mathbf{Q}_{\theta}
			=(\rho^{*},\,u_1^{*},\,u_2^{*},\,T^{*})^{T}$
			of Eq.~\eqref{eq:network_map}, from which the enhanced dimensional
			conservative field $\mathbf{U}^{\mathrm{PASSC}}$ is assembled according
			to Eq.~\eqref{eq:passc_reassembly}. The stored SUPG-YZ$\beta$ primitive
			snapshots $\mathbf{Q}^{h}$ enter the training through the data-consistency
			loss $\mathcal{L}_{\text{data}}$ (dashed arrow).}
		\label{fig:nn-architecture}
	\end{figure}
	
	\section{Test computations}
	\label{sec6}

	The problems considered are inviscid and unsteady. At each time step, the resulting nonlinear algebraic systems are solved with damped Newton--Raphson iterations within the adaptive time-stepping strategy described in Section~\ref{sec4}, and at each nonlinear iteration, the linear equation system is solved with the MUMPS sparse direct solver. 
	
	All finite element solvers are implemented within the open-source FEniCS computing environment, which provides high-level Python and C++ interfaces for the automated solution of PDEs~\cite{Alnaes2015,logg2012automated,abali2016computational}. As one of the most widely adopted open-source finite element platforms in computational science, the FEniCS Project has evolved into the next-generation FEniCSx framework, offering a modern software architecture and enhanced capabilities for high-performance scientific computing~\cite{dolfinx2023}. Further information about the project is available at \url{https://fenicsproject.org/}.
	
	The physics-informed neural network (PINN) models are implemented using the
	open-source PyTorch deep-learning framework~\cite{paszke2019pytorch}, which
	provides efficient tensor operations, GPU acceleration, and automatic
	differentiation capabilities for scientific machine learning. In the
	computations reported in this work, however, the governing-equation residual
	is evaluated using the derivative-free space--time control-volume formulation
	introduced in Section~\ref{sec:loss}; therefore, no spatial or temporal
	derivatives of the network outputs are computed by automatic differentiation
	in the physics loss. Automatic differentiation is used only for
	backpropagation of the total loss with respect to the trainable network
	parameters. Detailed documentation, installation instructions, and software
	releases are available from the official PyTorch website at
	\url{https://pytorch.org/}.
	
	All computations are performed on a workstation running Ubuntu 24.04 LTS, equipped with an Intel Core Ultra 9 275HX processor, 128~GB of RAM, and an NVIDIA GeForce RTX 5080 Laptop GPU with 16~GB of dedicated memory; the computational cost of the two stages is reported in Section~\ref{sec:cost}. The source code implementing both stages of the proposed PASSC framework is
	openly available at \url{https://github.com/scengizci/equilibrium_hypersonic}.
	The repository contains the SUPG-YZ$\beta$ stabilized finite element solver (FEniCS) for
	the compressible Euler equations, the
	physics-informed correction network (implemented in PyTorch), the unstructured
	mesh employed in the computations, the solver and training configuration files
	for all four free-stream Mach numbers considered, and the post-processing
	scripts used to generate the results reported in this section.

	\subsection{2D high-speed flow computations}
	We consider the canonical problem of inviscid high-speed flow past a circular
	cylinder, which has been widely employed as a benchmark for compressible-flow
	computations (see, e.g.,~\cite{Jiang1996}). The simulations are performed at
	four free-stream Mach numbers,
	$M_{\infty}=2.0$, $5.0$, $8.0$, and $12.0$, spanning the supersonic and
	hypersonic regimes. Although the computational geometry adopted here differs
	from that employed in the studies of Kirk et al.~\cite{kirk1,kirk2,kirk5}---in
	particular, the present domain includes the complete flow field around the
	cylinder rather than only the upstream region---their results provide useful
	reference data under comparable free-stream conditions and for related
	SUPG-based stabilized formulations. The numerical results are further assessed
	against the available normal-shock and stagnation relations
	(see, e.g.,~\cite{naca1135,Anderson2019}) and, where appropriate, the
	semi-empirical Billig correlation for bow-shock stand-off distance
	~\cite{billig1967}.

\begin{figure}[htb] 
		\centerline{\includegraphics[scale=0.55]{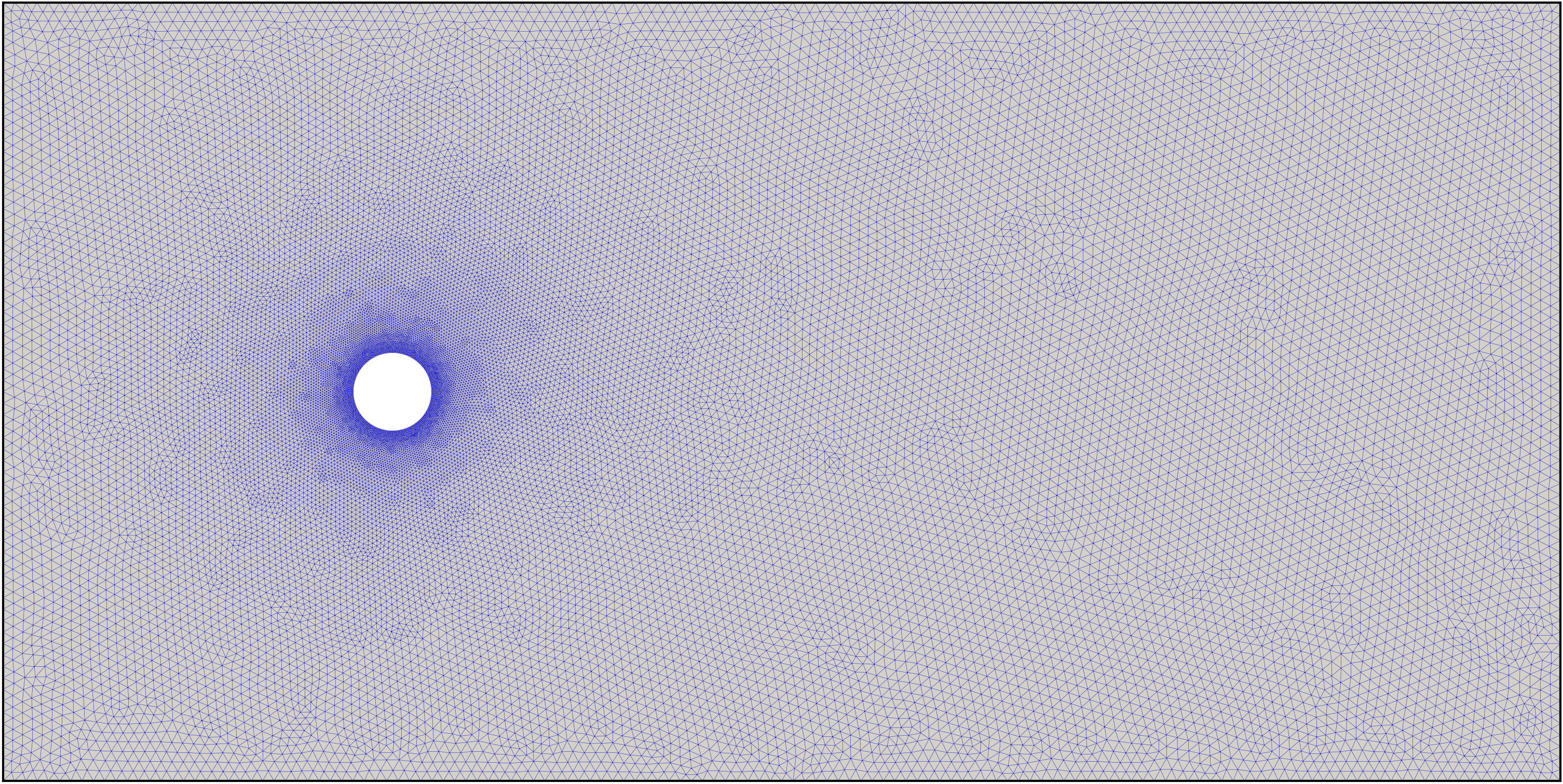}}
		\caption{Unstructured mesh used in computations. It has $23,934$ nodes and $47,264$ triangular elements.}
		\label{Mesh}
	\end{figure}

\begin{figure}[htb]
		\centerline{\includegraphics[scale=0.55]{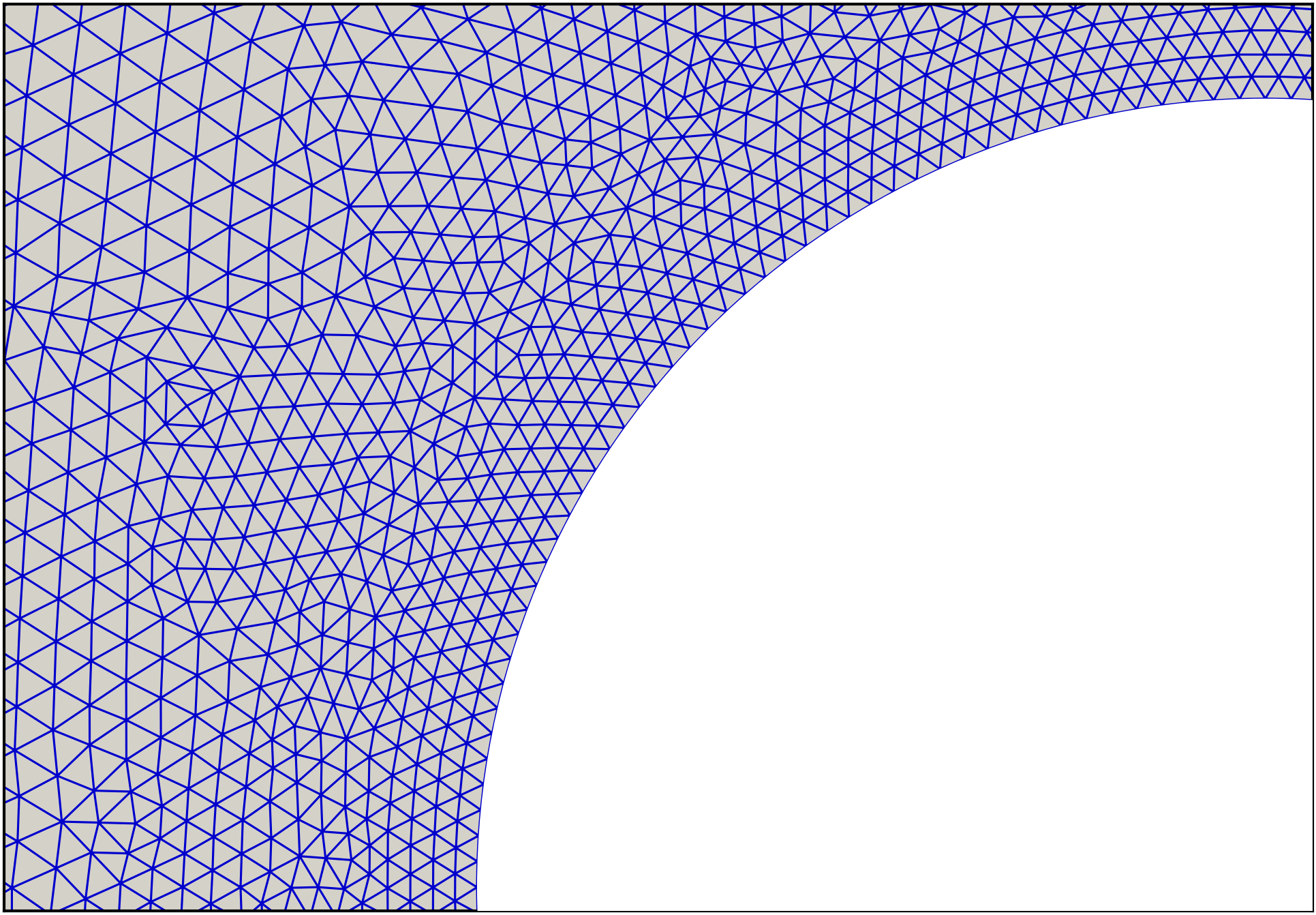}}
		\caption{Layers of constant-thickness elements near the cylinder.}
		\label{Meshzoomed}
	\end{figure}

	In all cases, $\rho_{\infty}=1.165$~kg/m$^{3}$ and $T_{\infty}=300.0$~K. In
	the YZ$\beta$ term, the sharpness parameter is $\beta=2$ (see
	Eq.~\eqref{shoc}). The Courant numbers employed are $C_{\varDelta t}=0.5$
	for $M_{\infty}=2.0$ and $M_{\infty}=5.0$, and $C_{\varDelta t}=0.4$ for
	$M_{\infty}=8.0$ and $M_{\infty}=12.0$. The cylinder has a radius of
	$R=0.05$~m (see Figure~\ref{fig:domain}), by which the shock stand-off
	distances reported below are normalized. The mesh has $23{,}934$ nodes and
	$47{,}264$ triangular elements (see Figure~\ref{Mesh}).
	Figure~\ref{Meshzoomed} shows the layers of constant-thickness elements
	near the cylinder. All computations are performed until
	$t_\text{f}=1.0\times10^{-3}$~s.
	
	All FEM simulations were completed without the need for Newton relaxation or time-step size reduction. These algorithmic safeguards were incorporated to enhance the robustness and generality of the proposed framework, particularly for potential extensions to even higher Mach numbers or more challenging flow problems.
	
	Table~\ref{tab:shock_metrics} collects the bow-shock stand-off
	distances and stagnation-pressure ratios obtained in all four cases,
	which are discussed individually below; the training behavior of the
	correction network, which follows the same pattern in every case, is
	discussed collectively in Section~\ref{sec:training_behavior}.
	\begin{table}[htb]
		\centering
		\caption{Bow-shock stand-off distances and stagnation-pressure ratios
			for the four free-stream Mach numbers considered. The stand-off
			distances are normalized by the cylinder radius $R$ and compared
			with the Billig correlation~\cite{billig1967}; the
			stagnation-pressure ratios are compared with the corresponding
			Rayleigh pitot values.}
		\label{tab:shock_metrics}
		\begin{tabular}{lcccc}
			\hline
			& $M_{\infty}=2.0$ & $M_{\infty}=5.0$ & $M_{\infty}=8.0$ & $M_{\infty}=12.0$ \\
			\hline
			$\varDelta/R$, SUPG-YZ$\beta$            & $1.030$ & $0.458$  & $0.424$ & $0.398$  \\
			$\varDelta/R$, PASSC                     & $1.086$ & $0.497$  & $0.424$ & $0.424$  \\
			$\varDelta/R$, Billig~\cite{billig1967}  & $1.241$ & $0.465$  & $0.415$ & $0.399$  \\
			\hline
			$p_{0}/p_{\infty}$, SUPG-YZ$\beta$       & $5.729$ & $32.135$ & $81.45$ & $182.44$ \\
			$p_{0}/p_{\infty}$, PASSC                & $5.778$ & $32.430$ & $81.10$ & $182.82$ \\
			$p_{0}/p_{\infty}$, Rayleigh pitot       & $5.640$ & $32.653$ & $82.9$  & $185.87$ \\
			\hline
		\end{tabular}
	\end{table}


	\subsubsection{Supersonic flow at $M_{\infty}=2.0$}
	\label{sec:mach2}
	
\begin{figure}[htb]
		\centering
		\includegraphics[width=1\linewidth]{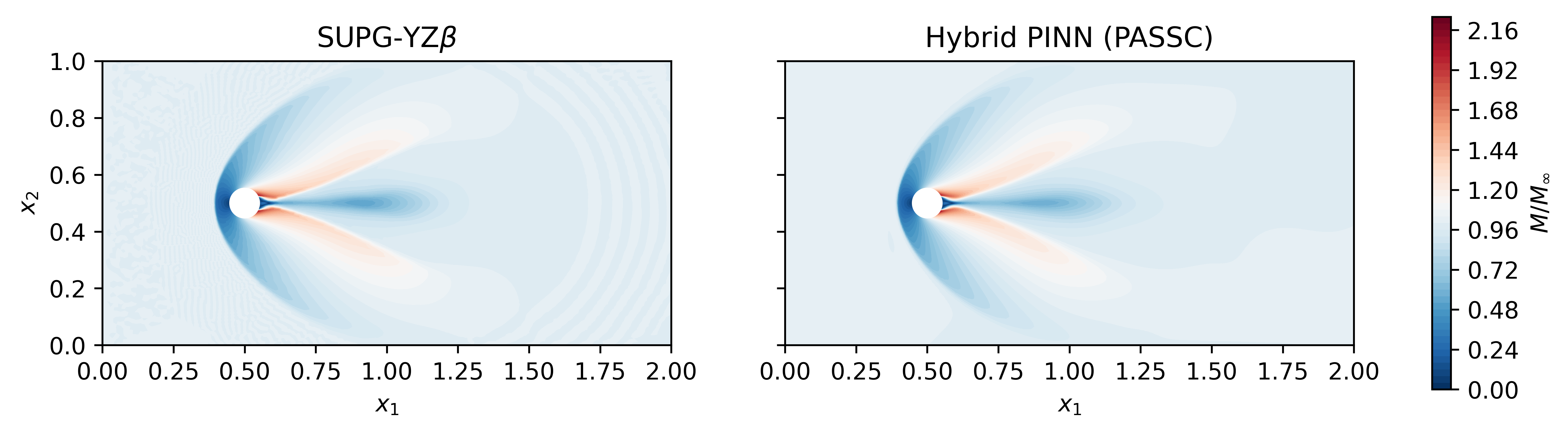}
		\caption{Contours of the scaled Mach number $(M/M_{\infty})$ for
			$M_{\infty}=2.0$, computed with the SUPG-YZ$\beta$ formulation and
			the PASSC framework.}
		\label{fig:ma2_mach}
	\end{figure}

\begin{figure}[htb]
		\centering
		\includegraphics[width=1\linewidth]{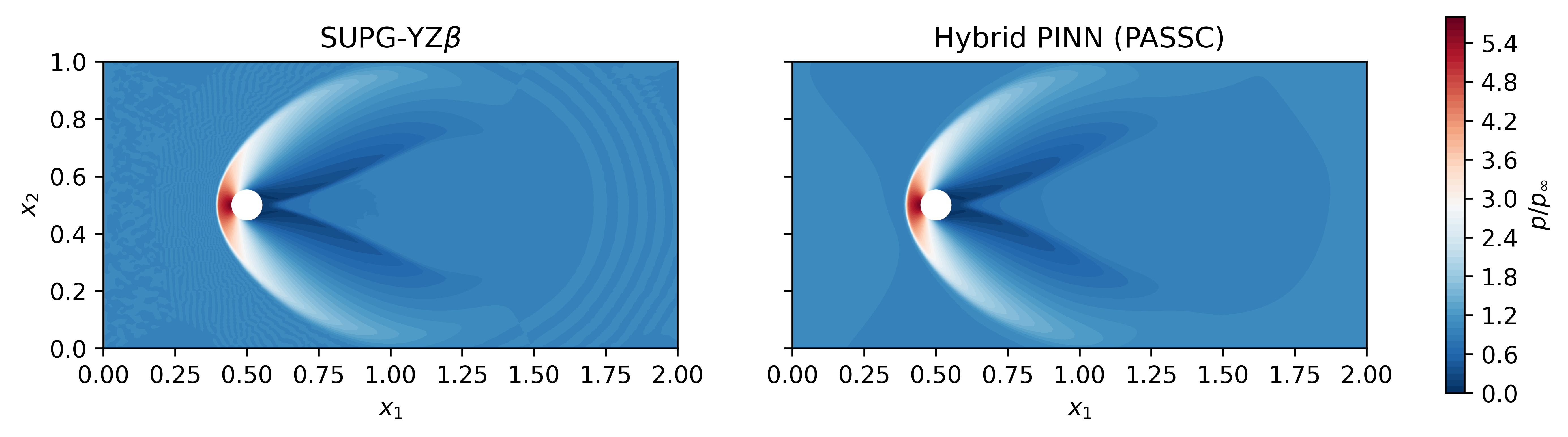}
		\caption{Contours of the scaled pressure $(p/p_{\infty})$ for
			$M_{\infty}=2.0$, computed with the SUPG-YZ$\beta$ formulation and
			the PASSC framework.}
		\label{fig:ma2_pressure}
	\end{figure}

\begin{figure}[htb]
		\centering
		\includegraphics[width=1\linewidth]{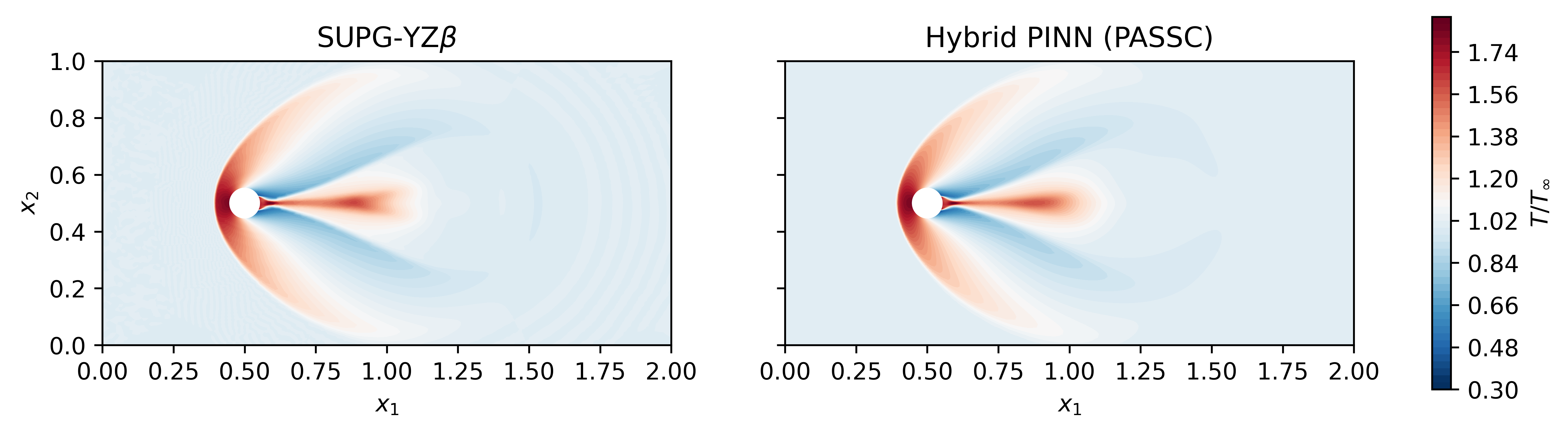}
		\caption{Contours of the scaled temperature $(T/T_{\infty})$ for
			$M_{\infty}=2.0$, computed with the SUPG-YZ$\beta$ formulation and
			the PASSC framework.}
		\label{fig:ma2_temp}
	\end{figure}

	We begin with the supersonic reference case, for which the free-stream
	velocity vector is
	$\mathbf{u}_{\infty}=(706.133,\,0.0)$~m/s.
	Figures~\ref{fig:ma2_mach}--\ref{fig:ma2_temp} show the contour plots of
	the scaled Mach number $(M/M_{\infty})$, pressure $(p/p_{\infty})$, and
	temperature $(T/T_{\infty})$ obtained with the SUPG-YZ$\beta$ formulation
	and the PASSC framework. Both approaches reproduce the principal features
	of the flow field: a detached bow shock upstream of the cylinder, the
	stagnation region at the front of the body where the Mach number decreases
	toward zero while the pressure and temperature increase, the expansion
	around the cylinder shoulders, and the low-speed, high-temperature wake
	developing downstream. The predicted bow-shock geometry is also broadly
	consistent between the two solutions.
	
	Compared with the SUPG-YZ$\beta$ solution, in which small-amplitude
	oscillatory structures are visible upstream of the shock and as concentric
	patterns in the downstream far field, the PASSC solution provides visibly
	smoother fields, with these localized oscillations substantially reduced.
	Similar small oscillations can also be observed in the density distribution
	reported in~\cite{kirk1}, indicating that such features are not uncommon in
	stabilized finite element approximations of this problem.
	
\begin{figure}[htb]
		\centering
		\includegraphics[width=1\linewidth]{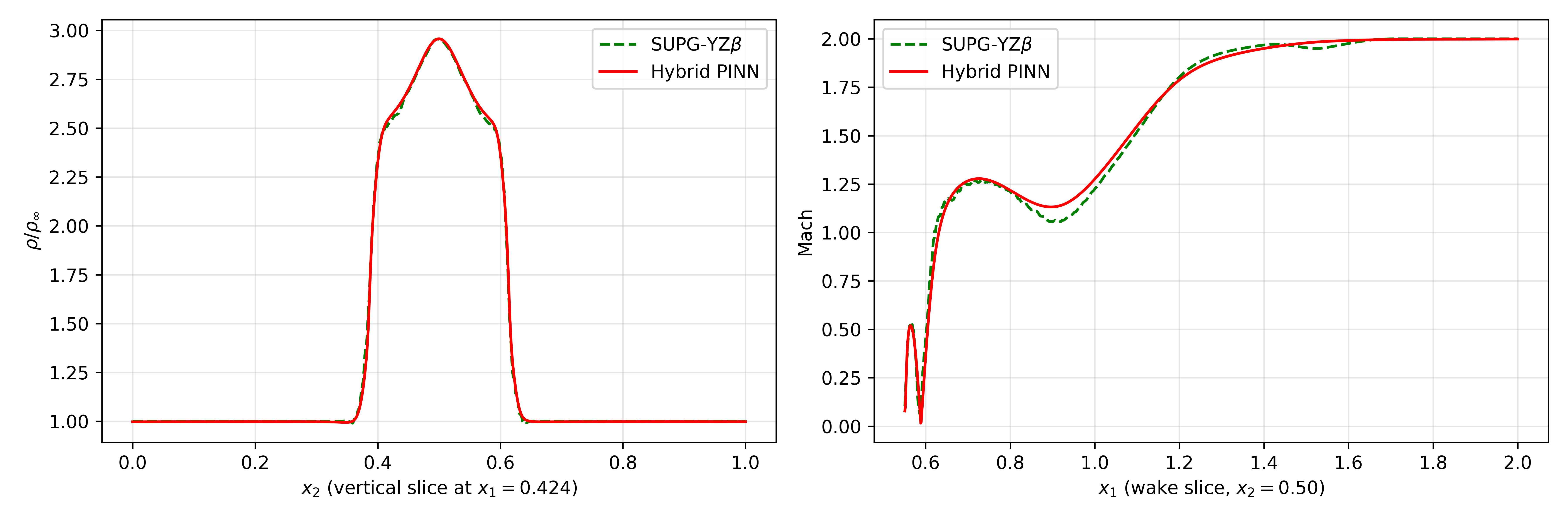}
		\caption{Comparison of the SUPG-YZ$\beta$ and PASSC solutions along
			selected cross sections for $M_{\infty}=2.0$: scaled density
			$(\rho/\rho_{\infty})$ along the vertical slice at $x_1=0.424$ and
			Mach number along the wake centerline at $x_2=0.50$.}
		\label{fig:ma2_cross}
	\end{figure}

	For $M_{\infty}=2.0$, the theoretical ratios immediately downstream of a
	normal shock are
	$p/p_{\infty}=4.5$,
	$\rho/\rho_{\infty}=2.66667$, and
	$T/T_{\infty}=1.6875$.
	Along the stagnation streamline, the post-shock flow subsequently
	decelerates toward the stagnation point, and the corresponding theoretical
	density ratio is
	$\rho_{0}/\rho_{\infty}\approx3.13$.
	This behavior is reflected in Figure~\ref{fig:ma2_cross}, which presents
	the scaled density along a vertical slice passing through the bow shock at
	$x_1=0.424$, together with the Mach number along the wake centerline at
	$x_2=0.50$. The density profile attains a peak value of approximately
	$\rho/\rho_{\infty}\approx2.95$ on the stagnation line, lying between the
	normal-shock value and the theoretical stagnation value. The two solutions
	remain in close agreement along the vertical slice, with the PASSC
	primarily smoothing the small stabilization-induced undulations flanking
	the density plateau.
	
	Along the wake centerline, the Mach-number recovery, including the local
	acceleration and recompression behind the cylinder, follows the
	SUPG-YZ$\beta$ solution closely. The PASSC solution renders the
	recompression dip near $x_1\approx0.9$ slightly shallower and smoother and
	reduces the residual undulation visible in the finite element profile
	around $x_1\approx1.5$, resulting in a smoother recovery toward the
	free-stream Mach number.
	
	The first column of Table~\ref{tab:shock_metrics} quantifies these
	observations. The SUPG-YZ$\beta$ solution places the bow shock at a
	stand-off distance of $\varDelta/R=1.030$, approximately $17\%$ below the
	value of $1.241$ predicted by the Billig correlation~\cite{billig1967},
	whereas the PASSC solution yields $\varDelta/R=1.086$, recovering roughly
	one quarter of this difference. The Billig correlation, which is based on
	steady open-flow data, is used here as a reference rather than as an exact
	target, since the present results are obtained in a finite computational
	domain and at a finite simulation time. Accordingly, the relevant
	observation is that the PASSC regularization shifts the detected shock
	position toward the correlation while preserving the overall flow
	structure. The stagnation-pressure ratios are
	$p_{0}/p_{\infty}=5.729$ for the SUPG-YZ$\beta$ solution and
	$p_{0}/p_{\infty}=5.778$ for the PASSC solution, both remaining within
	approximately $2.5\%$ of the Rayleigh pitot value
	$p_{0}/p_{\infty}=5.640$.
	
	The comparatively large discrepancy at $M_{\infty}=2.0$ admits a
	straightforward explanation in terms of temporal convergence toward the
	steady state. Since all four computations are advanced to the same physical
	time $t_{\mathrm{f}}=10^{-3}$~s, the number of elapsed convective
	flow-through times decreases with the free-stream velocity: based on the
	streamwise domain length of $2$~m, $t_{\mathrm{f}}$ corresponds to
	approximately $0.35$ flow-through times at $M_{\infty}=2.0$, compared with
	approximately $0.9$, $1.4$, and $2.1$ at $M_{\infty}=5.0$, $8.0$, and
	$12.0$, respectively. Moreover, the steady stand-off distance itself is
	largest at the lowest Mach number, so the detached bow shock must propagate
	farther upstream before settling. The $M_{\infty}=2.0$ field is therefore
	the farthest from its steady state at the common final time, and the
	residual underprediction of the Billig stand-off distance is attributable
	primarily to this incomplete temporal convergence rather than to the
	spatial discretization. Consistent with this interpretation, at the three
	hypersonic Mach numbers, for which the elapsed flow-through times approach
	or exceed unity, the stabilized stand-off distances already lie within
	approximately $2\%$ of the correlation.
	
\begin{figure}[htb]
		\centering
		\includegraphics[width=1\linewidth]{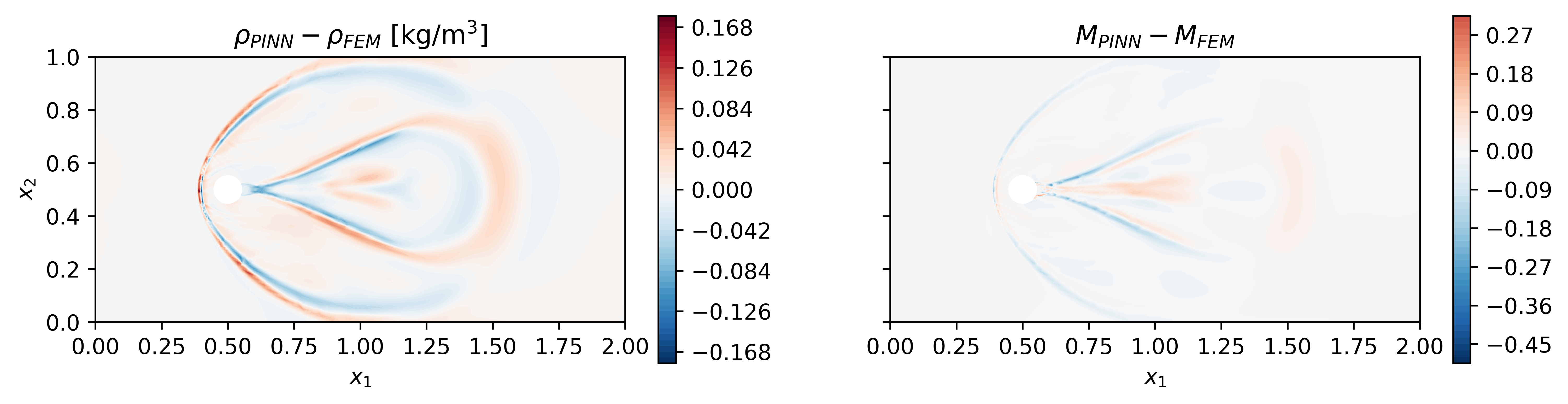}
		\caption{Pointwise differences between the PASSC and SUPG-YZ$\beta$
			solutions for $M_{\infty}=2.0$: density difference
			$\rho_{\mathrm{PASSC}}-\rho_{\mathrm{FEM}}$ in kg/m$^3$ and Mach-number
			difference $M_{\mathrm{PASSC}}-M_{\mathrm{FEM}}$.}
		\label{fig:ma2_diff}
	\end{figure}

	The pointwise difference maps in Figure~\ref{fig:ma2_diff} further
	illustrate the localized character of the PASSC modification. The largest
	differences are concentrated in thin dipole-like bands surrounding the bow
	shock and in the shear layers bounding the near wake, with peak density
	differences of approximately $0.17$~kg/m$^3$ (about $15\%$ of
	$\rho_{\infty}$) and Mach-number differences of up to approximately $0.45$
	in magnitude occurring across the shock front. Such localized differences
	are expected in regions of steep gradients, where even a small displacement
	or sharpening of a captured discontinuity can produce a comparatively
	large pointwise difference. A more diffuse, low-amplitude difference
	pattern is also visible in the downstream half of the domain, corresponding
	to regions in which the PASSC solution smooths the concentric oscillatory
	structures present in the finite element fields.
	
	Away from the shock and wake regions, the differences between the two
	solutions remain small throughout most of the computational domain,
	indicating that the PASSC solution remains closely anchored to the
	SUPG-YZ$\beta$ approximation while providing a smoother continuous
	representation of the flow field. As shown in the subsequent cases, the
	same qualitative pattern---localized differences concentrated around sharp
	flow features and substantially smaller differences elsewhere---persists
	at the higher Mach numbers. The interpretation is therefore discussed in
	detail here and only the case-specific features and amplitudes are
	emphasized subsequently.
	
	\subsubsection{Hypersonic flow at $M_{\infty}=5.0$}
	
\begin{figure}[!htb]
		\centering
		\includegraphics[width=1\linewidth]{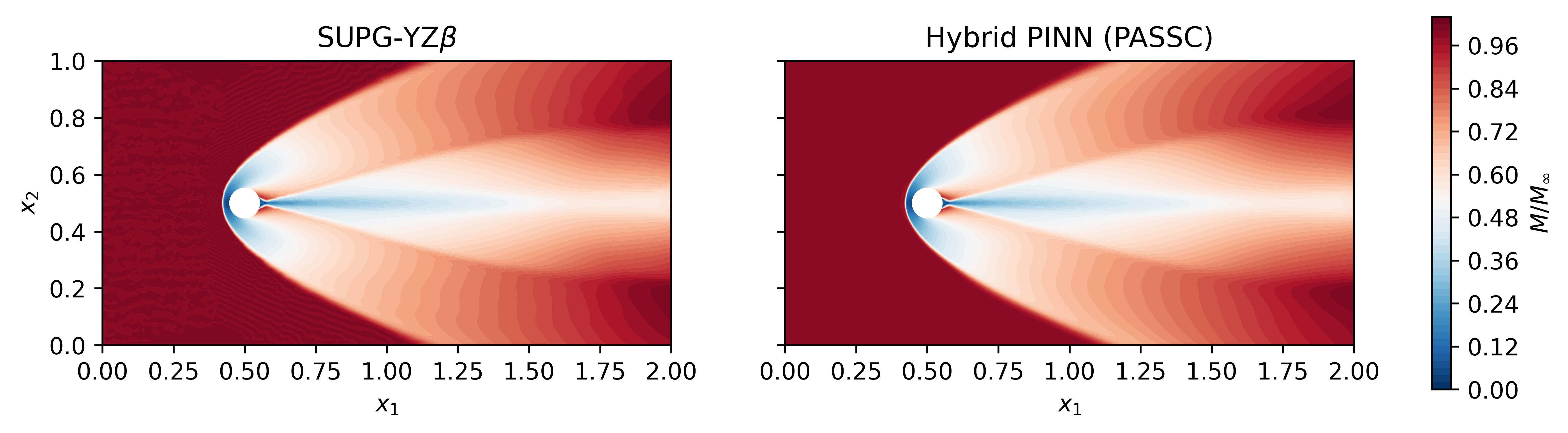}
		\caption{Contours of the scaled Mach number $(M/M_{\infty})$ for
			$M_{\infty}=5.0$, computed with the SUPG-YZ$\beta$ formulation and
			the PASSC framework.}
		\label{fig:ma5_mach}
	\end{figure}

\begin{figure}[htb]
		\centering
		\includegraphics[width=1\linewidth]{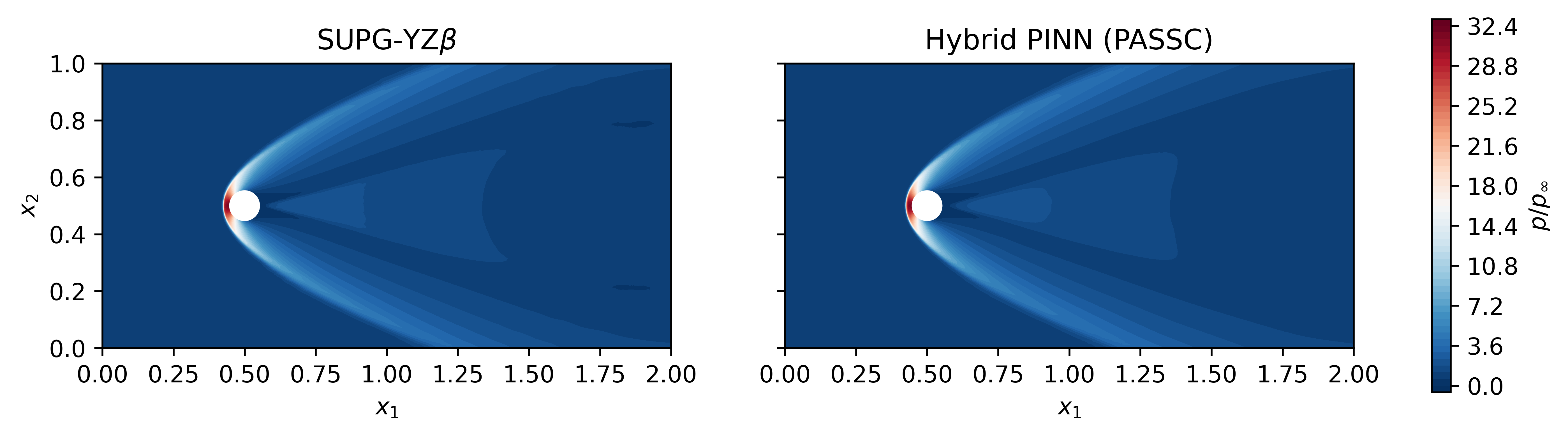}
		\caption{Contours of the scaled pressure $(p/p_{\infty})$ for
			$M_{\infty}=5.0$, computed with the SUPG-YZ$\beta$ formulation and
			the PASSC framework.}
		\label{fig:ma5_pressure}
	\end{figure}

\begin{figure}[htb]
		\centering
		\includegraphics[width=1\linewidth]{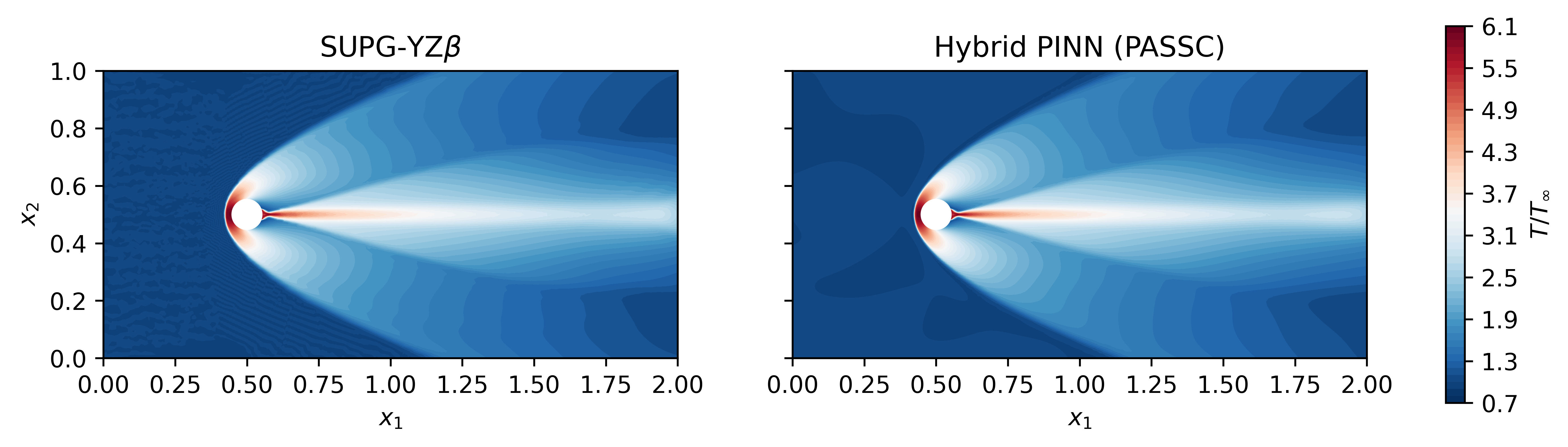}
		\caption{Contours of the scaled temperature $(T/T_{\infty})$ for
			$M_{\infty}=5.0$, computed with the SUPG-YZ$\beta$ formulation and
			the PASSC framework.}
		\label{fig:ma5_temp}
	\end{figure}

\begin{figure}[htb]
		\centering
		\includegraphics[width=1\linewidth]{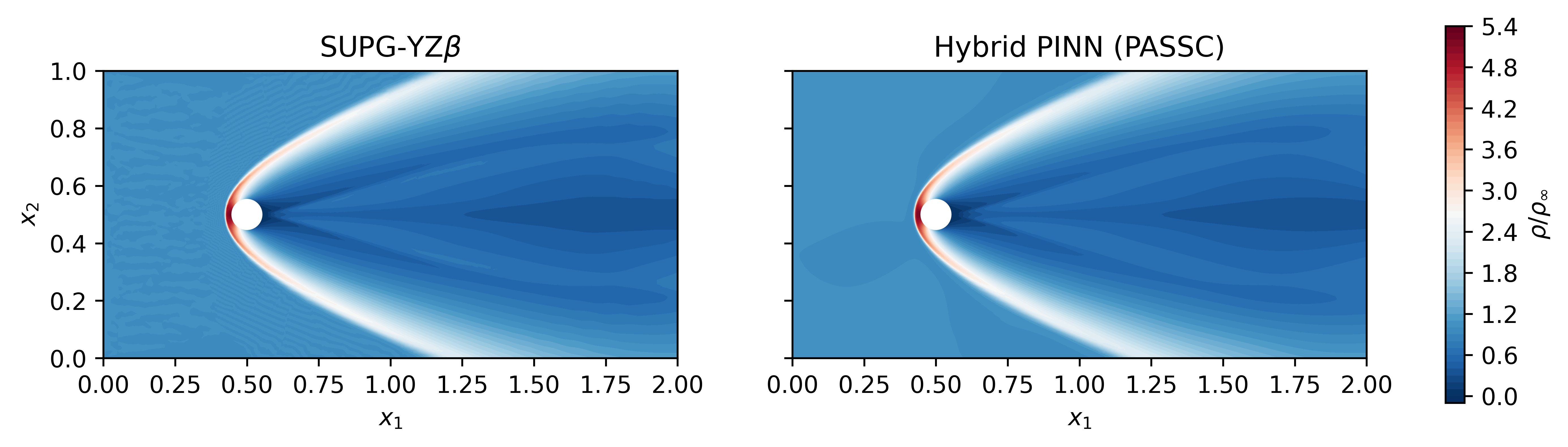}
		\caption{Contours of the scaled density $(\rho/\rho_{\infty})$ for
			$M_{\infty}=5.0$, computed with the SUPG-YZ$\beta$ formulation and
			the PASSC framework.}
		\label{fig:ma5_density}
	\end{figure}

	Raising the free-stream Mach number to $M_{\infty}=5.0$, i.e.,
	$\mathbf{u}_{\infty}=(1765.332,\,0.0)$~m/s, brings the flow into the
	hypersonic regime. As Figures~\ref{fig:ma5_mach}--\ref{fig:ma5_density}
	show, the bow shock moves markedly closer to the body and its opening
	angle narrows, while the post-shock temperature rise becomes considerably
	stronger, particularly along the stagnation line and in the hot wake
	region behind the cylinder. The two approaches predict a similar overall
	bow-shock shape and wake structure, although a difference in the detected
	stand-off distance is observed and quantified below. The mesh-scale
	speckle visible upstream of the shock in the SUPG-YZ$\beta$ fields,
	together with the wavy, scalloped contour bands along the shock wings and
	in the downstream region, is largely absent from the PASSC solution.
	The remaining variations in the free stream consist primarily of faint,
	smooth, low-amplitude modulations, most visible in the temperature field,
	while the overall far-field distribution remains closely tied to the
	stabilized solution.
	
\begin{figure}[htb]
		\centering
		\includegraphics[width=1\linewidth]{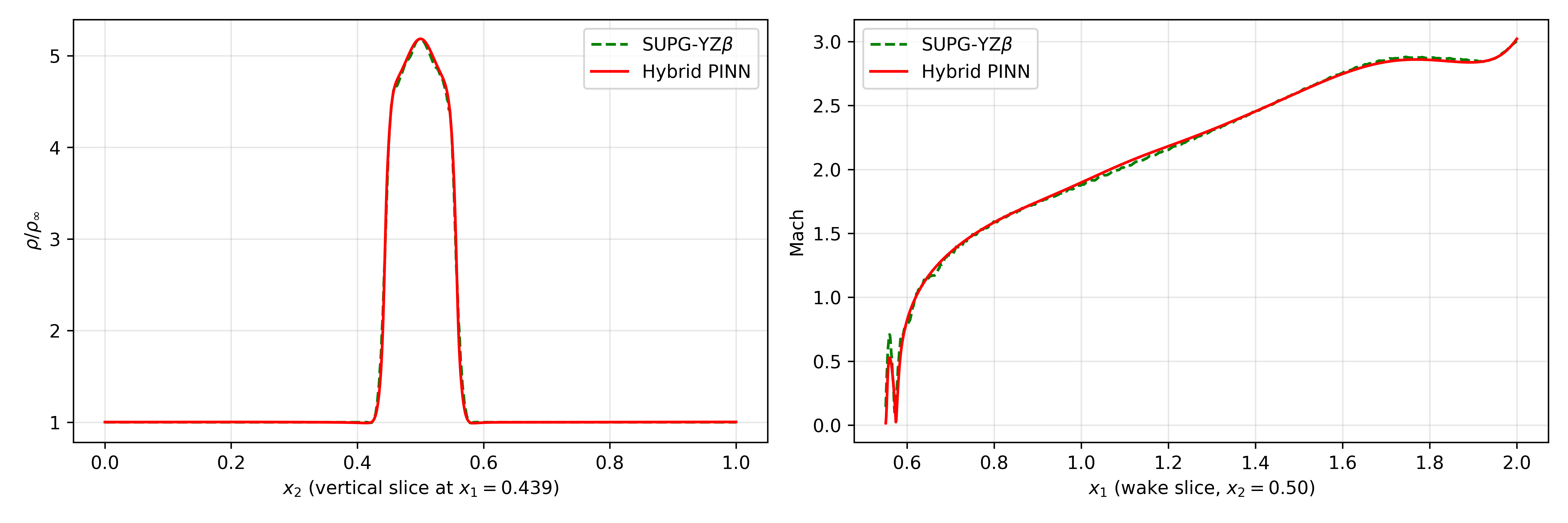}
		\caption{Comparison of the SUPG-YZ$\beta$ and PASSC solutions along
			selected cross sections for $M_{\infty}=5.0$: scaled density
			$(\rho/\rho_{\infty})$ along the vertical slice at $x_1=0.439$ and
			Mach number along the wake centerline at $x_2=0.50$.}
		\label{fig:ma5_cross}
	\end{figure}

	For $M_{\infty}=5.0$, the theoretical ratios immediately downstream of a
	normal shock are
	$p/p_{\infty}=29.0$,
	$\rho/\rho_{\infty}=5.0$, and
	$T/T_{\infty}=5.8$.
	Along the stagnation streamline, the post-shock flow subsequently
	decelerates toward the stagnation point, for which the corresponding
	theoretical ratios are
	$p_{0}/p_{\infty}\approx32.65$,
	$\rho_{0}/\rho_{\infty}\approx5.44$, and
	$T_{0}/T_{\infty}=6.0$.
	The peak values observed in
	Figures~\ref{fig:ma5_pressure}--\ref{fig:ma5_density} and in the density
	profile of Figure~\ref{fig:ma5_cross}, which attains
	$\rho/\rho_{\infty}\approx5.17$ along the vertical slice at $x_1=0.439$,
	lie between the corresponding normal-shock and stagnation values. The
	temperature in the stagnation region and near wake rises to approximately
	six times the free-stream value, i.e., about $1{,}800$~K for the present
	free-stream conditions. Along the wake centerline, the Mach number
	recovers steadily behind the cylinder, reaching $M\approx3$ near the
	outflow. The two solutions remain in close agreement over the entire
	slice, with the PASSC primarily smoothing the noisy double peak
	immediately behind the body and the small stair-like undulations of the
	finite element profile during the early recovery.
	
	The second column of Table~\ref{tab:shock_metrics} provides a quantitative
	comparison. The SUPG-YZ$\beta$ solution places the bow shock at
	$\varDelta/R=0.458$, within approximately $2\%$ of the value
	$\varDelta/R=0.465$ predicted by the Billig correlation~\cite{billig1967}.
	The PASSC solution yields $\varDelta/R=0.497$, corresponding to an
	upstream displacement of the detected shock front and a value
	approximately $7\%$ above the correlation. This difference should be
	interpreted with some caution because the Billig correlation is an
	empirical open-flow relation, whereas the present computations are
	performed in a finite domain and evaluated at a finite simulation time.
	
	The stagnation-pressure comparison shows a different trend. The
	SUPG-YZ$\beta$ value,
	$p_{0}/p_{\infty}=32.135$, is approximately $1.6\%$ below the Rayleigh
	pitot value of $32.653$, whereas the PASSC solution gives
	$p_{0}/p_{\infty}=32.430$, reducing this discrepancy to approximately
	$0.7\%$. Thus, although the detected bow-shock position is shifted slightly
	beyond the Billig reference, the corrected stagnation pressure moves closer
	to its analytical value. Taken together with the $M_{\infty}=2.0$ results,
	these observations indicate that the PASSC regularization can modify the
	localized shock representation while preserving the global flow structure;
	whether a particular scalar metric moves toward or slightly beyond its
	corresponding reference depends on the initial bias of the stabilized
	solution and on the finite spatial resolution of the shock detection.
	
\begin{figure}[htb]
		\centering
		\includegraphics[width=1\linewidth]{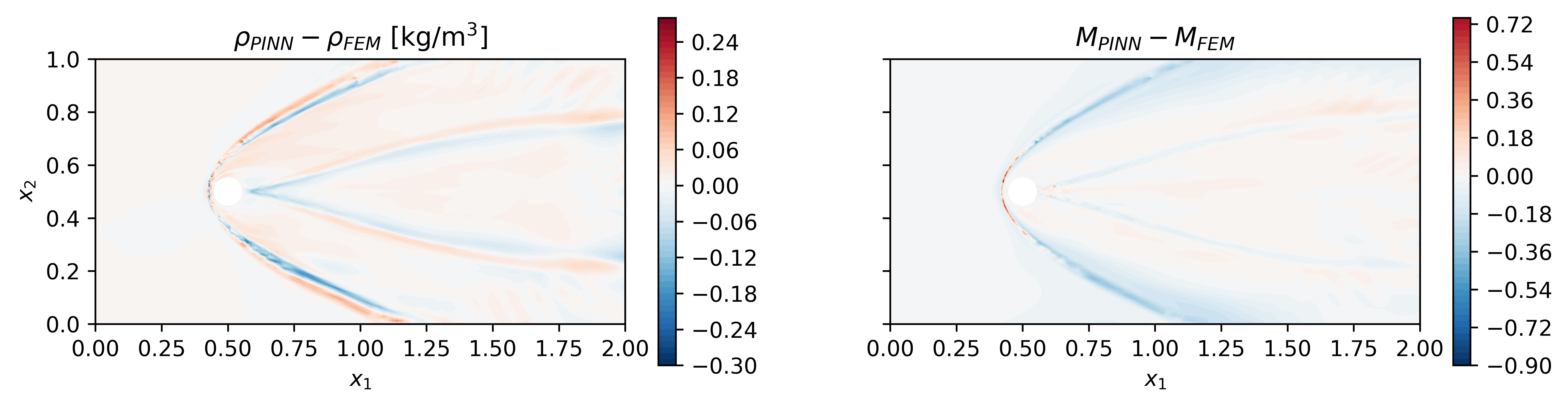}
		\caption{Pointwise differences between the PASSC and SUPG-YZ$\beta$
			solutions for $M_{\infty}=5.0$: density difference
			$\rho_{\mathrm{PASSC}}-\rho_{\mathrm{FEM}}$ in kg/m$^3$ and
			Mach-number difference
			$M_{\mathrm{PASSC}}-M_{\mathrm{FEM}}$.}
		\label{fig:ma5_diff}
	\end{figure}

	In the difference maps of Figure~\ref{fig:ma5_diff}, the localized pattern
	already observed at $M_{\infty}=2.0$ reappears with larger amplitudes.
	Density differences of approximately $0.3$~kg/m$^3$ and Mach-number
	differences of up to approximately $0.9$ in magnitude occur directly
	across the shock front. These larger pointwise differences are expected
	for the stronger and thinner shock at $M_{\infty}=5.0$, since small
	changes in the location or thickness of a steep numerical transition can
	produce comparatively large local differences. The dipole-like structure
	along the shock wings, with density increased on the upstream side and
	decreased on the downstream side, is consistent with the upstream shift
	of the detected shock front. Broad, low-amplitude Mach-number differences
	behind the shock wings likewise reflect the slightly modified extent of
	the post-shock region. Away from these localized structures, the
	differences remain substantially smaller, confirming that the PASSC
	solution remains strongly anchored to the stabilized finite element
	approximation.
	
\begin{figure}[!htb]
		\centering
		\includegraphics[width=1\linewidth]{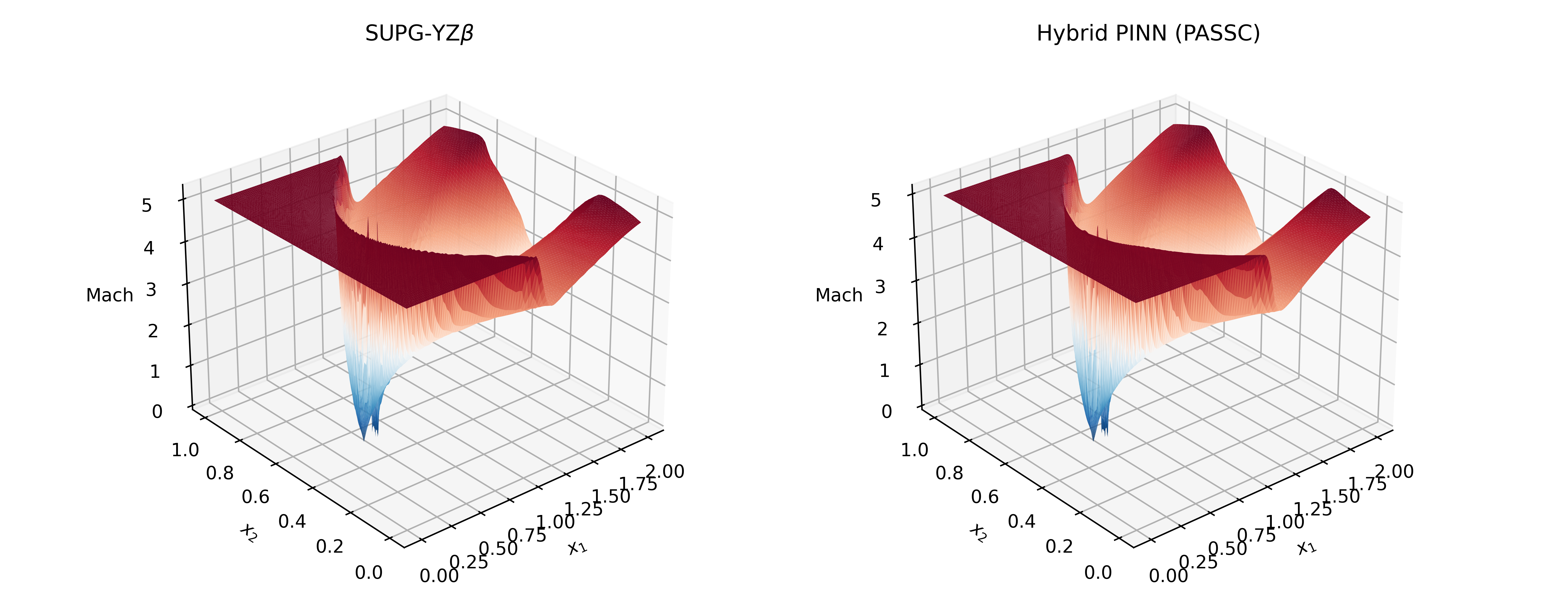}
		\caption{Three-dimensional surface representation of the Mach-number
			field at $M_{\infty}=5.0$ for the SUPG-YZ$\beta$ and PASSC solutions.}
		\label{fig:ma5_3dsurface}
	\end{figure}

	Figure~\ref{fig:ma5_3dsurface} presents a three-dimensional surface
	representation of the Mach-number field at $M_{\infty}=5.0$, providing
	an additional view of the localized differences between the two
	solutions. In the SUPG-YZ$\beta$ solution, the bow shock appears as a
	steep transition whose crest exhibits fine mesh-scale serrations,
	whereas the free-stream plateau remains essentially uniform at the scale
	of the surface plot. The weaker dispersive structures visible in the
	planar contour plots are too small to be clearly resolved in this
	representation. The PASSC field preserves the height and overall topology
	of the shock structure, together with the wake deficit downstream of the
	cylinder and the subsequent recovery toward the outflow, while producing
	a somewhat displaced and visibly smoother shock surface with substantially
	reduced serration along much of its crest.
	
	\subsubsection{Hypersonic flow at $M_{\infty}=8.0$}
	
\begin{figure}[htb]
		\centering
		\includegraphics[width=1\linewidth]{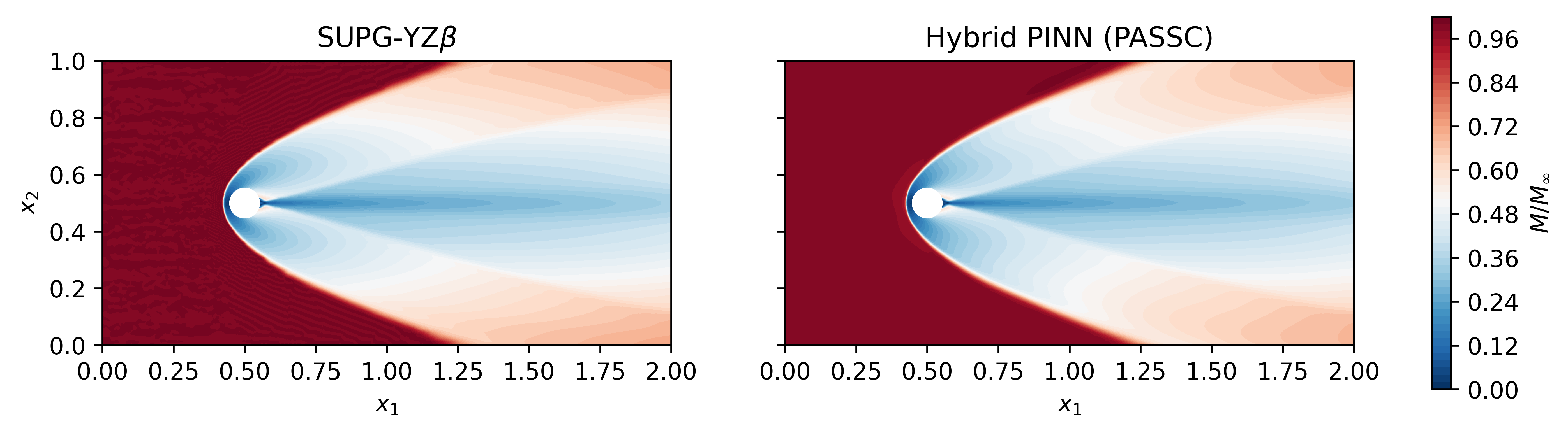}
		\caption{Contours of the scaled Mach number $(M/M_{\infty})$ for
			$M_{\infty}=8.0$, computed with the SUPG-YZ$\beta$ formulation and
			the PASSC framework.}
		\label{fig:ma8_mach}
	\end{figure}

\begin{figure}[htb]
		\centering
		\includegraphics[width=1\linewidth]{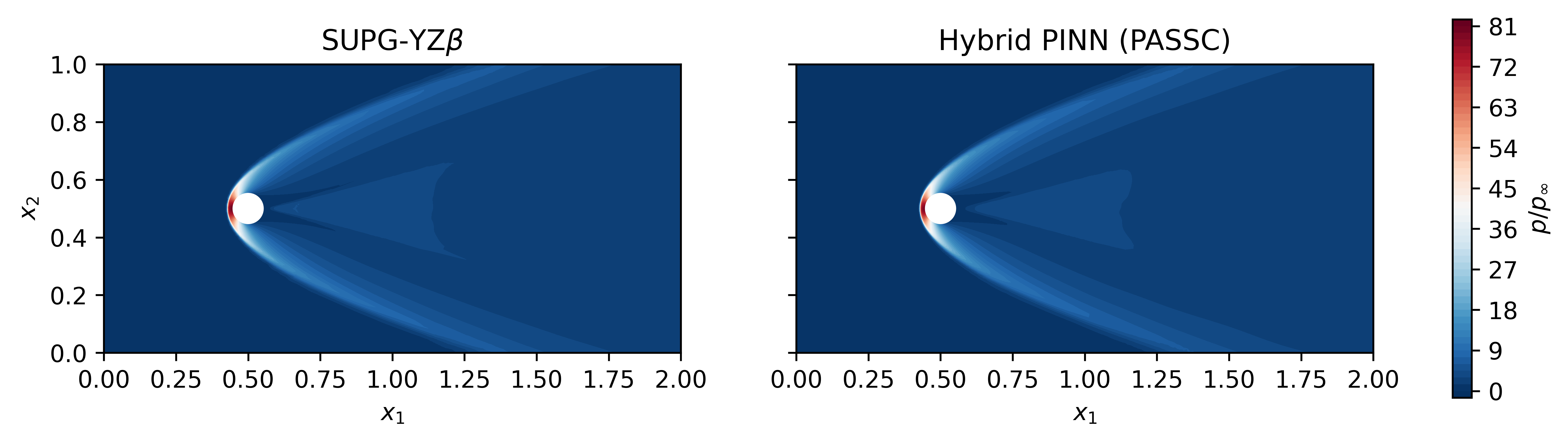}
		\caption{Contours of the scaled pressure $(p/p_{\infty})$ for
			$M_{\infty}=8.0$, computed with the SUPG-YZ$\beta$ formulation and
			the PASSC framework.}
		\label{fig:ma8_pressure}
	\end{figure}

\begin{figure}[htb]
		\centering
		\includegraphics[width=1\linewidth]{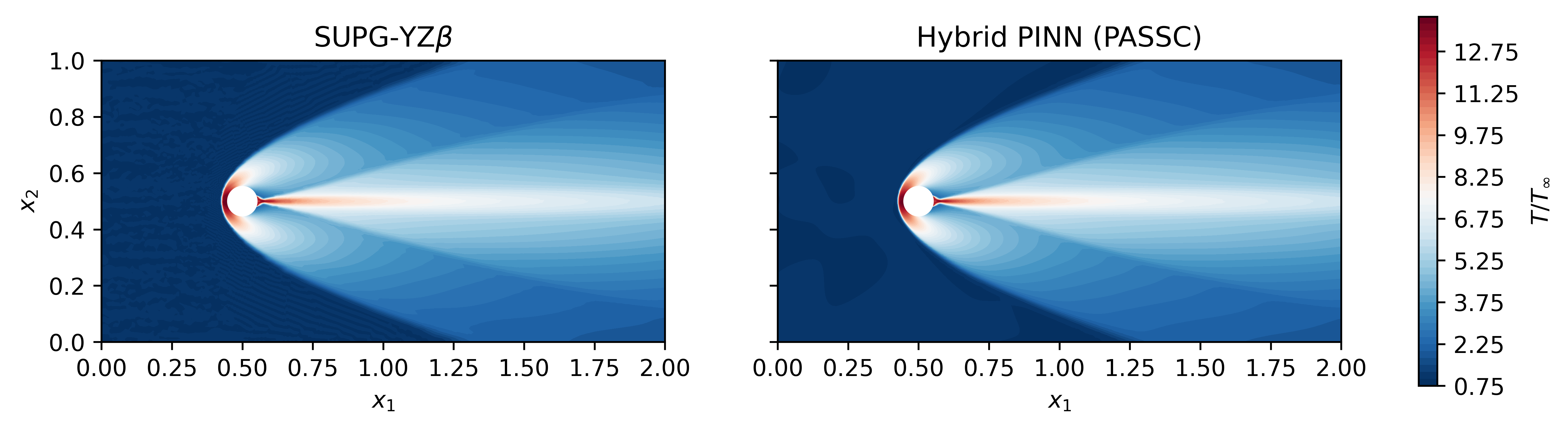}
		\caption{Contours of the scaled temperature $(T/T_{\infty})$ for
			$M_{\infty}=8.0$, computed with the SUPG-YZ$\beta$ formulation and
			the PASSC framework.}
		\label{fig:ma8_temp}
	\end{figure}

\begin{figure}[htb]
		\centering
		\includegraphics[width=1\linewidth]{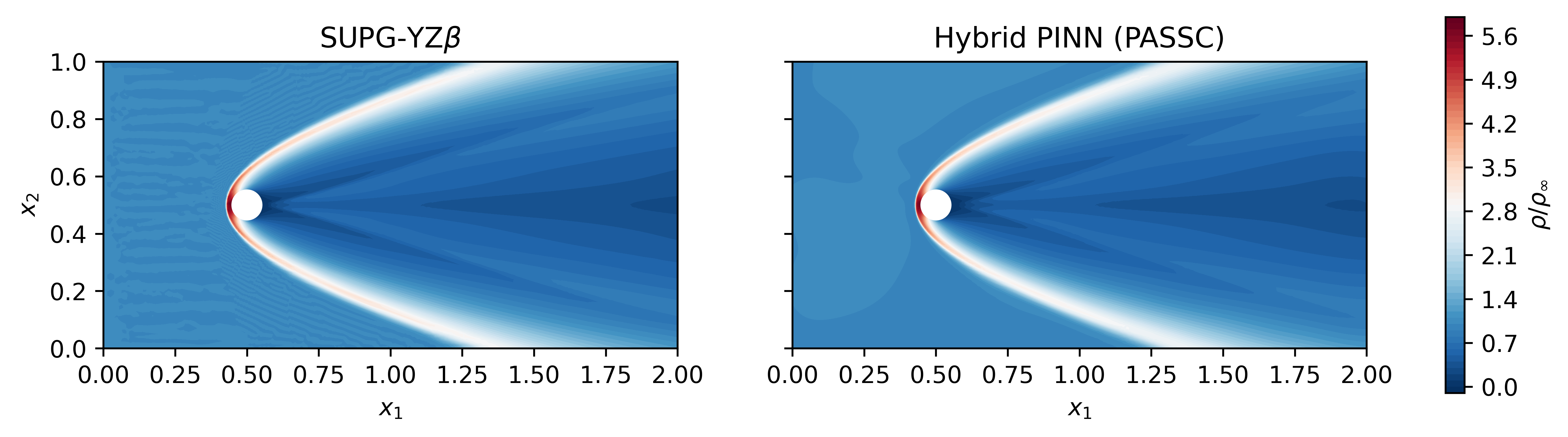}
		\caption{Contours of the scaled density $(\rho/\rho_{\infty})$ for
			$M_{\infty}=8.0$, computed with the SUPG-YZ$\beta$ formulation and
			the PASSC framework.}
		\label{fig:ma8_density}
	\end{figure}

	At $M_{\infty}=8.0$, i.e.,
	$\mathbf{u}_{\infty}=(2824.532,\,0.0)$~m/s, the bow shock wraps still more
	tightly around the cylinder and its opening angle narrows further, while
	the shock layer and the thermal wake become considerably hotter
	(Figures~\ref{fig:ma8_mach}--\ref{fig:ma8_density}). The computed flow
	structures are also broadly consistent with those reported
	in~\cite{kirk2,kirk5}. The SUPG-YZ$\beta$ and PASSC solutions predict the
	same detected shock stand-off distance and similar overall shock and wake
	structures. Compared with the stabilized solution, the PASSC field
	substantially reduces the mesh-scale speckle, the faint dispersive striping
	upstream of the shock, and the scalloped contour patterns in the downstream
	region while retaining a sharply localized shock transition.
	
	For $M_{\infty}=8.0$, the theoretical ratios immediately downstream of a
	normal shock are
	$p/p_{\infty}=74.5$,
	$\rho/\rho_{\infty}=5.56522$, and
	$T/T_{\infty}=13.38672$.
	Along the stagnation streamline, the post-shock flow subsequently
	decelerates toward the stagnation point, for which the temperature ratio
	approaches the theoretical value
	$T_{0}/T_{\infty}=13.8$, while the corresponding stagnation-pressure ratio
	is given by the Rayleigh pitot relation as
	$p_{0}/p_{\infty}\approx82.9$.
	The peak values observed in Figures~\ref{fig:ma8_pressure}
	and~\ref{fig:ma8_temp} are consistent with these reference values; in
	particular, the temperature immediately in front of the cylinder rises to
	nearly $14$ times the free-stream value, corresponding to approximately
	$4{,}100$~K under the present free-stream conditions.
	
	The third column of Table~\ref{tab:shock_metrics} shows that the
	SUPG-YZ$\beta$ solution places the bow shock at
	$\varDelta/R=0.424$, approximately $2\%$ above the value
	$\varDelta/R=0.415$ predicted by the Billig
	correlation~\cite{billig1967}. The PASSC solution yields the same detected
	stand-off distance to the resolution of the present measurement. The
	stagnation-pressure ratios are
	$p_{0}/p_{\infty}=81.45$ for the SUPG-YZ$\beta$ solution and
	$p_{0}/p_{\infty}=81.10$ for the PASSC solution, both remaining within
	approximately $2\%$ of the Rayleigh pitot value of $82.9$. Thus, in
	contrast to the measurable shock-position shifts observed at
	$M_{\infty}=2.0$ and $M_{\infty}=5.0$, the correction at
	$M_{\infty}=8.0$ leaves the detected stand-off distance unchanged while
	producing only a small change in the stagnation-pressure ratio. The main
	differences between the two solutions therefore appear in the localized
	structure and smoothness of the computed fields rather than in these
	global scalar metrics.
	
\begin{figure}[htb]
		\centering
		\includegraphics[width=1\linewidth]{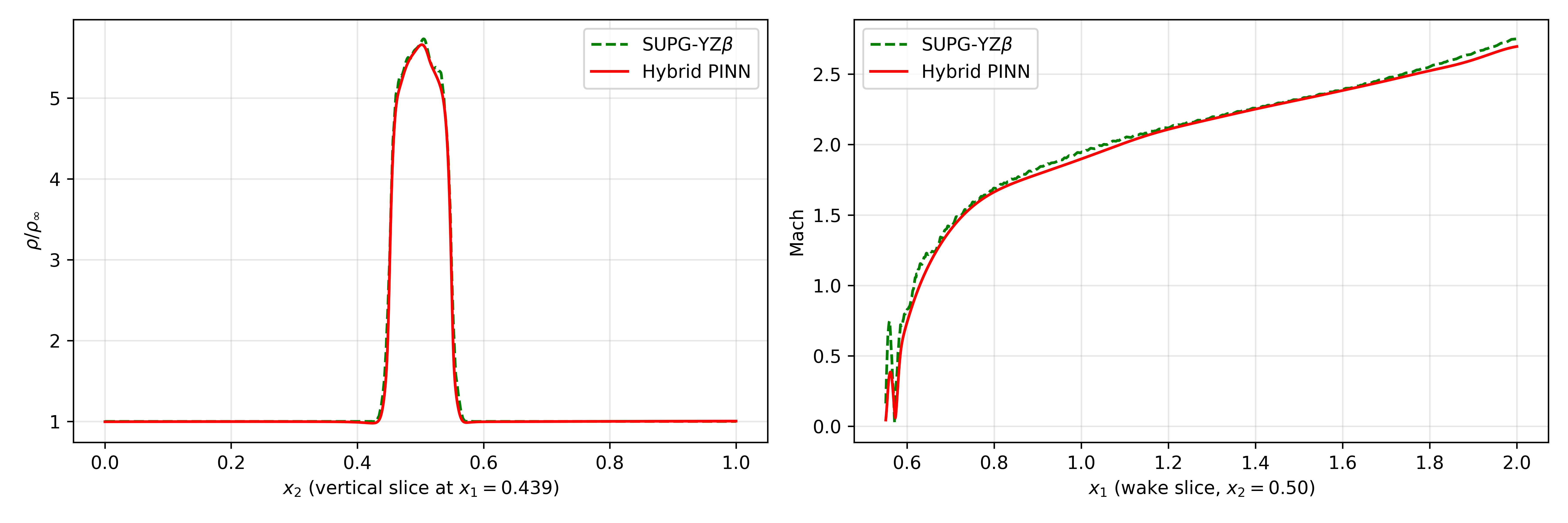}
		\caption{Comparison of the SUPG-YZ$\beta$ and PASSC solutions along
			selected cross sections for $M_{\infty}=8.0$: scaled density
			$(\rho/\rho_{\infty})$ along the vertical slice at $x_1=0.439$ and
			Mach number along the wake centerline at $x_2=0.50$.}
		\label{fig:ma8_cross}
	\end{figure}

	Figure~\ref{fig:ma8_cross} presents the scaled density along the vertical
	slice at $x_1=0.439$ and the Mach number along the wake centerline at
	$x_2=0.50$. The density profile attains a peak value of approximately
	$\rho/\rho_{\infty}\approx5.7$ on the stagnation line, lying above the
	normal-shock value and below the limiting stagnation-region value expected
	for this high-Mach-number calorically perfect-gas flow. The two solutions
	remain in close agreement along both cross sections. Small localized
	oscillations are visible in the finite element density profile near the
	shock crossing; similar nonphysical oscillations have also been reported
	for related stabilized computations (see, e.g.,~\cite{kirk1}). The PASSC
	solution reduces these local variations while leaving the detected shock
	position and the principal jump amplitude essentially unchanged. Along the
	wake centerline, the PASSC solution smooths the oscillatory behavior of the
	finite element profile immediately behind the body and remains marginally
	below it during the subsequent recovery, with the difference remaining
	small relative to the local Mach number up to the outflow.
	
\begin{figure}[htb]
		\centering
		\includegraphics[width=1\linewidth]{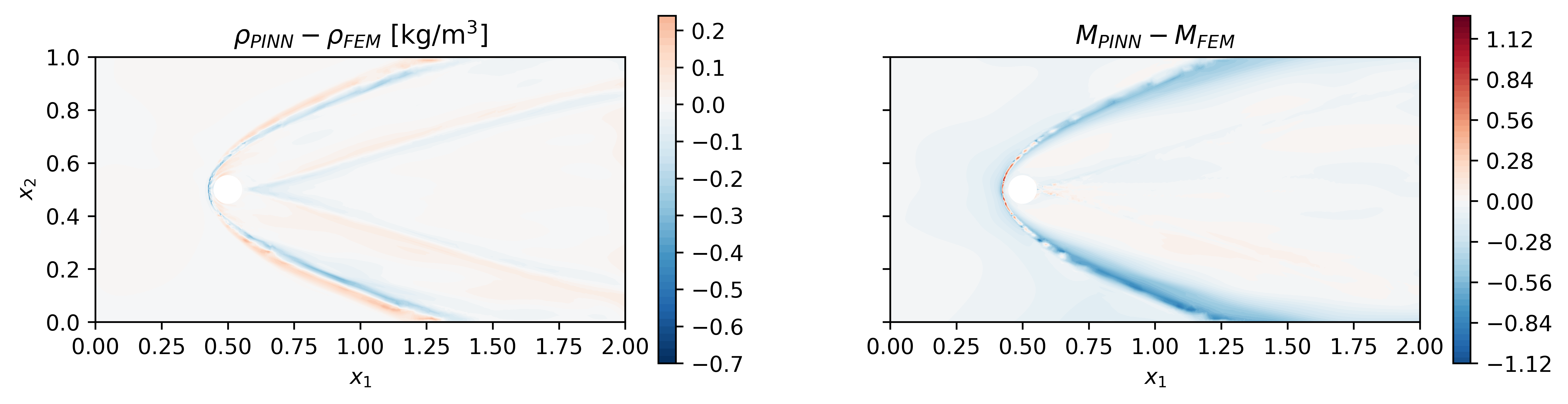}
		\caption{Pointwise differences between the PASSC and SUPG-YZ$\beta$
			solutions for $M_{\infty}=8.0$: density difference
			$\rho_{\mathrm{PASSC}}-\rho_{\mathrm{FEM}}$ in kg/m$^3$ and
			Mach-number difference
			$M_{\mathrm{PASSC}}-M_{\mathrm{FEM}}$.}
		\label{fig:ma8_diff}
	\end{figure}

	The difference maps of Figure~\ref{fig:ma8_diff} continue the localized
	pattern observed in the preceding cases. Density differences reach
	approximately $0.7$~kg/m$^3$ in a narrow region near the shock crest,
	while Mach-number differences of up to approximately $1.1$ in magnitude
	occur across the shock front. The increase in these peak pointwise
	differences relative to the $M_{\infty}=2.0$ and $M_{\infty}=5.0$ cases
	is consistent with the progressively stronger and thinner shock, for which
	small changes in the local position or thickness of the numerical
	transition produce larger pointwise differences. Near the nose, the
	difference field forms a compact dipole-like structure with little net
	displacement of the detected front, consistent with a localized
	modification of the shock profile rather than a measurable change in its
	stand-off distance. Broader, lower-amplitude Mach-number differences are
	also visible downstream of the shock wings.
	
\begin{figure}[!htb]
		\centering
		\includegraphics[width=1\linewidth]{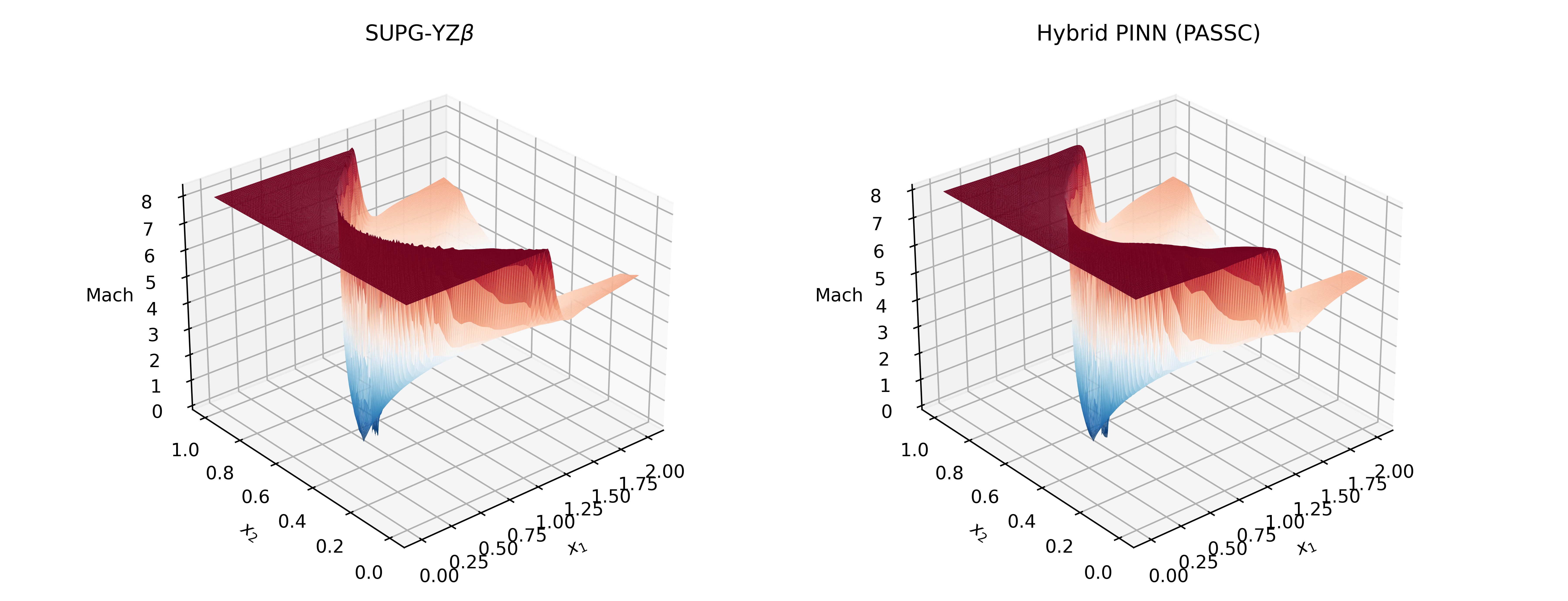}
		\caption{Three-dimensional surface representation of the Mach-number
			field at $M_{\infty}=8.0$ for the SUPG-YZ$\beta$ and PASSC solutions.}
		\label{fig:ma8_3dsurface}
	\end{figure}

	The three-dimensional surface representation in
	Figure~\ref{fig:ma8_3dsurface} provides a complementary view of these
	features. In the SUPG-YZ$\beta$ solution, low-amplitude dispersive
	variations are visible on the nominally uniform free-stream plateau
	upstream of the steep shock surface, together with fine serrations along
	the shock crest. The PASSC field preserves the overall location, height,
	and topology of the principal shock and wake structures while producing a
	smoother free-stream plateau and substantially reducing the serration of
	the shock surface; mild residual corrugation remains along portions of the
	far shock wings.
	
	\subsubsection{Hypersonic flow at $M_{\infty}=12.0$}
	
\begin{figure}[htb]
		\centering
		\includegraphics[width=1\linewidth]{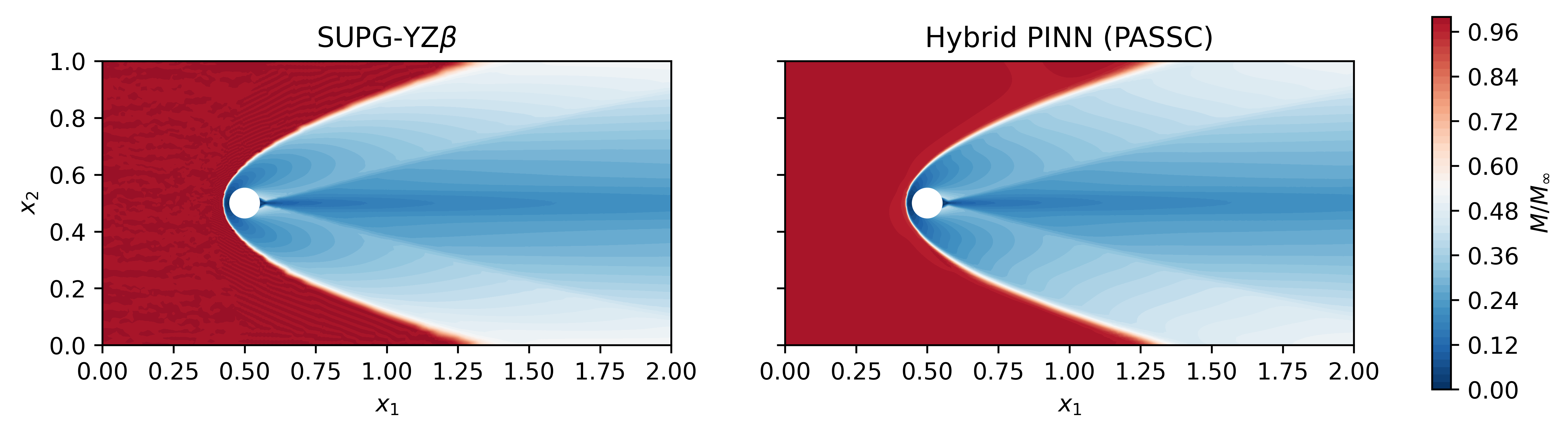}
		\caption{Contours of the scaled Mach number $(M/M_{\infty})$ for
			$M_{\infty}=12.0$, computed with the SUPG-YZ$\beta$ formulation and
			the PASSC framework.}
		\label{fig:ma12_mach}
	\end{figure}

\begin{figure}[htb]
		\centering
		\includegraphics[width=1\linewidth]{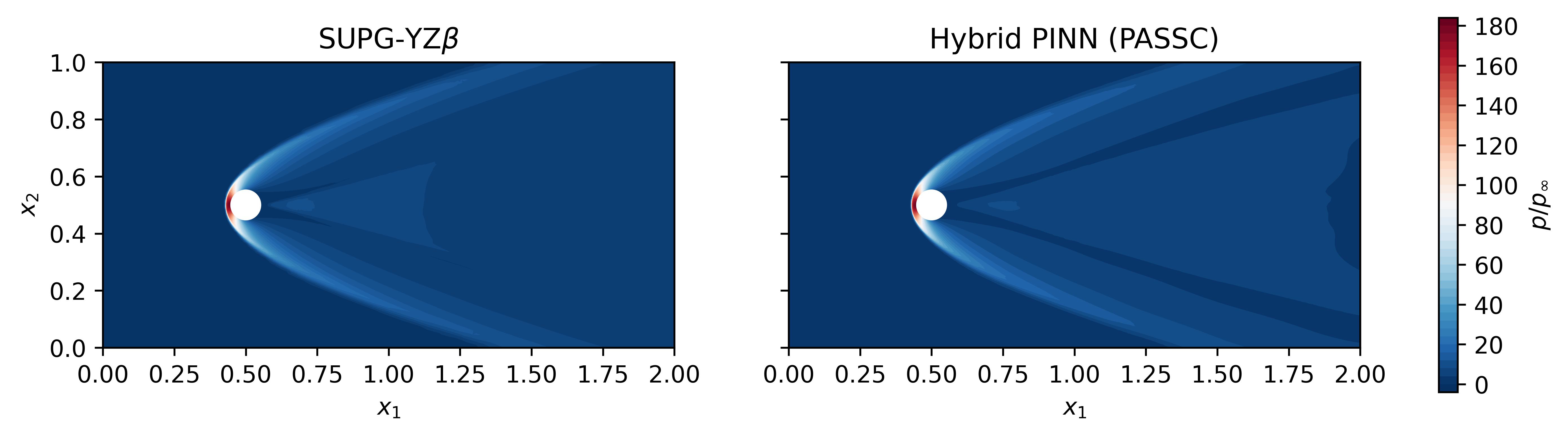}
		\caption{Contours of the scaled pressure $(p/p_{\infty})$ for
			$M_{\infty}=12.0$, computed with the SUPG-YZ$\beta$ formulation and
			the PASSC framework.}
		\label{fig:ma12_pressure}
	\end{figure}

\begin{figure}[htb]
		\centering
		\includegraphics[width=1\linewidth]{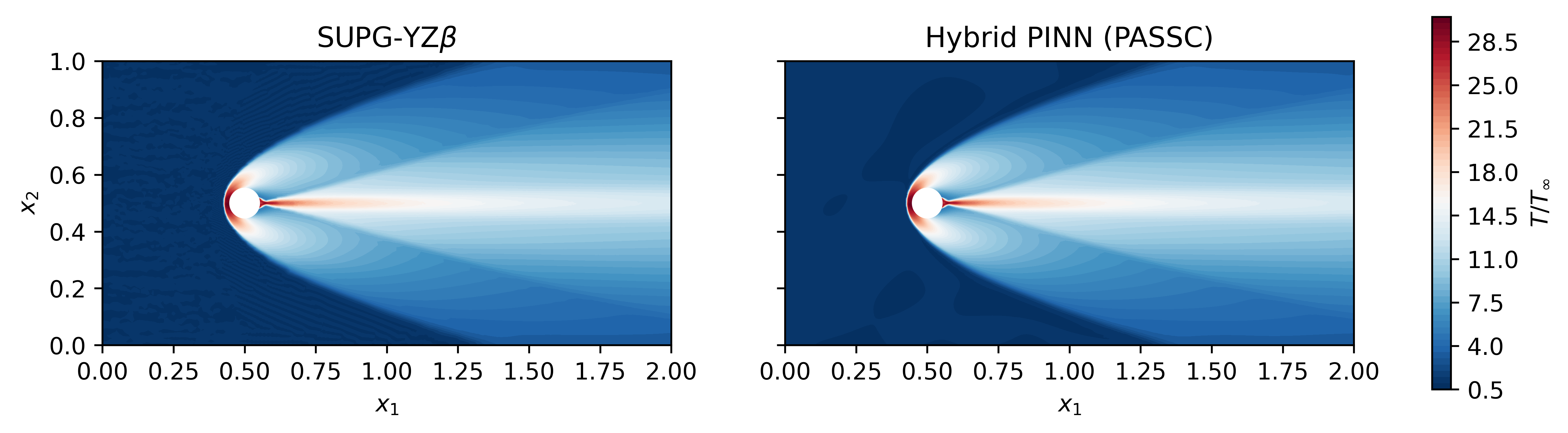}
		\caption{Contours of the scaled temperature $(T/T_{\infty})$ for
			$M_{\infty}=12.0$, computed with the SUPG-YZ$\beta$ formulation and
			the PASSC framework.}
		\label{fig:ma12_temp}
	\end{figure}

\begin{figure}[htb]
		\centering
		\includegraphics[width=1\linewidth]{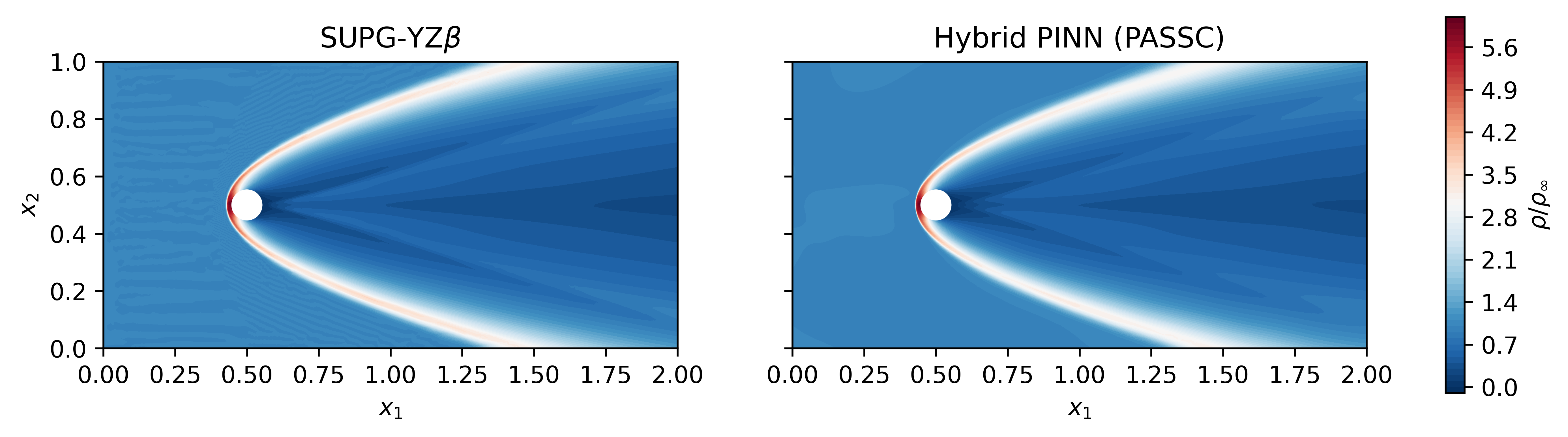}
		\caption{Contours of the scaled density $(\rho/\rho_{\infty})$ for
			$M_{\infty}=12.0$, computed with the SUPG-YZ$\beta$ formulation and
			the PASSC framework.}
		\label{fig:ma12_density}
	\end{figure}

	The final and most demanding case, $M_{\infty}=12.0$, with
	$\mathbf{u}_{\infty}=(4236.797,\,0.0)$~m/s, places the flow deep in the
	hypersonic regime and brings the normal-shock density ratio close to the
	strong-shock limit
	$(\gamma+1)/(\gamma-1)=6$ for $\gamma=1.4$.
	As Figures~\ref{fig:ma12_mach}--\ref{fig:ma12_density} show, the resulting
	bow shock is very strong and tightly wrapped around the cylinder, while the
	post-shock region and thermal wake remain strongly heated throughout the
	domain. The two approaches predict similar overall shock and wake structures,
	although a small difference in the detected stand-off distance is observed
	and quantified below. Compared with the SUPG-YZ$\beta$ solution, the PASSC
	field substantially reduces the upstream mesh-scale speckle and dispersive
	striping while retaining a sharply localized high-density region along the
	shock front. Only faint, smooth, low-amplitude variations remain in parts of
	the nominally uniform free stream, most visibly in the temperature field.
	
	For $M_{\infty}=12.0$, the theoretical ratios immediately downstream of a
	normal shock are
	$p/p_{\infty}=167.83333$,
	$\rho/\rho_{\infty}=5.79866$, and
	$T/T_{\infty}=28.94348$.
	Along the stagnation streamline, the post-shock flow subsequently decelerates
	toward the stagnation point, for which the pressure ratio approaches the
	Rayleigh pitot value
	$p_{0}/p_{\infty}\approx185.9$ and the temperature ratio approaches the
	theoretical stagnation value
	$T_{0}/T_{\infty}=29.8$.
	The peak values observed in Figures~\ref{fig:ma12_pressure}
	and~\ref{fig:ma12_temp} are consistent with these reference values. In
	particular, the temperature immediately in front of the cylinder rises to
	approximately $30$ times the free-stream value, corresponding to about
	$8{,}900$~K under the present free-stream conditions.
	
	At such temperatures, the calorically perfect-gas assumption is no longer
	physically representative of real high-temperature nitrogen. In particular,
	the pure-N$_2$ model employed in the present computations neglects
	temperature-dependent thermodynamic properties, vibrational excitation, and
	nitrogen dissociation. For atmospheric air, additional thermochemical
	processes, including oxygen dissociation and, at sufficiently high
	temperatures, ionization, would also have to be considered. The present
	$M_{\infty}=12.0$ computation should therefore be interpreted primarily as a
	demanding numerical test of the proposed PASSC framework rather than as a
	quantitatively predictive representation of the corresponding
	high-temperature physical flow.
	
	As the last column of Table~\ref{tab:shock_metrics} shows, the
	SUPG-YZ$\beta$ solution places the bow shock at
	$\varDelta/R=0.398$, essentially coincident with the value
	$\varDelta/R=0.399$ predicted by the Billig correlation~\cite{billig1967}.
	The PASSC solution yields $\varDelta/R=0.424$, corresponding to an outward
	shift of the detected shock front by one resolution increment of the
	stand-off measurement and a value approximately $6\%$ above the correlation.
	The stagnation-pressure ratios are
	$p_{0}/p_{\infty}=182.44$ for the SUPG-YZ$\beta$ solution and
	$p_{0}/p_{\infty}=182.82$ for the PASSC solution, both remaining within
	approximately $2\%$ of the Rayleigh pitot value of $185.87$, with the latter
	lying slightly closer to the reference.
	
	Viewed across the four Mach-number cases, the correction changes the detected
	shock position only by a few millimeters, corresponding approximately to one
	or two near-wall element layers. At $M_{\infty}=2.0$, where the stabilized
	stand-off distance exhibits the largest discrepancy from the Billig
	correlation, this shift moves the detected front appreciably toward the
	reference value. At the hypersonic Mach numbers, where the stabilized
	stand-off distances are already close to the correlation, the same spatial
	granularity appears either as no measurable change or as a small overshoot.
	The largest discrepancy occurring precisely at $M_{\infty}=2.0$ is
	consistent with the temporal-convergence argument of
	Section~\ref{sec:mach2}: at the common final time, the lowest-Mach case has
	completed the fewest flow-through times and its detached shock, whose
	steady stand-off distance is the largest of the four cases, is still
	approaching its steady position.
	These differences should be interpreted with some caution because the Billig
	correlation is an empirical open-flow relation, whereas the present
	computations are performed in a finite computational domain and evaluated at
	a finite simulation time.
	
\begin{figure}[htb]
		\centering
		\includegraphics[width=1\linewidth]{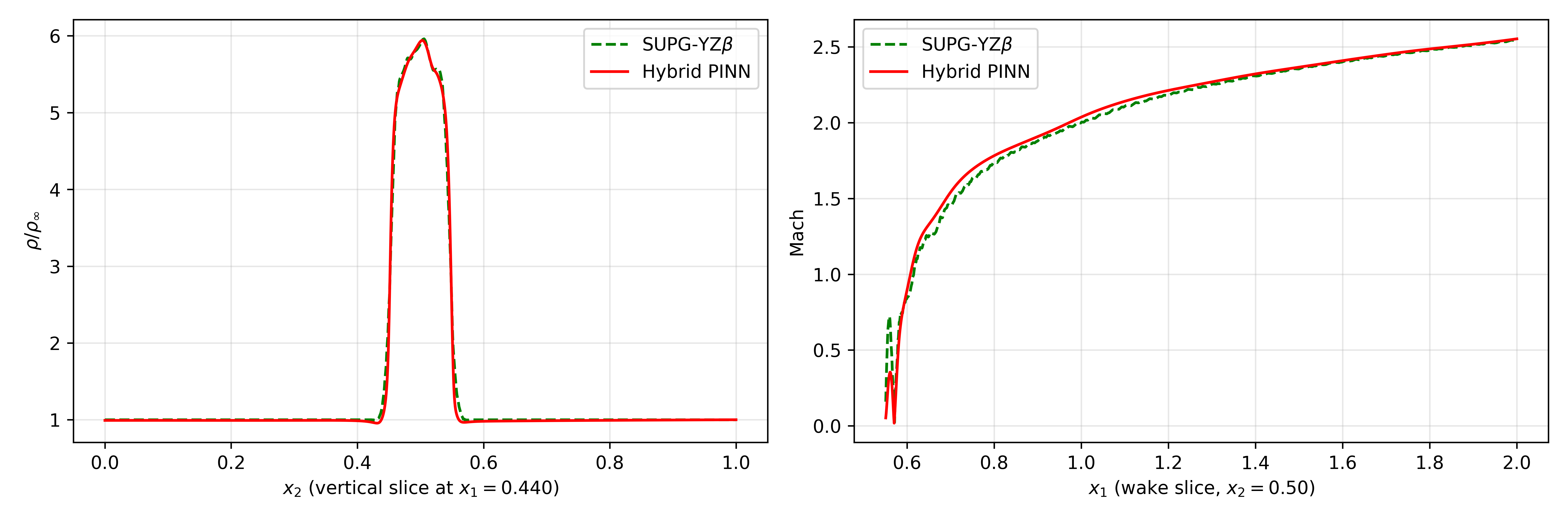}
		\caption{Comparison of the SUPG-YZ$\beta$ and PASSC solutions along
			selected cross sections for $M_{\infty}=12.0$: scaled density
			$(\rho/\rho_{\infty})$ along the vertical slice at $x_1=0.440$ and
			Mach number along the wake centerline at $x_2=0.50$.}
		\label{fig:ma12_cross}
	\end{figure}

	Figure~\ref{fig:ma12_cross} presents the scaled density along the vertical
	slice at $x_1=0.440$ and the Mach number along the wake centerline at
	$x_2=0.50$. The density profile attains a peak value of approximately
	$\rho/\rho_{\infty}\approx5.95$ on the stagnation line, slightly above the
	normal-shock value and close to the strong-shock density-ratio limit of six.
	The PASSC profile reproduces essentially the same peak amplitude as the
	stabilized solution while smoothing the small oscillatory variations around
	the crest and the flanks of the transition. Along the wake centerline, the
	PASSC solution smooths the oscillatory behavior of the finite element profile
	immediately behind the body and remains marginally above it during the early
	recovery, with the two profiles converging well before the outflow.
	
\begin{figure}[htb]
		\centering
		\includegraphics[width=1\linewidth]{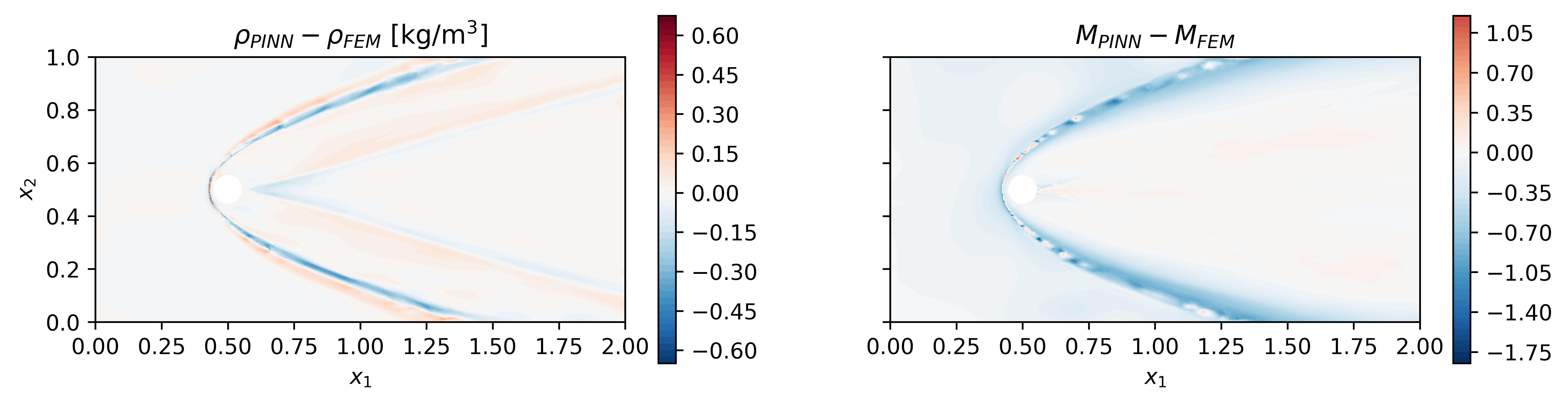}
		\caption{Pointwise differences between the PASSC and SUPG-YZ$\beta$
			solutions for $M_{\infty}=12.0$: density difference
			$\rho_{\mathrm{PASSC}}-\rho_{\mathrm{FEM}}$ in kg/m$^3$ and
			Mach-number difference
			$M_{\mathrm{PASSC}}-M_{\mathrm{FEM}}$.}
		\label{fig:ma12_diff}
	\end{figure}

	The difference maps of Figure~\ref{fig:ma12_diff} remain concentrated in thin
	bands surrounding the bow shock. Peak density differences are approximately
	$0.6$~kg/m$^3$, while Mach-number differences reach approximately $1.8$ in
	magnitude directly across the shock front. The peak Mach-number differences
	therefore increase across the Mach-number sequence, from about $0.45$ at
	$M_{\infty}=2.0$ to approximately $0.9$, $1.1$, and $1.8$ in the subsequent
	cases. This trend is consistent with the increasing strength and decreasing
	numerical thickness of the shock, for which small local changes in the
	position or profile of the captured transition generate increasingly large
	pointwise differences.
	
	At $M_{\infty}=12.0$, the shock-associated difference band also develops a
	more spatially modulated structure along the front. This pattern is consistent
	with the comparison between the smoother PASSC representation and the more
	serrated finite element shock surface. Broader, lower-amplitude Mach-number
	differences remain visible downstream of the shock wings. These pointwise
	differences should therefore be interpreted primarily as measures of the
	sensitivity to local shock position and profile rather than as direct measures
	of field accuracy: across a nearly discontinuous transition, even a small
	shift of the captured front can produce a large local difference, whereas
	away from the shock region the two solutions remain in close agreement.
	
\begin{figure}[!htb]
		\centering
		\includegraphics[width=1\linewidth]{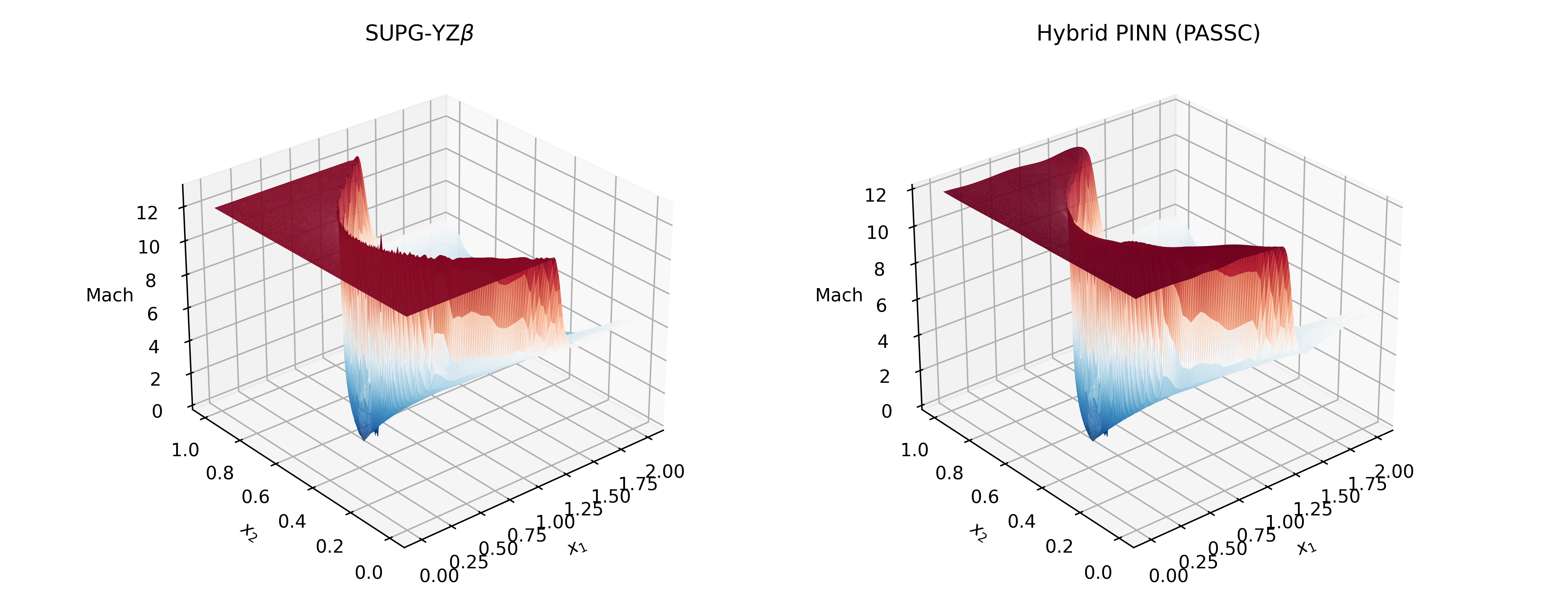}
		\caption{Three-dimensional surface representation of the Mach-number
			field at $M_{\infty}=12.0$ for the SUPG-YZ$\beta$ and PASSC solutions.}
		\label{fig:ma12_3dsurface}
	\end{figure}

	In the three-dimensional surface representation of
	Figure~\ref{fig:ma12_3dsurface}, the bow shock appears in the
	SUPG-YZ$\beta$ solution as an extremely steep surface with pronounced
	mesh-scale corrugations, together with an isolated spike near the shock crest
	and low-amplitude variations on the nominally uniform free-stream plateau.
	The PASSC field preserves the overall height and topology of the principal
	shock and wake structures while producing a smoother free-stream plateau,
	removing the isolated crest spike, and substantially reducing the serration of
	the shock surface. Mild corrugation remains along portions of the far shock
	wings. The principal shock amplitude is essentially preserved, consistent
	with the cross-section comparison, while the small outward displacement of
	the detected front is difficult to distinguish at the scale of the surface
	representation.
	
	\subsubsection{Training behavior across the Mach number series}
	\label{sec:training_behavior}
	
\begin{figure}[htb]
		\centering
		\includegraphics[width=1\linewidth]{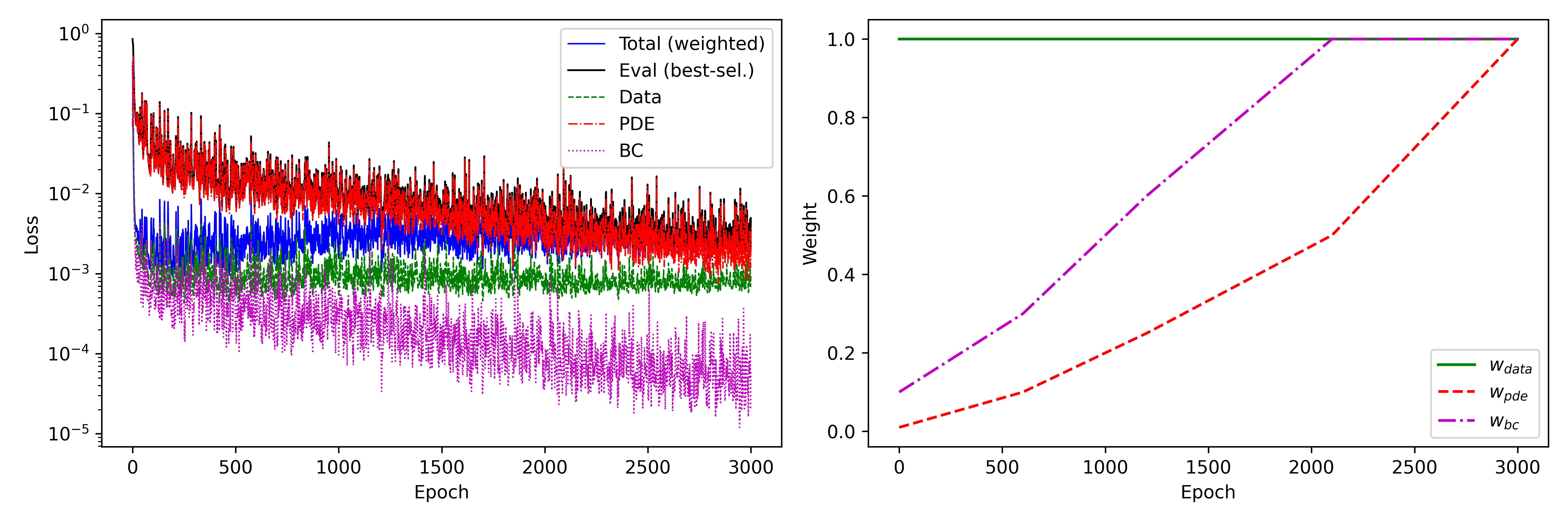}
		\caption{Training diagnostics of the PASSC framework for
			$M_{\infty}=2.0$: evolution of the weighted total loss, the unweighted
			evaluation metric of Eq.~\eqref{eq:eval_metric} used for model selection,
			and the data, PDE, and boundary-condition loss components (left), together
			with the continuously interpolated loss-weight schedule
			$(w_\text{data},\,w_\text{pde},\,w_\text{bc})$ of
			Eq.~\eqref{eq:weight_schedule} (right) over the training epochs.}
		\label{fig:ma2_training}
	\end{figure}

\begin{figure}[htb]
		\centering
		\includegraphics[width=1\linewidth]{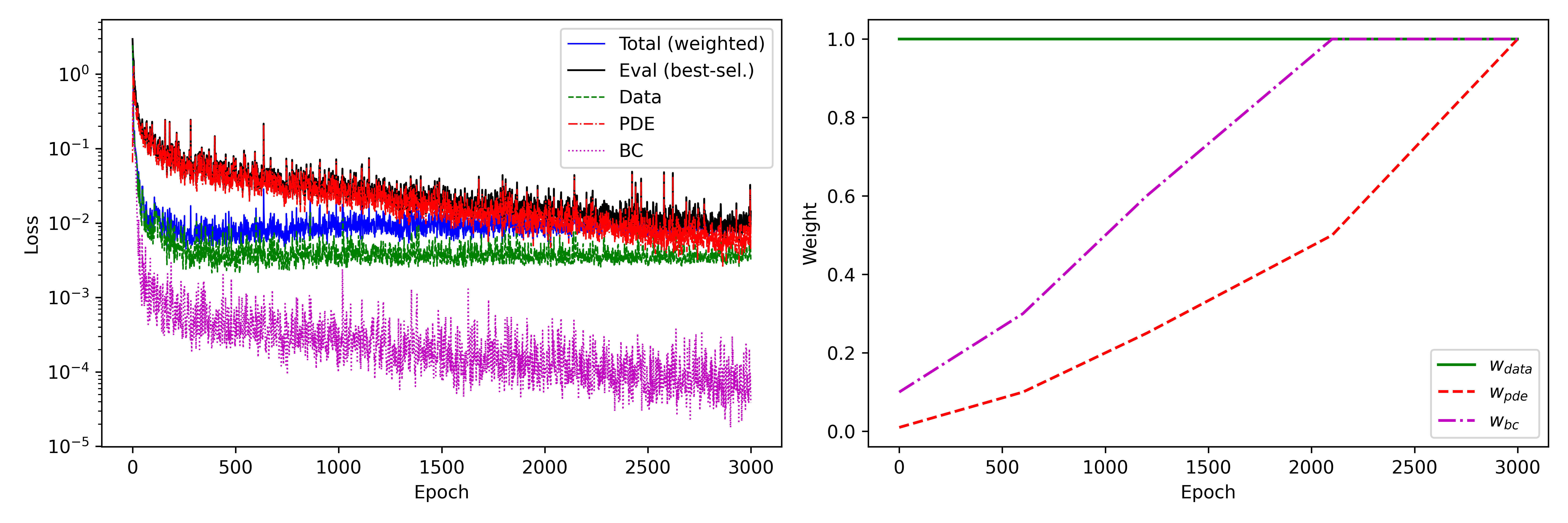}
		\caption{Training diagnostics of the PASSC framework for
			$M_{\infty}=5.0$; the layout is the same as in
			Figure~\ref{fig:ma2_training}.}
		\label{fig:ma5_training}
	\end{figure}

\begin{figure}[htb]
		\centering
		\includegraphics[width=1\linewidth]{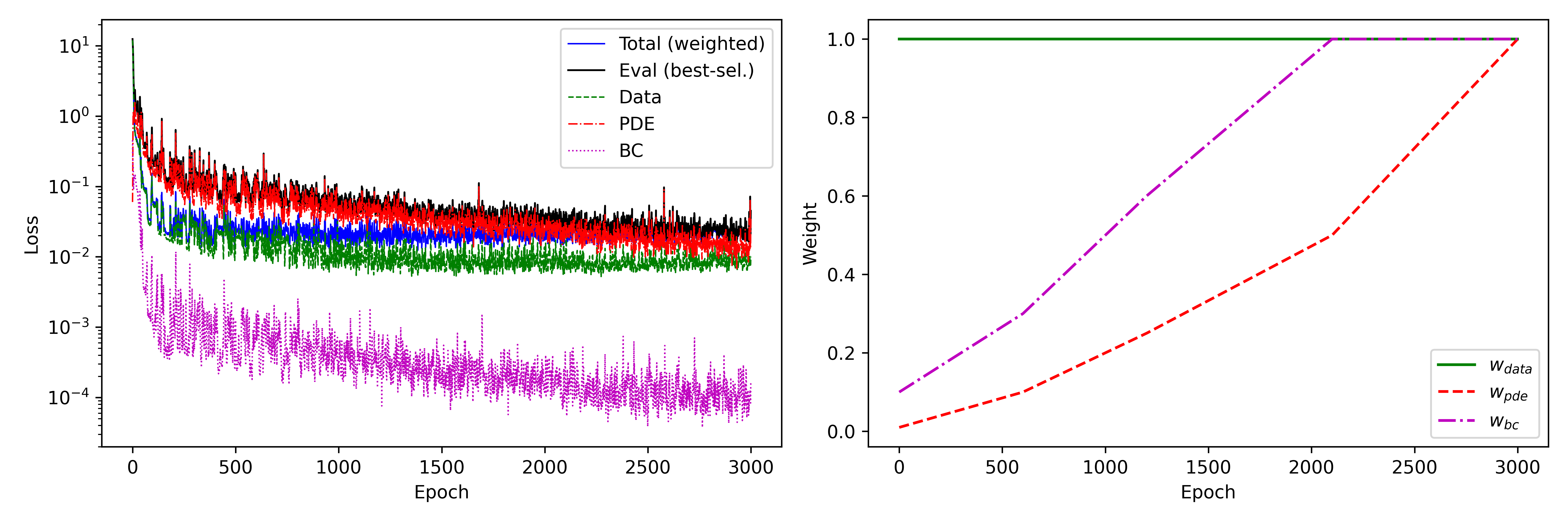}
		\caption{Training diagnostics of the PASSC framework for
			$M_{\infty}=8.0$; the layout is the same as in
			Figure~\ref{fig:ma2_training}.}
		\label{fig:ma8_training}
	\end{figure}

\begin{figure}[htb]
		\centering
		\includegraphics[width=1\linewidth]{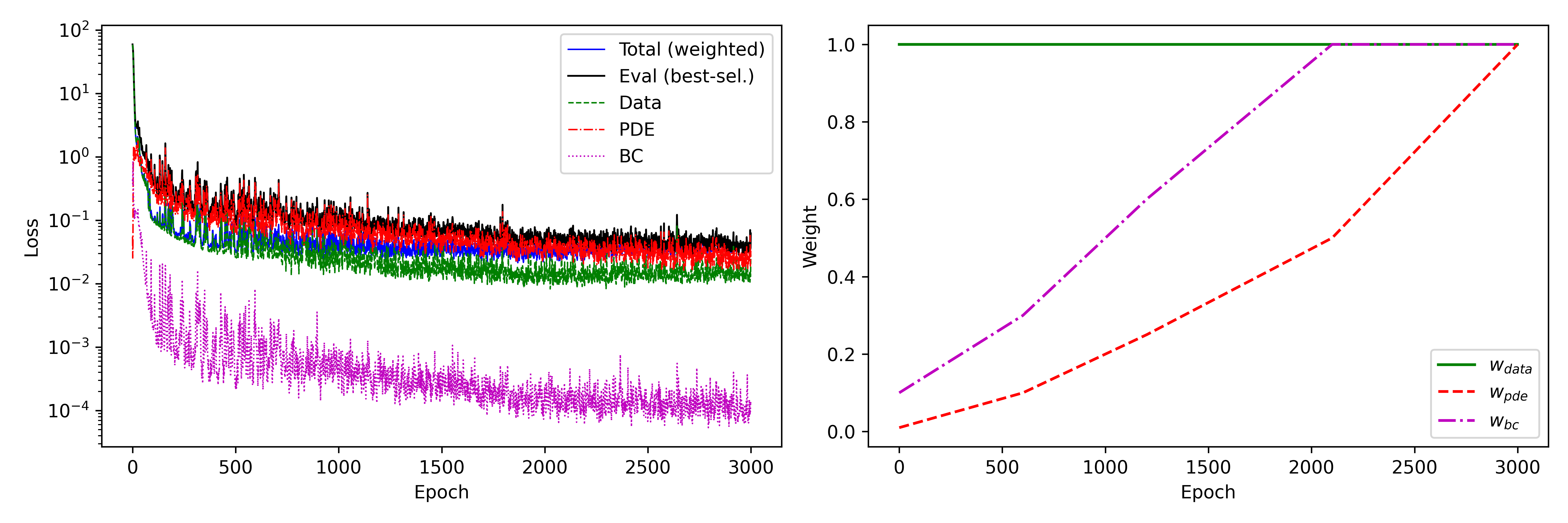}
		\caption{Training diagnostics of the PASSC framework for
			$M_{\infty}=12.0$; the layout is the same as in
			Figure~\ref{fig:ma2_training}.}
		\label{fig:ma12_training}
	\end{figure}

	Since the same network architecture, epoch budget, and loss-weight
	schedule are employed in all four cases, the training behavior of the
	correction network is discussed collectively.
	Figures~\ref{fig:ma2_training}, \ref{fig:ma5_training},
	\ref{fig:ma8_training}, and~\ref{fig:ma12_training} show, for each
	free-stream Mach number, the evolution of the weighted total loss, the
	unweighted evaluation metric of Eq.~\eqref{eq:eval_metric} used for model
	selection, and the data, PDE, and boundary-condition loss components over
	the $3{,}000$ training epochs, together with the continuously interpolated
	loss-weight schedule of Eq.~\eqref{eq:weight_schedule}. The data weight is
	held at $w_\text{data}=1$ throughout, while the physics and boundary weights
	are increased smoothly, without discontinuous jumps. Specifically,
	$w_\text{pde}$ grows from $0.01$ at the beginning of training to parity
	with the data weight at the end, while $w_\text{bc}$ reaches $1$ at
	$70\%$ of the epoch budget. The early part of the optimization therefore
	places greater emphasis on consistency with the finite element solution,
	after which the governing-equation and boundary-condition contributions
	are progressively strengthened. In contrast to a piecewise-constant staged
	schedule, the continuously interpolated weights avoid abrupt changes in the
	training objective.
	
	The overall qualitative behavior is similar in all four cases. The loss
	components decrease rapidly during the first few hundred epochs and
	subsequently evolve more gradually, with increasingly pronounced
	fluctuations in the physics component at the higher Mach numbers. The
	characteristic late-training levels increase with the strength of the
	captured shock. At $M_{\infty}=2.0$, the data loss levels off just below
	$10^{-3}$, the physics residual of Eq.~\eqref{eq:pde_total} decreases to
	approximately $2\times10^{-3}$, and the boundary-condition residual reaches
	the order of $10^{-5}$. At $M_{\infty}=5.0$, the corresponding levels are
	approximately $3\times10^{-3}$, $8\times10^{-3}$, and $10^{-4}$. At
	$M_{\infty}=8.0$, the data loss settles at approximately
	$8\times10^{-3}$ while the physics residual fluctuates in a band around
	$1.5\times10^{-2}$; at $M_{\infty}=12.0$, the data loss settles slightly
	above $10^{-2}$ and the physics residual fluctuates around
	$3\times10^{-2}$. The boundary-condition residual remains of the order of
	$10^{-4}$ in both of the latter cases.
	
	In every case, the physics residual constitutes the largest contribution
	to the unweighted evaluation metric of Eq.~\eqref{eq:eval_metric}, on which
	the retained model is selected. This behavior is expected because the
	control-volume balance need not vanish exactly within the finite-thickness
	numerical representation of a strong shock. Nevertheless, the physics
	residual decreases substantially over the course of training even as its
	loss weight increases by two orders of magnitude, from
	$w_\text{pde}=0.01$ to $w_\text{pde}=1$. This reduction indicates an
	increasingly physics-consistent correction and is consistent with the
	localized modification and smoothing of the shock structure observed in
	the flow-field comparisons.
	
	The spikes and band-like fluctuations that become more visible in the
	physics loss at $M_{\infty}=8.0$ and $M_{\infty}=12.0$ are associated
	primarily with control volumes sampling the immediate neighborhood of the
	increasingly strong and thin shock. At $M_{\infty}=12.0$, for example, the
	density ratio across the shock approaches the strong-shock limit of six, so
	different members of the multiscale stochastic control-volume batch probe
	slightly different portions of a very steep numerical transition. These
	fluctuations are therefore not, by themselves, evidence of unstable
	optimization. Their envelopes remain bounded and generally decrease, while
	the data loss remains near its converged level and the boundary-condition
	loss stays small. Taken together, these trends indicate that the network
	remains anchored to the stabilized finite element solution while continuing
	to reduce the physics residual.
	
	Across the four cases, the characteristic late-training data and physics
	loss levels increase substantially with the free-stream Mach number, by
	approximately an order of magnitude from $M_{\infty}=2.0$ to
	$M_{\infty}=12.0$. This trend is consistent with the increasing strength
	and decreasing numerical thickness of the captured shock. The
	boundary-condition contribution remains markedly smaller than the data and
	physics components throughout the Mach-number range. Importantly, the same
	loss-weight schedule is used without case-specific adjustment in all four
	computations. The resulting networks remain closely anchored to the
	stabilized finite element solutions while reducing the physics and
	boundary-condition residuals and producing the smoother flow fields
	documented in the preceding sections.
	
	\subsection{Computational cost}
	\label{sec:cost}
	
	Table~\ref{tab:cost} summarizes the wall-clock cost of the two stages of the
	proposed framework on the workstation described at the beginning of this
	section. For the finite element stage, the per-step wall-clock cost is measured
	directly for the $M_\infty=2.0$ case and used to estimate the corresponding
	costs at the higher Mach numbers, as detailed below.
	
	\begin{table}[ht]
		\centering
		\caption{Computational cost of the two stages of the PASSC framework for the
			four free-stream Mach numbers considered. The finite element (FEM) stage is
			executed on the CPU of the workstation described in Section~7, whereas the
			PASSC training stage runs on its GPU. For the FEM stage, the per-step
			wall-clock time is measured directly at $M_\infty=2.0$ using in-code
			instrumentation (Newton solves $\approx5.8$~s, stabilization-parameter
			updates $\approx1.3$~s, and diagnostics and I/O $\approx0.55$~s per step).
			The higher-Mach FEM costs are estimates obtained using the corresponding
			step counts and the per-step cost observed at $M_\infty=2.0$, with a small
			allowance for the observed variation in the number of Newton iterations.
			The reported step counts $N_{\varDelta t}$ correspond to the actual accepted
			time steps in the simulations; the adaptive strategy of Section~4 did not
			trigger any time-step reduction in any of the four cases.}
		\label{tab:cost}
		\begin{tabular}{lcccc}
			\toprule
			& $M_\infty = 2.0$ & $M_\infty = 5.0$ & $M_\infty = 8.0$ & $M_\infty = 12.0$ \\
			\midrule
			Courant number $C_{\varDelta t}$
			& 0.5 & 0.5 & 0.4 & 0.4 \\
			Nominal step $\varDelta t_0$ (s)
			& $7.723\times10^{-7}$ & $3.861\times10^{-7}$
			& $2.059\times10^{-7}$ & $1.426\times10^{-7}$ \\
			Number of time steps $N_{\varDelta t}$
			& 1{,}295 & 2{,}590 & 4{,}856 & 7{,}014 \\
			FEM wall time per step (s)
			& 7.6 & $\approx7.7$ & $\approx7.8$ & $\approx7.8$ \\
			FEM stage, total wall time (s)
			& $\approx9{,}840$ & $\approx19{,}900$
			& $\approx37{,}900$ & $\approx54{,}700$ \\
			PASSC training, total wall time (s)
			& $\approx4{,}200$ & $\approx4{,}200$
			& $\approx4{,}200$ & $\approx4{,}200$ \\
			\midrule
			Training-to-FEM cost ratio
			& 0.43 & 0.21 & 0.11 & 0.08 \\
			\bottomrule
		\end{tabular}
	\end{table}
	
	The PASSC training stage requires approximately $140$~s per $100$ epochs on
	the GPU, corresponding to about $4{,}200$~s ($\approx1.2$~h) for the full
	budget of $3{,}000$ epochs. Since the same network architecture, epoch budget,
	sampling strategy, and loss-weight schedule are used in all four cases, the
	observed training cost is approximately independent of the free-stream Mach
	number. A substantial portion of the per-epoch cost arises from the repeated
	network evaluations required by the space--time control-volume quadratures.
	As described in Section~\ref{sec:loss}, these edge and temporal-face
	quadratures are evaluated without computing spatial or temporal derivatives of
	the network outputs by automatic differentiation.
	
	The cost of the stabilized finite element stage, in contrast, increases with
	$M_\infty$ primarily through the CFL-based time-step size of
	Eq.~\eqref{dtnominal}. As the nominal step $\varDelta t_0$ decreases from
	$7.72\times10^{-7}$~s at $M_\infty=2.0$ to $1.43\times10^{-7}$~s at
	$M_\infty=12.0$, the number of time steps required to reach
	$t_\text{f}=10^{-3}$~s increases from about $1{,}300$ to about $7{,}000$.
	The solver logs show that, after the initial transient, each accepted step
	typically requires $3$--$4$ Newton iterations. In-code wall-clock
	instrumentation for the $M_\infty=2.0$ computation gives an average cost of
	approximately $7.6$~s per accepted step, of which about $76\%$ is spent in
	the Newton solves, dominated by the MUMPS factorization of the
	$95{,}736\times95{,}736$ Jacobian at each nonlinear iteration, about $17\%$
	in the per-step update of the frozen stabilization and shock-capturing data,
	and the remainder in diagnostic projections and output.
	
	Because these dominant costs are determined primarily by the fixed mesh and
	linear-system size, the measured $M_\infty=2.0$ per-step cost provides the
	basis for estimating the FEM wall-clock costs at the higher Mach numbers.
	A small allowance is included for the observed variation between three and
	four Newton iterations per accepted step. The resulting higher-Mach FEM
	wall-clock times in Table~\ref{tab:cost} should therefore be interpreted as
	estimates rather than as independent timing measurements.
	
	The adaptive time-stepping safeguards of Section~4 were not activated in any
	of the four computations: no adaptive time-step reduction or Newton-relaxation
	fallback was required, including in the $M_\infty=12.0$ case. Apart from any
	final adjustment required to terminate exactly at $t_\text{f}$, the accepted
	steps therefore use the nominal CFL-based time-step size. The step counts
	reported in Table~\ref{tab:cost} correspond to the actual simulations. Based
	on the estimated FEM costs, the relative wall-clock cost of the
	physics-informed correction decreases from approximately $43\%$ of the FEM
	cost at $M_\infty=2.0$ to approximately $8\%$ at $M_\infty=12.0$. Thus,
	within the present implementation and hardware configuration, the relative
	wall-clock overhead of the correction decreases substantially as the
	CFL-driven cost of the finite element stage increases.
	
	We emphasize that the two stages run on different processing units (CPU and
	GPU, respectively); hence, the reported ratios compare wall-clock times on
	the same workstation rather than intrinsic algorithmic complexities. Once
	trained, the network can be evaluated directly at arbitrary space--time
	locations at comparatively low inference cost, providing a continuous,
	mesh-independent representation of the corrected field without requiring an
	additional finite element solve.
	
	Finally, the computational times reported here are specific to the hardware
	and implementation used in the present study and should therefore be
	interpreted primarily as an indication of the relative wall-clock cost of the
	two stages rather than as hardware-independent performance measures.

	\section{Concluding remarks and future work}
	\label{sec7}
	
	We have presented a hybrid stabilized FEM--PINN framework for the
	simulation of inviscid supersonic and hypersonic flows around a circular
	cylinder. The working fluid is modeled as chemically nonreactive nitrogen.
	The compressible-flow SUPG formulation is complemented with the
	residual-based YZ$\beta$ shock-capturing technique, and the resulting
	stabilized finite element solution is subsequently enhanced through the
	PASSC correction stage. The neural network is anchored to the stabilized
	solution through a shock-weighted data-consistency loss while the governing
	equations are enforced in a space--time control-volume form, supplemented
	by macroscopic conservation windows, an entropy-admissibility penalty, and
	the boundary conditions of the underlying problem.
	
	The numerical results show that the PASSC solution remains closely anchored
	to the stabilized finite element approximation over the entire range
	$M_{\infty}=2.0$--$12.0$, while substantially reducing the residual
	oscillatory features associated with the underlying discretization and
	providing a smooth, continuous representation of the corrected flow field.
	The detected bow-shock position changes by no more than approximately one
	to two near-wall element layers in the cases considered. At
	$M_{\infty}=2.0$, where the stabilized solution underpredicts the Billig
	stand-off distance by approximately $17\%$---a discrepancy attributable
	primarily to the smaller number of flow-through times elapsed at the common
	final simulation time, so that this case remains the farthest from its
	steady state---the correction recovers roughly one quarter of this
	discrepancy. At the three hypersonic Mach numbers, the
	stabilized stand-off distances already lie within approximately $2\%$ of
	the Billig correlation, and the PASSC solutions introduce only small
	absolute changes in the detected shock position, with the resulting
	stand-off distances remaining within approximately $7\%$ of the
	correlation. The stagnation-pressure ratios obtained with both approaches
	remain within approximately $2.5\%$ of the corresponding Rayleigh pitot
	values for all four Mach numbers. The computed shock and stagnation-region
	quantities are also broadly consistent with the theoretical normal-shock
	and stagnation relations and with previously reported solutions for
	comparable configurations.
	
	The training histories further show that the same network architecture and
	loss-weight schedule can be employed over the complete Mach-number range
	without case-specific adjustment. For the fixed architecture and epoch
	budget used here, the observed PASSC training time is approximately
	independent of the Mach number. On the hardware employed in this study, the
	training wall-clock time corresponds to approximately $43\%$ of the measured
	finite element cost at $M_{\infty}=2.0$ and approximately $8\%$ of the
	estimated finite element cost at $M_{\infty}=12.0$. Once trained, the network
	provides a continuous, mesh-independent representation that can be evaluated
	directly at arbitrary space--time locations without requiring an additional
	finite element solve.
	
	Despite these results, several limitations should be acknowledged. The
	correction stage contains a number of user-defined quantities, including
	the loss weights and their schedule, the network architecture and
	optimization parameters, the Fourier-feature scale, the residual saturation
	threshold, the control-volume sampling and quadrature parameters, the
	scale-balancing parameters, and the entropy and anchor-snapshot settings.
	The underlying stabilized formulation likewise involves the
	shock-capturing parameters $C_{\mathrm{YZ}\beta}$ and $\beta$. These
	quantities were selected through numerical experimentation rather than a
	systematic optimization or sensitivity analysis. In addition, the corrected
	solution is assessed primarily relative to the stabilized finite element
	approximation and analytical or empirical scalar reference quantities,
	rather than against an independent high-resolution numerical solution.
	Consequently, the reported pointwise differences characterize the
	modification introduced by the PASSC stage but should not, by themselves,
	be interpreted as direct measures of an accuracy improvement.
	The present coupling is also one-directional and offline, and the numerical
	experiments are restricted to two-dimensional inviscid flow around a single
	geometry. Furthermore, the calorically perfect, single-species gas model
	does not account for the high-temperature thermochemical processes that
	become increasingly important in the higher-Mach-number cases. This
	limitation is particularly significant at $M_{\infty}=12.0$, for which the
	predicted stagnation-region temperatures approach $9{,}000$~K. The
	highest-Mach-number results should therefore be interpreted primarily as
	demanding numerical tests of the proposed correction methodology rather
	than as quantitatively complete descriptions of the corresponding
	high-temperature physical flows.
	
	These limitations motivate several directions for future work. A primary
	extension is the treatment of multicomponent reacting mixtures in
	thermochemical nonequilibrium, including temperature-dependent
	thermodynamic properties, vibrational excitation, finite-rate chemistry,
	and dissociation. Extension to the compressible Navier--Stokes equations,
	three-dimensional configurations, turbulent high-speed flows, adaptive
	mesh refinement, and alternative spatial discretizations such as
	isogeometric analysis will further establish the behavior of the framework
	in more realistic and computationally demanding settings.
	
	From the scientific machine learning perspective, an important direction is
	the systematic reduction of manually prescribed parameters. Self-adaptive,
	gradient-based, neural-tangent-kernel-based, multi-objective, or
	uncertainty-based loss-balancing strategies could replace the present
	manually prescribed weight schedule. Similarly, the network architecture,
	Fourier-feature parameters, optimization settings, and selected parameters
	of the stabilized formulation could be determined through systematic
	hyperparameter optimization rather than numerical experimentation. Adaptive
	and residual-guided control-volume sampling may also reduce the computational
	effort required to enforce the physics in the vicinity of strong shocks.
	
	Another important direction concerns the form in which the physical
	constraints are imposed. The present framework enforces conservation and
	entropy admissibility through soft control-volume penalties. Future
	formulations could investigate discrete conservation, positivity, and
	entropy consistency imposed more directly by construction, as well as
	richer variational test spaces beyond the piecewise-constant control
	volumes employed here. A systematic comparison against independently
	resolved reference solutions is also needed to quantify the accuracy--cost
	trade-off between PASSC correction and conventional mesh refinement,
	preferably together with repeated training runs to characterize the
	statistical variability of the learned correction. Finally, parametric
	networks covering a range of free-stream conditions and iterative
	FEM--PINN coupling, in which information is exchanged between the two
	stages rather than only from FEM to PINN, constitute natural extensions of
	the present methodology.
	
	Overall, the results demonstrate that a physics-informed correction can be
	combined with stabilized finite element shock capturing without replacing
	the underlying numerical solver or sacrificing its robust global flow
	structure. Within the scope of the present two-dimensional inviscid tests,
	the PASSC framework reduces discretization-induced oscillatory features,
	preserves the principal shock and stagnation-region characteristics, and
	provides a continuous corrected representation while remaining closely
	anchored to the stabilized finite element solution. In line with the
	central objective of the framework, the correction improves the shock
	representation---removing the localized zigzagging, speckle, and serration
	introduced by the discretization---at no measurable cost in the global
	scalar accuracy metrics, which remain within the same few-percent band of
	the analytical and empirical references as the stabilized solution and, in
	several instances (the stagnation-pressure ratio at $M_{\infty}=5.0$ and
	$12.0$ and the detected stand-off distance at $M_{\infty}=2.0$), move
	closer to them. These findings provide
	a basis for extending the methodology to more general
	convection-dominated and high-speed compressible-flow problems.
	
	\section*{Acknowledgments}
	The finite element component of this work is based primarily on the first
	author's Ph.D. dissertation~\cite{Cengizci_thesis}, submitted to the
	Institute of Applied Mathematics, Middle East Technical University, in 2022.
	The PINN component was developed under Grant No.~225M468 from the Scientific
	and Technological Research Council of Turkey (T\"{U}B\.{I}TAK), whose
	financial support the authors gratefully acknowledge.

	\section*{Author Declarations}
	
	\subsection*{Conflict of interest}
	The authors declare that they have no conflict of interest.
	
	\subsection*{Data and code availability}
	The code, mesh, configuration files, and post-processing scripts supporting the
	findings of this study are openly available at
	\url{https://github.com/scengizci/equilibrium_hypersonic}.

	\subsection*{Author contributions}
	\textbf{Süleyman Cengizci}: Conceptualization (equal); Formal analysis (equal); Investigation (equal); Methodology (equal); Software (lead); Validation (lead); Visualization (lead); Writing -- original draft (lead); Writing -- review \& editing (equal). \textbf{Ömür Uğur}: Conceptualization (equal); Formal analysis (equal); Investigation (equal); Methodology (equal); Software (supporting); Supervision (lead); Writing -- review \& editing (equal).

	\bibliographystyle{elsarticle-num-names}
	\bibliography{myBiblio}
	
\end{document}